\documentclass[useamsfonts]{pasj01}
\usepackage{natbib}
\usepackage{aas_macros}
\usepackage{url}
\usepackage{bm}
\usepackage{lineno}

\newcommand{\simgt}{\lower.5ex\hbox{$\; \buildrel > \over \sim \;$}}
\newcommand{\simlt}{\lower.5ex\hbox{$\; \buildrel < \over \sim \;$}}

\begin{document}
\SetRunningHead{T. Hamana et al.}{Cosmological constraints from cosmic shear
  COSEBIs with HSC survey 1-year data}
\Received{2022/1/30}
\Accepted{2022/5/27}
\Published{2022/6/17}

\title{E/B mode decomposition of HSC-Y1 cosmic shear
  using COSEBIs: cosmological constraints and comparison with other
  two-point statistics}

%
%
\author{Takashi \textsc{Hamana}\altaffilmark{1}}
\author{Chiaki \textsc{Hikage}\altaffilmark{2}}
\author{Masamune \textsc{Oguri}\altaffilmark{3,4,2}}
\author{Masato \textsc{Shirasaki}\altaffilmark{1,5}}
\author{Surhud \textsc{More}\altaffilmark{6,2}}
\altaffiltext{1}{National Astronomical Observatory of Japan, Mitaka,
  Tokyo 181-8588, Japan}
\altaffiltext{2}{Kavli Institute for the Physics and Mathematics of the Universe (Kavli IPMU, WPI), University of Tokyo, Chiba 277-8582, Japan}
\altaffiltext{3}{Research Center for the Early Universe, University of Tokyo, Tokyo 113-0033, Japan}
\altaffiltext{4}{Center for Frontier Science, Chiba University, 1-33 Yayoi-cho, Inage-ku, Chiba 263-8522, Japan}
\altaffiltext{5}{The Institute of Statistical Mathematics, Tachikawa, Tokyo 190-8562, Japan}
\altaffiltext{6}{The Inter-University Center for Astronomy and Astrophysics, Post bag 4, Ganeshkhind, Pune, 411007, India}
%
%


\maketitle

\begin{abstract}
We perform a cosmic shear analysis of Hyper Suprime-Cam Subaru Strategic
Program first-year data (HSC-Y1) using Complete Orthogonal Sets of
E/B-Integrals (COSEBIs) to derive cosmological constraints.
We compute E/B-mode COSEBIs from cosmic shear two-point
correlation functions measured on an angular range of
$4\arcmin<\theta<180\arcmin$. 
We perform the standard Bayesian likelihood analysis for 
cosmological inference from the measured E-mode COSEBIs, including 
contributions from intrinsic alignments of galaxies as well as 
systematic effects from point spread function model errors, shear
calibration uncertainties,
and source redshift distribution errors.
We adopt a covariance matrix derived from realistic mock catalogs
constructed from full-sky gravitational lensing simulations that fully
take account of the survey geometry and measurement noise.
For a flat $\Lambda$ cold dark matter model, we find 
$S_8 \equiv \sigma_8\sqrt{\Omega_m/0.3}=0.809_{-0.026}^{+0.036}$.
We carefully check the robustness of the cosmological results
against astrophysical modeling uncertainties and systematic
uncertainties in measurements, and find that none of them has 
a significant impact on the cosmological constraints.
We also find that the measured B-mode COSEBIs are
consistent with zero.
We examine, using mock HSC-Y1 data, the consistency of our $S_8$
constraints with those derived from the other cosmic shear two-point
statistics, the power spectrum
analysis by Hikage et al (2019) and the two-point correlation function
analysis by Hamana et al (2020), which adopt the same HSC-Y1 shape catalog, 
and find that all the $S_8$ constraints are consistent with each other,
although expected
correlations between derived $S_8$ constraints are weak.
\end{abstract}

%
%
\section{Introduction}
\label{sec:intro}
The cosmic shear is measured from the coherent distortion of the shapes
of distant galaxies caused by gravitational lensing of intervening large-scale
structures, and is one of the most powerful tools for cosmology, 
as it provides a unique means of studying the matter distribution in the
Universe, including the dark matter component.
Statistical measures of cosmic shear, such as 
the two-point correlation function (TPCF) or the power
spectrum (PS), depend both on the time evolution of the cosmic
structure and on the cosmic expansion history at relatively recent
epochs ($z < 1$), and thus serve as a unique late-time cosmological probe. 
Aiming to place useful constraints on cosmological parameters
independently from early-time probes, currently three weak lensing projects with
wide-field imaging surveys are underway; the Dark Energy Survey 
\citep[DES,][]{2016MNRAS.460.1270D}, the Kilo-Degree 
survey \citep[KiDS,][]{2013ExA....35...25D}, and the 
Hyper Suprime-Cam Subaru Strategic Program 
\citep[hereafter the HSC survey;][]{2018PASJ...70S...4A}.
All the three projects have published initial/mid-term cosmic shear results
from early data, yielding 
constraints with 2-8 percent precision
on $S_8 = \sigma_8 (\Omega_m/0.3)^{0.5}$, where $\sigma_8$ is
the amplitude of matter fluctuations in scales of 8Mpc$/h$
and $\Omega_m$ is the mean matter density parameter
\citep{2018PhRvD..98d3528T, 2022PhRvD.105b3514A, 2017MNRAS.465.1454H,
  2017MNRAS.471.4412K, 2020A&A...633A..69H, 2021A&A...645A.104A,
  2019PASJ...71...43H, 2020PASJ...72...16H}.

Cosmic shear analyses with 
the HSC survey first-year weak lensing
shape catalog \citep[HSC-Y1;][]{2018PASJ...70S...8A,2018PASJ...70S..25M}
were conducted in Fourier space using the power spectra
\citep[PS; ][hereafter H19]{2019PASJ...71...43H}, and in
configuration space using the two-point
correlation functions \citep[TPCF; ][hereafter H20]{2020PASJ...72...16H}.
Although those two analyses are both based on two-point statistics with
the same data set, they are complementary to each other for the following
three reasons, among others.
Firstly, masking and finite field effects affect PS/TPCF analyses in
different ways:
The Fourier space measurements are directly affected by those effects,
although a technique aimed at correcting those effects
(the pseudo-$C_l$ method) was
adopted in H19. The pseudo-$C_l$ method was also adopted in
\citet{2021arXiv211107203C}, while \citet{2017MNRAS.471.4412K} adopted the
quadratic estimator.
On the other hand, the configuration space measurements are not affected
by the effects, although the effects need to be taken into account in
their covariance estimation. 
Secondly, E/B-mode decomposition of cosmic shear is properly conducted in
Fourier space PS measurements, while is not
feasible for configuration space TPCFs as it requires a measurement of TPCFs with
angular separations ranging from zero to infinity.
This is a disadvantage of TPCF analyses for the following reason:
gravitational lensing by a scalar gravitational field generates
only E-mode shear at a leading order, and B-mode signal generated by
lensing is expected to be very small unless there are any systematics in
the measurements \citep[see ][for a review and references
  therein]{2015RPPh...78h6901K}.
Therefore, null tests of B-mode signals can be used to verify a standard
assumption of a scalar gravitational field, and to check systematics in
cosmic shear measurements.
Thus, a clean E/B-mode decomposition is of vital importance for
cosmological analyses.
Finally, the PS and TPCF probe different multipole ranges, as the kernels
linking the TPCFs to the PS have very broad shapes.
The HSC-Y1 PS analysis by H19 adopted the multipole range of
$300<\ell<1900$, whereas a large part of the contribution to TPCFs on 
a range of angular separations adopted in H20
comes from $\ell < 300$.
Constraints on $S_8$ derived from those two studies appear to differ
significantly ($S_8=0.780_{-0.033}^{+0.030}$ for PS, whereas
$S_8=0.823_{-0.028}^{+0.032}$ for TPCF).
In order to check whether the difference can be explained simply by a
statistical fluctuation, a cross correlation analysis of realistic HSC
mock catalogs was conducted:
H19 and H20 performed the cosmological inference on the same 100 mock catalogs
with softwares used in their cosmological analyses on the real
HSC-Y1 data, and derived constraints were used as a statistical sample
of cross-correlation analysis of differences in derived $S_8$ values
between PS and TPCF analyses.
Then, it was found that the differences in derived $S_8$ was explained by
a statistical fluctuation at $\sim 1.6\sigma$ level, and thus those
$S_8$ constraints are consistent with each other (H20, and see section
\ref{sec:comparison_Hikage}).

Complete Orthogonal Sets of E/B-Integrals
\citep[COSEBIs:][]{2010A&A...520A.116S} are an another
configuration space measure of cosmic shear two-point statistics.
COSEBIs are complementary to TPCF in the sense that 
they eliminate
the two disadvantages of the TPCF \citep[see][for a closely related
  discussion]{2021A&A...645A.104A}: 
First, COSEBIs are defined by an integration of TPCFs weighted by kernel
functions over a finite angular separations, and
COSEBIs' kernels linking to the PS have more compact shapes in
multipole space than TPCF's kernels. 
Accordingly, it offers a better control over multipole ranges
contributing to COSEBIs signals. 
Notably, COSEBIs are less sensitive
to low-multipoles where only a few independent modes are available from
finite-area survey data.
Second, COSEBIs enable a clean E/B-mode separation over a range of
finite angular separations available from a finite survey area.

In this paper, we present a cosmic shear analysis of the HSC-Y1 weak lensing
shape catalog using E/B-mode COSEBIs.
We use the same data set as that used in H19 and H20 along with
the same tomographic
redshift binning (four bins with $0.3<z<0.6$, $0.6<z<0.9$, $0.9<z<1.2$
and $1.2<z<1.5$). 
We also use realistic HSC-Y1 mock catalogs constructed
from full-sky gravitational lensing simulations
\citep{2017ApJ...850...24T} with fully taking account of the survey geometry
and measurement noise \citep{2019MNRAS.486...52S}, from which we derive
E/B-mode covariance matrices.

The structure of this paper is as follows.
In Section~\ref{sec:data}, we briefly summarize the HSC-Y1 shear catalog
and the photometric redshift data used in this study.
In Section~\ref{sec:theory}, 
we present an overview of the
theoretical modeling
of COSEBIs and define a scale-cut adopted in our analyses. 
In Section~\ref{sec:measurements}, we describe the method to measure the
cosmic shear COSEBIs, and present our measurements.
We also present covariance matrices derived from the realistic HSC-Y1
mock shape catalogs. The result of B-mode null test is also presented.
In Section~\ref{sec:analyses}, we describe a method for cosmological
inference along with methods to take into account 
various systematics in our cosmological analysis, for which we closely
follow the analysis framework adopted in H20.
In Section~\ref{sec:results}, we present results of our cosmological
constraints and tests for systematics.
We compare our cosmological constraints with other cosmic shear results
and the {\it Planck} cosmic microwave background (CMB) result.
We also compare our results with those from the HSC-Y1 cosmic shear PS
and TPCF analyses, and examine the consistency among them using mock HSC-Y1
data.
Finally, we summarize and discuss our results in
Section~\ref{sec:summary}.
In Appendix~\ref{sec:cosebis_systematics}, we examine COSEBIs expected
from errors in measurements and modeling of the point spread functions
(PSFs), and from constant shears over survey fields,
and describe empirical models for those systematics.
In Appendix~\ref{sec:supplementary_figures}, we present supplementary
figures.

Throughout this paper we quote 68\% credible intervals for 
parameter uncertainties unless otherwise stated.

%
%
\section{HSC-Y1 data set}
\label{sec:data}

We use the HSC-Y1 data set that is exactly the same as the one used in
H19 and H20, and thus here we focus on aspects that are directly relevant
to this study.
We refer the readers to those two papers and references therein for
details.

We use the HSC first-year shape catalog \citep{2018PASJ...70S..25M},
which covers 136.9 deg$^2$, and contains $\sim$12.1M galaxies 
that pass selection criteria; among others, the four major criteria are, 
\begin{enumerate}
\renewcommand{\labelenumi}{(\arabic{enumi})}
\item {\it full-color and full-depth cut}: the object should be located in regions reaching 
approximately full survey depth in all five ($grizy$) broad bands,
\item {\it magnitude cut}: $i$-band cmodel magnitude (corrected for
  extinction) should be brighter than 24.5 AB mag, 
\item {\it resolution cut}: the galaxy size normalized by the PSF size
  defined by the re-Gaussianization method should be larger than a given
  threshold of {\tt ishape\_hsm\_regauss\_resolution $\ge$ 0.3}, 
\item {\it bright object mask cut}: the object should 
not be located within the bright object masks.
\end{enumerate}
See Table~4 of \citet{2018PASJ...70S..25M} for the full description of
the selection criteria.
The shapes of galaxies are estimated on the $i$-band coadded image
using the re-Gaussianization PSF correction method 
\citep{2003MNRAS.343..459H}.
Following Appendix~A of \citet{2018PASJ...70S..25M}, 
an estimator for the shear, denoted by $\bm{\hat{\gamma}} =
(\hat{\gamma}_1, \hat{\gamma}_2)$, is obtained for each galaxy as 
\begin{equation}
  \label{eq:gamma-e}
        \hat{\gamma}_i = {1\over {1+\bar{m}}}
        \left[ {{e_i}\over{2 \cal{R}}}-c_i\right], 
\end{equation}
where $e_i$ is the two-component distortion representing the shape of
each galaxy image, $m$ is multiplicative bias, and $c_i$ is additive bias.
The multiplicative and additive biases of individual galaxy shapes are
estimated using simulations of HSC images of the Hubble Space
Telescope COSMOS galaxy sample \citep{2018PASJ...70S..25M}.
In addition, we 
take into account
of two additional multiplicative biases
arising from the tomographic redshift galaxy selection (to be specific
$m_{\rm sel}$ and $m_{\rm R}$ in H19's terminology) in the same manner
used in H19 (see their section 5.7).

We use photometric redshift (hereafter photo-$z$) information to divide
galaxies into tomographic redshift bins \citep[see][for details of
  photo-$z$'s of HSC-Y1 data]{2018PASJ...70S...9T}.
To do this, we adopt the same procedure as one adopted in H19 and H20:
Adopting the {\tt best} estimate of a neural network code {\tt Ephor AB}
\citep{2018PASJ...70S...9T}
for the point estimator of photo-$z$'s, we select galaxies with
$0.3 < z_{\rm best} < 1.5$, and divide them into four 
tomographic redshift bins with equal redshift width of $\Delta z =0.3$.
We choose the above redshift range because photo-$z$ estimations
  with the HSC-Y1 data are most accurate in that redshift range
  \citep{2018PASJ...70S...9T}.
The final numbers of galaxies are 3.0M, 3.0M, 2.3M, and 1.3M galaxies
respectively from the lowest to highest redshift bins.
We adopt the redshift distributions of galaxies in individual bins
derived with the reweighting method based on the HSC's five-band 
photometry and COSMOS 30-band photo-$z$ catalog
\citep{2009ApJ...690.1236I,2016ApJS..224...24L}.
See Figure~1 of H20 for derived redshift distributions.
We refer the readers to Section~5.2 of H19 and
references therein for 
full details of the method.

%
%
\section{Theoretical models}
\label{sec:theory}

\subsection{COSEBIs} 
\label{sec:theory_cosebis}

The E/B-mode COSEBIs are defined as integrals over TPCFs on a finite
range of angular separations $\theta_{\rm min}<\theta <\theta_{\rm max}$
\citep{2010A&A...520A.116S},
\begin{eqnarray}
\label{eq:cosebis-e}
E_n^{ab} &=& {1\over 2} 
\int_{\theta_{\rm min}}^{\theta_{\rm max}} 
d\theta~\theta \left[ 
T_{+n}(\theta) \xi_+^{ab}(\theta) + T_{-n}(\theta) \xi_-^{ab}(\theta) 
\right],\\
\label{eq:cosebis-b}
B_n^{ab} &=& {1\over 2} 
\int_{\theta_{\rm min}}^{\theta_{\rm max}} 
d\theta~\theta \left[ 
T_{+n}(\theta) \xi_+^{ab}(\theta) - T_{-n}(\theta) \xi_-^{ab}(\theta) 
\right],
\end{eqnarray}
where $\xi_{\pm}^{ab}(\theta)$ are TPCFs for two tomographic redshift
bins $a$ and $b$, a natural number $n$, starting from 1, is the order of
COSEBIs modes, 
and $T_{\pm n}$ are the COSEBIs filter functions.
Alternatively, the E/B-mode COSEBIs can be expressed as
a function of E/B-mode cosmic shear PS denoted by $P_{E/B}^{ab}(\ell)$,
\begin{eqnarray}
\label{eq:cosebis-ps}
E_n^{ab} &=&
{1\over {2\pi}}\int_0^\infty
d \ell~\ell P_E^{ab}(\ell) W_n(\ell),\\
B_n^{ab} &=& 
{1\over {2\pi}}\int_0^\infty
d \ell~\ell P_B^{ab}(\ell) W_n(\ell),
\end{eqnarray}
where $W_n(\ell)$ are the Hankel transform of $T_{\pm n}(\theta)$,
\begin{eqnarray}
\label{eq:wn}
W_n &=&
\int_{\theta_{\rm min}}^{\theta_{\rm max}} 
d\theta~\theta 
T_{+n}(\theta) J_0(\ell \theta), \nonumber \\
{} &=&
\int_{\theta_{\rm min}}^{\theta_{\rm max}} 
d\theta~\theta 
T_{-n}(\theta) J_4(\ell \theta),
\end{eqnarray}
with $J_{0/4}$ being the zeroth- or fourth-order
Bessel functions of the first kind.
Note that TPCFs are also related to power spectra but mix E- and B-modes,
\begin{equation}
\label{eq:xipm}
\xi_{\pm}^{ab}(\theta) = {1\over {2\pi}}\int_0^\infty
d\ell~\ell J_{0/4} (\ell \theta)
\left[
P_E^{ab}(\ell) \pm P_B^{ab}(\ell)
\right].
\end{equation}

\citet{2010A&A...520A.116S} introduced two sets of COSEBIs filter
functions, the Lin- and Log-COSEBIs, which are given in terms of
polynomials in $\theta$ and $\ln \theta$,
respectively \citep[see section 3 of
  ][for the explicit equations for those filter functions]{2010A&A...520A.116S}.
It is found in \citet{2010A&A...520A.116S} that the Log-COSEBIs require
the first five modes to essentially capture all the cosmological
information, whereas the Lin-COSEBIs require $\sim 25$ modes to capture
the equivalent information.
We adopt the Log-COSEBIs, as they require much fewer modes compared to
Lin-COSEBIs.
Following previous studies
\citep{2020A&A...634A.127A,2021A&A...645A.104A}, we use the first five
COSEBIs modes (i.e., $1\leq n\leq 5$) in the following analyses.

\subsection{Cosmic shear power spectra} 
\label{sec:powerspectra}

In this study, we consider the standard cold dark matter (CDM) cosmological
model and thus we assume no B-mode shear generated by gravitational lensing.
Note that we find the B-mode COSEBIs measured from the HSC-Y1 are
consistent with zero as shown in the latter
section~\ref{sec:bmode_nulltest} (see also H19 for the power spectrum
analysis of the B-mode shear).
Therefore, for theoretical model computations in this study, we set
$P_B^{ab}(\ell) = 0$. 

In computing theoretical models of cosmic shear power spectra, we
follow the framework adopted in H20, and we
refer the readers to the section 4 of the paper and references therein
for expressions and details.
In short, for 
the computation of the shear-shear
power spectrum (also known as the convergence power spectrum),
$P_{\rm GG}^{ab}(\ell)$, we adopt the standard expression which relates it to the
matter power spectrum.
For the computation of the linear matter power spectrum, we use {\tt CAMB}
\citep{2011PhRvD..84d3516C}. 
In order to model the nonlinear matter power spectrum, we 
employ the fitting function by \citet{2012MNRAS.420.2551B}, 
which is based on the {\tt halofit} model
\citep{2003MNRAS.341.1311S,2012ApJ...761..152T} but is modified 
so as to include the effect of non-zero neutrino mass.
It is well known that the evolution of the 
nonlinear matter power spectrum, especially on small scales, 
is affected by baryon physics such as 
gas cooling, star formation, 
and supernova and active galactic nuclei (AGN) feedbacks
\citep{2010MNRAS.402.1536S,2011MNRAS.415.3649V,2015MNRAS.454.1958M,2016MNRAS.461L..11H,2017MNRAS.465.2936M,2018MNRAS.475..676S,2018MNRAS.480.3962C}.
When we evaluate baryonic feedback effects on our analyses, we adopt an
extreme model, the AGN feedback model by \citet{2015MNRAS.450.1212H}
that is based on the OverWhelming Large Simulations 
\citep{2010MNRAS.402.1536S,2011MNRAS.415.3649V}. 
Our treatment of baryon feedback effects in our cosmological
analyses is described in Section~\ref{sec:astrophysical_parameters}.

\subsection{Scale-cut} 
\label{sec:scalecut}

We set the scale-cut of COSEBIs (see eq.~\ref{eq:cosebis-e}) as
$\theta_{\rm min} =4\arcmin$ and $\theta_{\rm max} =180\arcmin$, for the
following reasons.
First, to determine the minimum scale, we impose a requirement that changes
of E-mode COSEBIs due to baryon feedback effects are less than 2~percents.
In evaluating it, we adopt the AGN feedback model by
\citet{2015MNRAS.450.1212H} as an extreme model, and find that
$\theta_{\rm min} >4\arcmin$ meets the requirement.
Second, for the large-scale cut, we follow the HSC-Y1 TPCF analysis
(H20), and set $\theta_{\rm max}=180\arcmin$, 
which is the largest angular scale used in H20
determined based on
  the condition that the signal-to-noise ratio per individual angular
  bin of the measured TPCFs is greater than 1 (see section 5.1 of H20).

%
%
\begin{figure*}
\begin{center}
  \includegraphics[width=82mm]{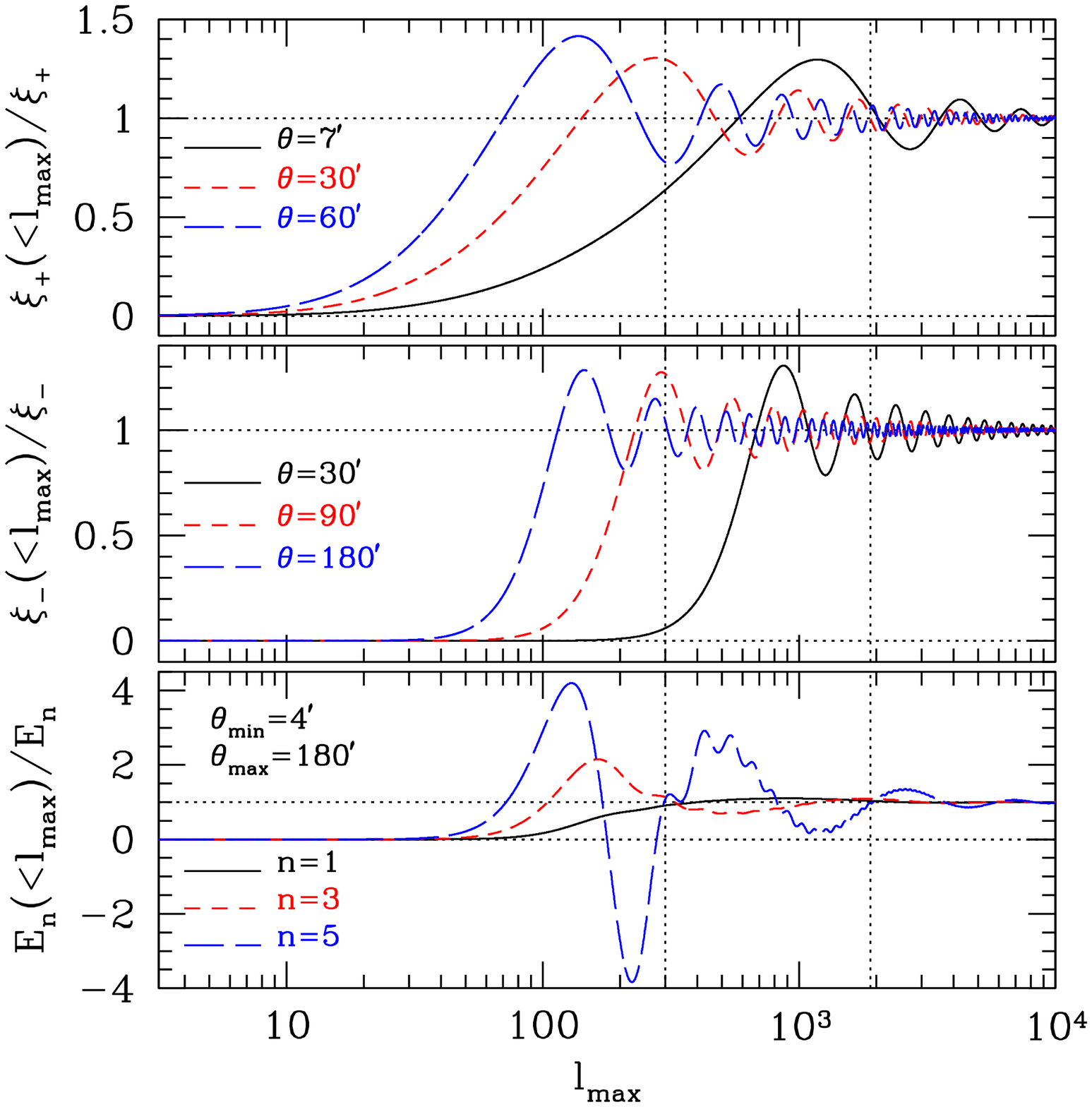}
  \includegraphics[width=82mm]{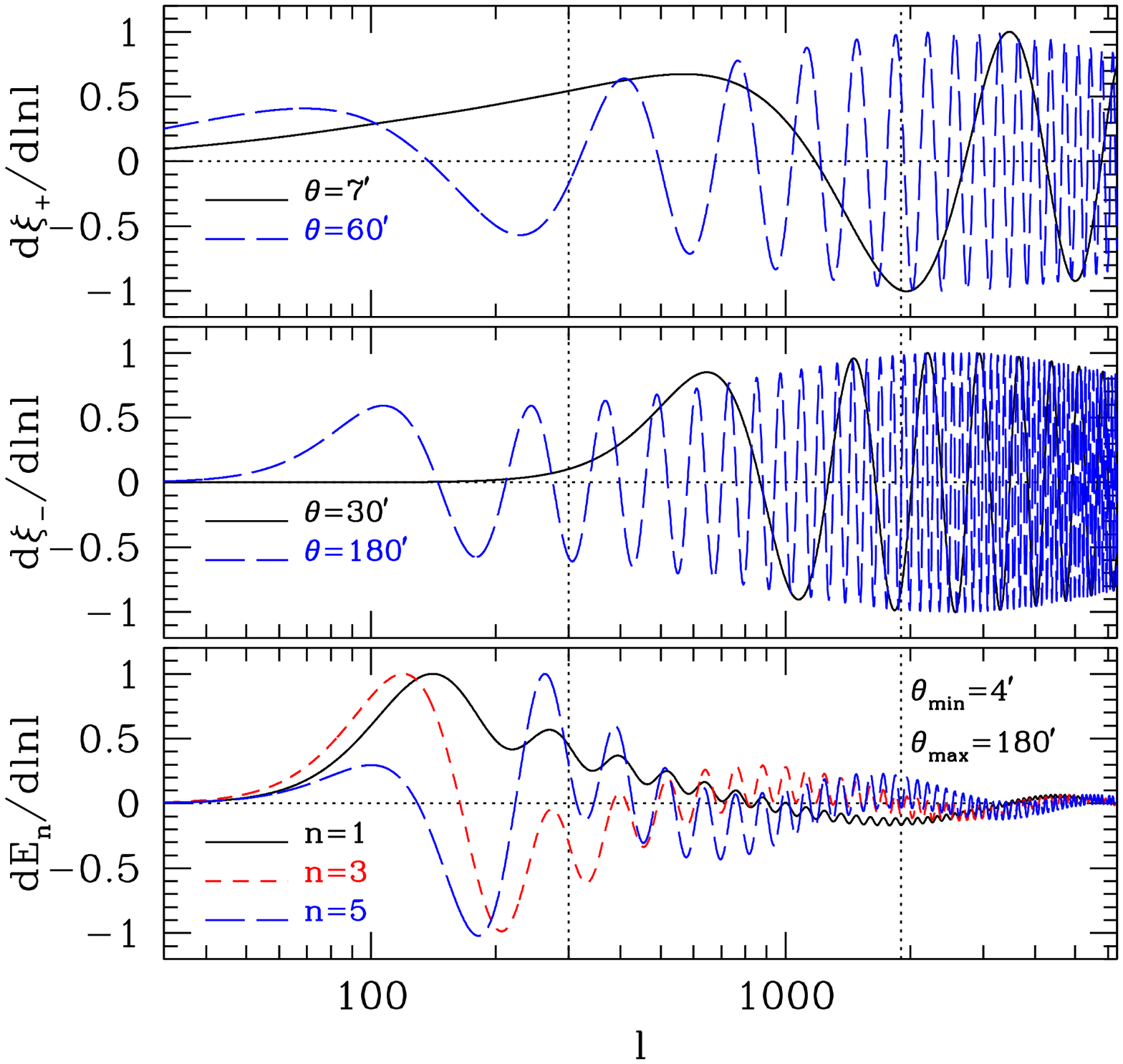}
\end{center}
\caption{{\it Left}: A cumulative contribution of power spectra to $X$ from a
  limited range of
  $\ell<\ell_{\rm max}$ as a function of $\ell_{\rm max}$ (to be
  specific, the upper limit of the $\ell$-integration in
  equations~(\ref{eq:xipm}) and (\ref{eq:cosebis-ps}) are truncated at
  $\ell_{\rm max}$), where $X$ 
  stands for TPCF $\xi_+$ (top panel), $\xi_-$ (middle panel), or
  COSEBIs $E_n$ (bottom panel). All the results are normalized by their
  unlimited value. Here a combination of tomographic redshift bins of 
  $z_3 \times z_3$, and the WMAP9
  cosmological model \citep{2013ApJS..208...19H} are adopted, but the
  results are insensitive to those choices. For TPCFs, three angular
  scales roughly corresponding to the minimum, intermediate, and maximum
  scales of H20 are shown. Vertical dotted lines
  show the $\ell$-range used in the Fourier space PS analysis by H19.
  {\it Right}: Integrands of the transformation between the power
  spectra to $\xi_\pm$ (top and middle panels, see equation (\ref{eq:xipm})), or $E_n$
  (bottom panel, see equation (\ref{eq:cosebis-ps})). All integrands
  are normalized by their maximum value. The same angular scales shown
  in the left panels are shown except for $\xi_{\pm}$ where the intermediate
  case is omitted for clarity.
  \label{fig:dcosen_dl}}
\end{figure*}

Right panels of Figure~\ref{fig:dcosen_dl} compares differential
  contributions of power spectra to three two-point statistics ($\xi_+$,
  $\xi_-$, and $E_n$ from the top to bottom panels) as a function of
  $\ell$, to be specific, integrands of the relevant equations
  (equation (\ref{eq:xipm}) for $\xi_\pm$, and equation
  (\ref{eq:cosebis-ps}) for $E_n$).
Left panels of Figure~\ref{fig:dcosen_dl} compares 
cumulative contributions of power spectra to three two-point statistics ($\xi_+$,
$\xi_-$, and $E_n$ from the top to bottom panels) from a limited range of
$\ell<\ell_{\rm max}$ as a function of $\ell_{\rm max}$, to be specific,
the upper 
limits of the $\ell$-integration in the relevant equations
(equation~(\ref{eq:xipm}) and (\ref{eq:cosebis-ps})) are truncated at
$\ell_{\rm max}$.
For TPCFs, 
the three angular scales shown in the plots roughly correspond
to the minimum, intermediate and maximum scales of H20. Vertical dotted
lines show the $\ell$-range ($300<\ell<1900$)
used in the Fourier space PS analysis by H19.
One can see, from the figure,  noticeable differences in $\ell$-ranges 
contributing to each three HSC-Y1 cosmic shear two-point statistics.
Notably, $\xi_+$ has the broadest sensitivity range
(non-zero ranges without the rapid oscillation in the
  differential contributions shown in the right panels of
  Figure~\ref{fig:dcosen_dl}), especially toward very low-$\ell$ modes. 
Compared to the TPCFs' broad sensitivity range, our scale-cut of COSEBIs
suppresses the sensitivity on both low- and high-$\ell$ modes, but a
large part of contribution comes from modes with $\ell<300$.
Therefore, we expect weak correlations in resulting cosmological
parameter constraints between COSEBIs and TPCFs as well as between
COSEBIs and PS, 
as in the case between PS and TPCFs found in H20, in which the two
  methods probe the different $\ell$-ranges and a 1$\sigma$ level
  difference in resulting $S_8$ constraints between PS and TPCFs was
  found (H20).

%
%
\section{Measurements}
\label{sec:measurements}

\subsection{Measurements of COSEBIs from the HSC-Y1 data}
\label{sec:measurements_cosebis}

%
%
\begin{figure}[t]
\begin{center}
  \includegraphics[width=82mm]{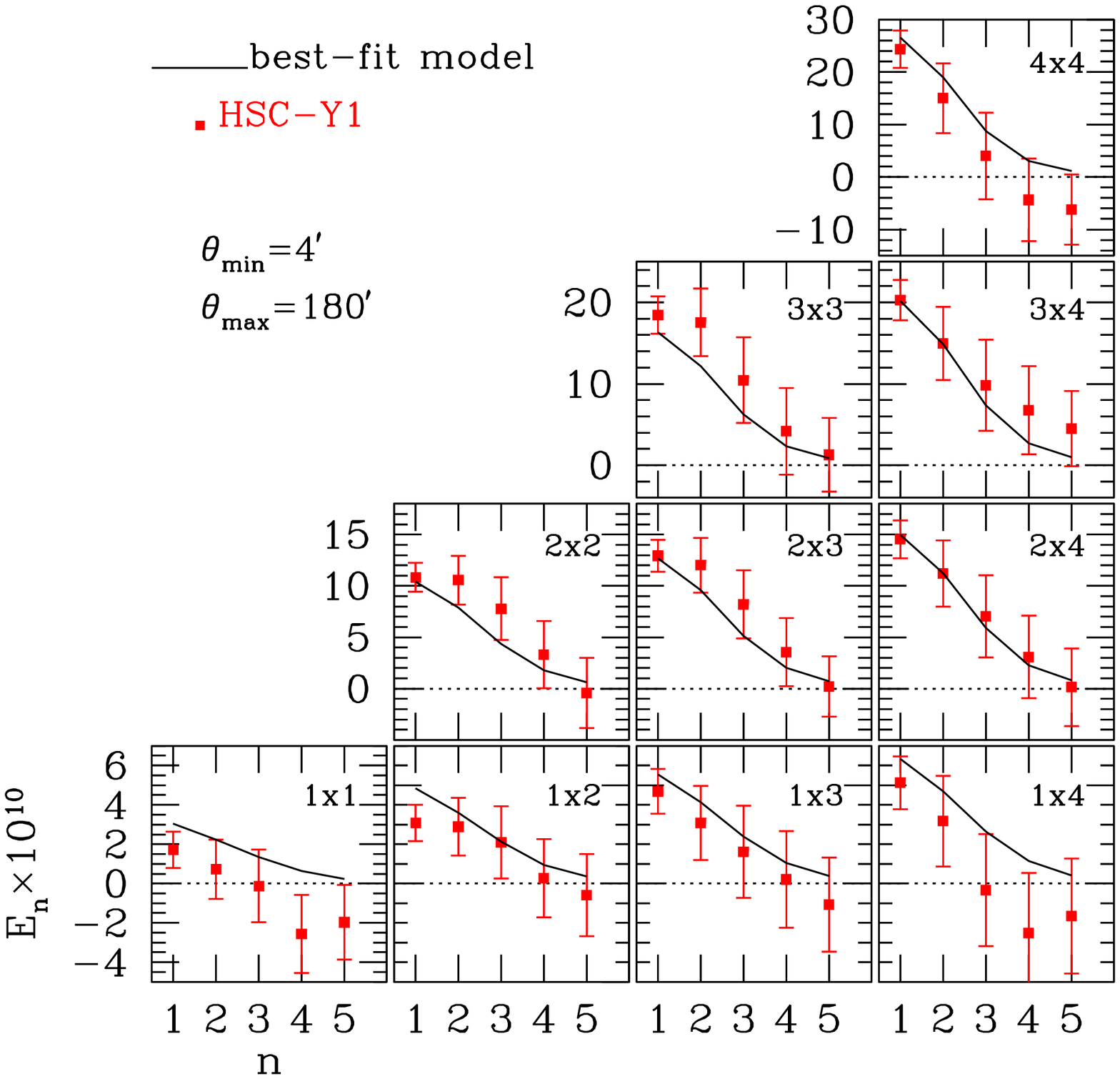}\\
  \includegraphics[width=82mm]{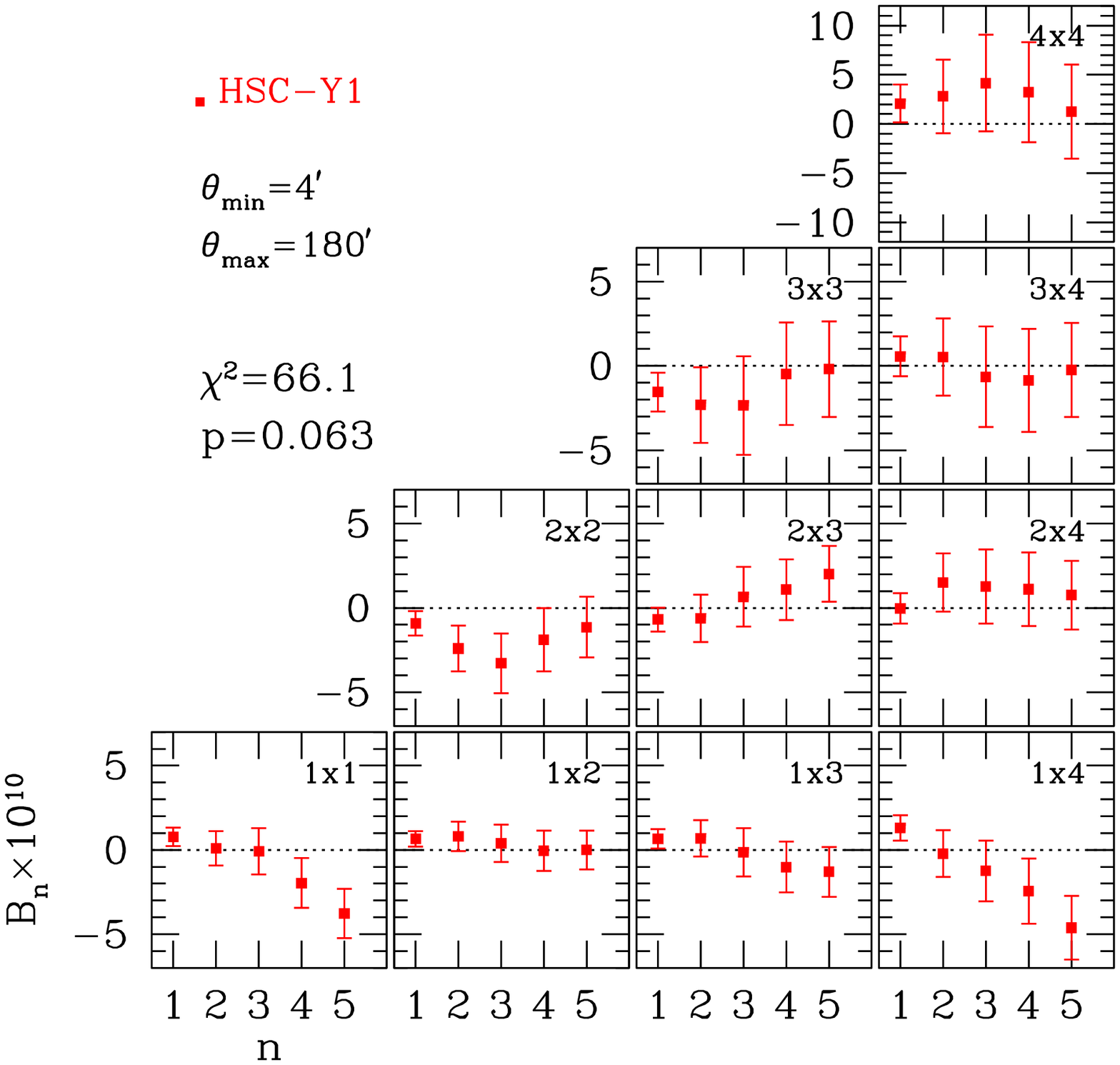}
\end{center}
\caption{Comparison of the HSC-Y1 tomographic cosmic shear COSEBIs with
  the best-fitting theoretical model for the fiducial
  flat $\Lambda$CDM model. Top and bottom triangular-tiled panels show
  $E_n^{ab}$ and $B_n^{ab}$, respectively.
  Error bars represent the
  square-root of the diagonal elements of the covariance matrix. The
  solid lines in the top panels correspond to the best-fit (maximum
  likelihood) fiducial model.
  \label{fig:cosebis_best}}
\end{figure}

The method of measuring COSEBIs precisely was developed and tested by
\citet{2017MNRAS.464.1676A,2021A&A...645A.104A}.
We confirmed the accuracy of the method adopted in
\citet{2021A&A...645A.104A} with HSC-Y1 mock catalogs
(see Appendix~\ref{sec:test_measurement} for details of this
  test), 
and we adopt it in this study.
We first measure the cosmic shear TPCFs for two tomographic
redshift bins $a$ and $b$ taking the shape
weight into account as
\begin{equation}
  \label{eq:estimator_xipm}
  \hat{\xi}_\pm^{ab}(\theta)={
    {\sum_{ij} w_i w_j \left[\hat{\gamma}_{i,t}^{a}(\vec{\theta}_i)
        \hat{\gamma}_{j,t}^{b}(\vec{\theta}_j) \pm \hat{\gamma}_{i,\times}^{a}(\vec{\theta}_i)
        \hat{\gamma}_{j,\times}^{b}(\vec{\theta}_j)\right]}
    \over
    {\sum_{ij} w_i w_j }
    },
\end{equation}
where the tangential ($\hat{\gamma}_t$) and cross ($\hat{\gamma}_{\times}$)
components of shear are defined with respect to the direction
connecting a pair of galaxies under consideration ($i$ and $j$), $w_i$
is shape weight for each galaxy, and the summation runs over pairs of
galaxies with their  angular separation
$\theta=|\vec{\theta}_i-\vec{\theta}_j|$ within an interval
$\Delta \theta $ around $\theta$.
Following \citet{2021A&A...645A.104A}, we adopt a fine
$\log \theta$-binning, to be specific, 4000 bins over
$0.5\arcmin<\theta<300\arcmin$ with an equal $\log\theta$-bin width,
although we only use the limited range of $4\arcmin<\theta<180\arcmin$.
For actual measurements of the TPCFs, we used the public software {\tt
  Athena}\footnote{http://www.cosmostat.org/software/athena}
\citep{2002A&A...396....1S}.
We then perform the linear transformation in
equations~(\ref{eq:cosebis-e}) and (\ref{eq:cosebis-b}) to compute
E/B-mode COSEBIs.
We use the first five COSEBIs modes in the following analyses following
the previous studies \citep{2010A&A...520A.116S,2012A&A...542A.122A}.

The measured signals are shown in Figure~\ref{fig:cosebis_best} with 
error bars representing the square-root of the diagonal elements of the
covariance matrix (see Section~\ref{sec:covariance}).
For comparison, the best-fit theoretical model obtained from the cosmological
inference of our fiducial $\Lambda$CDM model (described in
Section~\ref{sec:fiducial_cosmology}) is shown in 
the E-mode plots. 
The E-mode signals shown in Figure~\ref{fig:cosebis_best} form the data
vector of our cosmological inference (see Section \ref{sec:analyses}).
We define the data vector as 
$d_{E,i} = (E_1^{11},E_2^{11},,,E_5^{11},E_1^{12},,,E_5^{44})$.
Since there are 5 modes for each of 10 combinations of tomographic
redshift bins, the data vector consists of 50 elements.

\subsection{Covariance from HSC-Y1 mock catalogs}
\label{sec:covariance}

%
%
\begin{figure}[t]
\begin{center}
  \includegraphics[height=82mm,angle=-90]{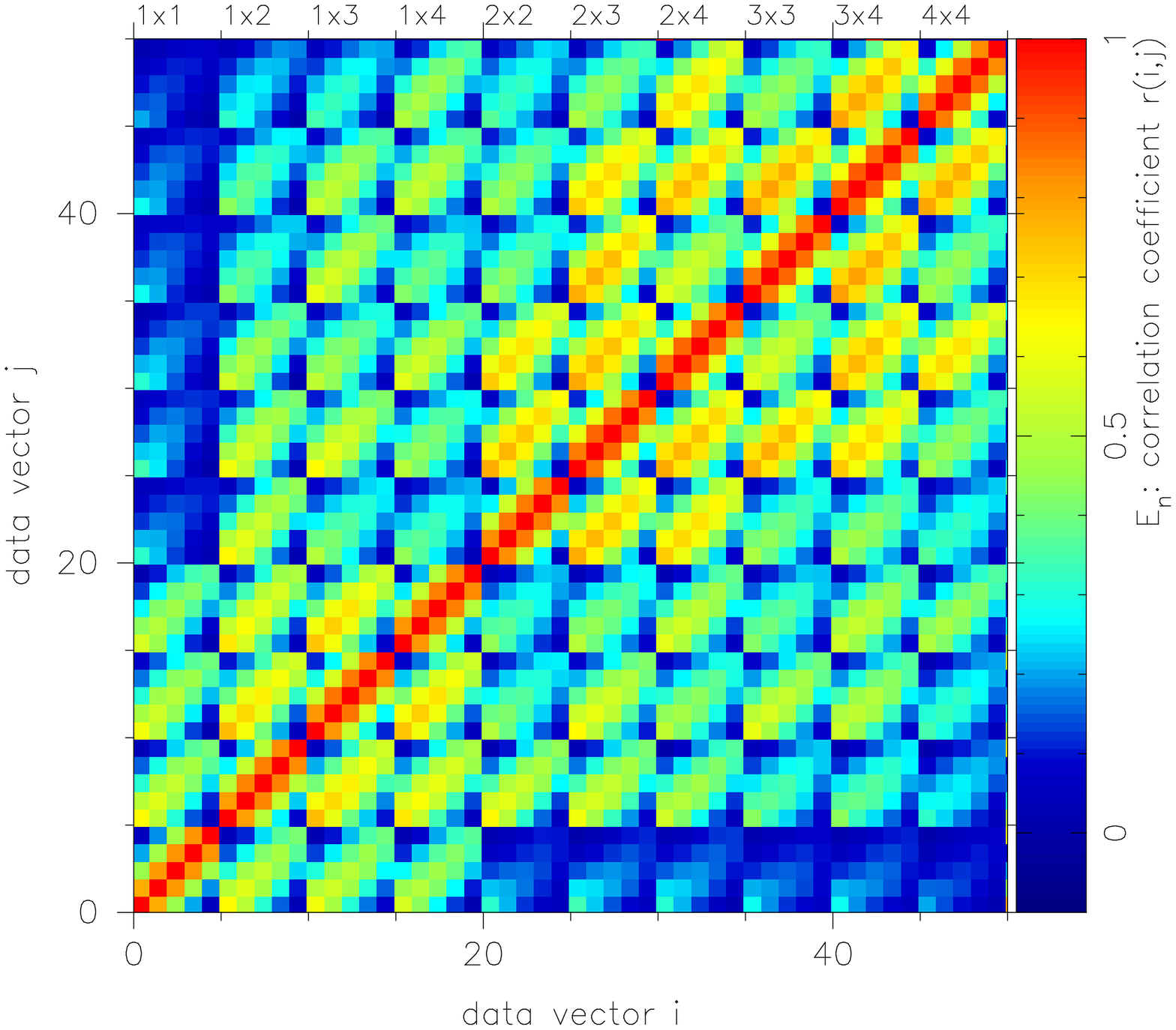}\\
  \vspace{5mm}
  \includegraphics[height=82mm,angle=-90]{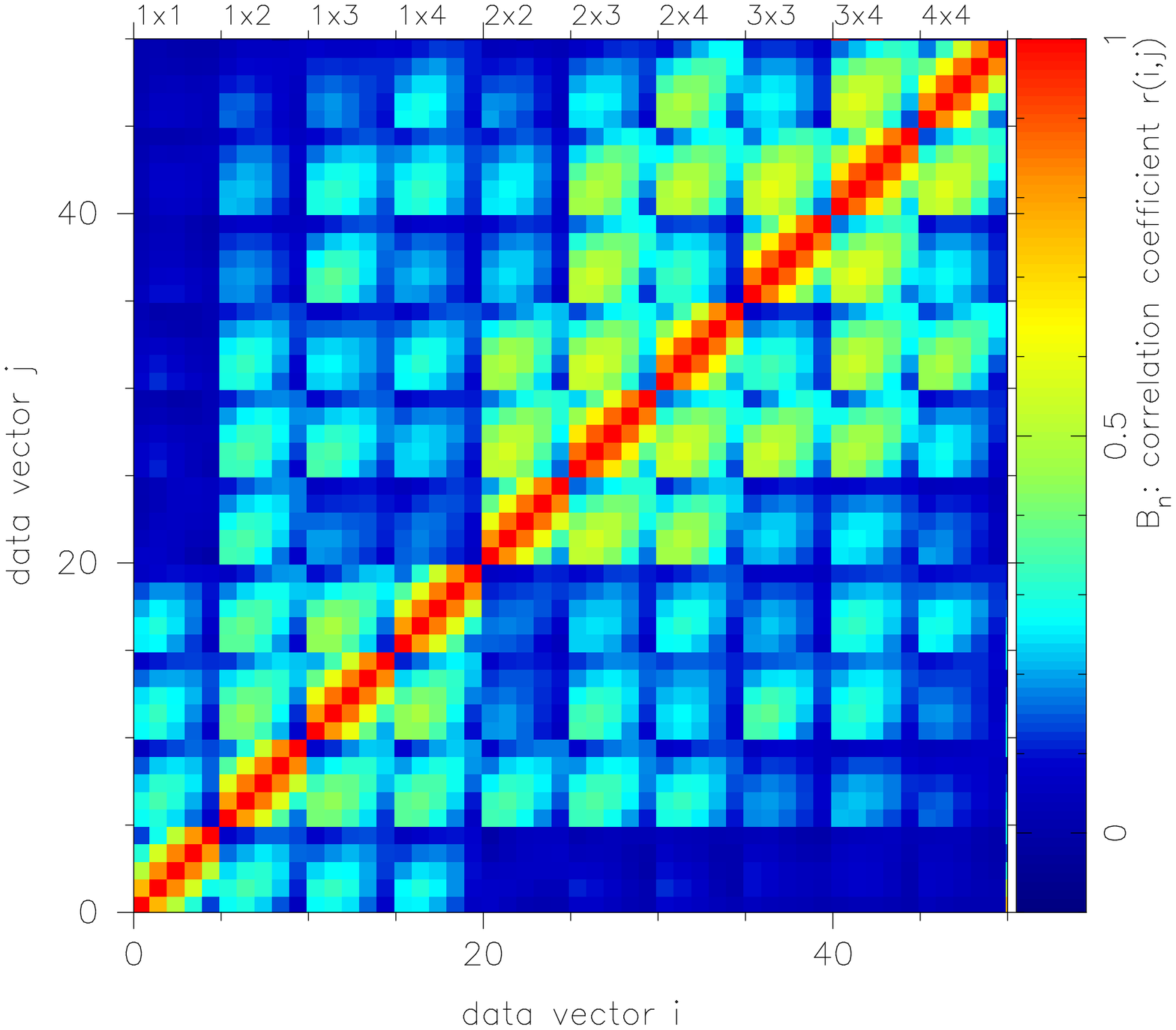}
\end{center}
\caption{Covariance matrices normalized by the diagonal components, i.e.,
  ${\rm Cov}_{ij}/\sqrt{{\rm Cov}_{ii}{\rm Cov}_{jj}}$, are shown.
  Top panel is for $E_n$, and bottom panel is for $B_n$. 
  The x- and y-axis indicate elements of the data vector, 
which is defined as $d_{E,i} =
  (E_1^{11},E_2^{11},,,E_5^{11},E_1^{12},,,E_5^{44})$ and has
  50 elements consisting of
  10 tomographic combinations of COSEBIs (indicated as top label) each
  with 5 modes.
  \label{fig:cov2dimg}}
\end{figure}

%
%
\begin{figure}
\begin{center}
\includegraphics[width=82mm]{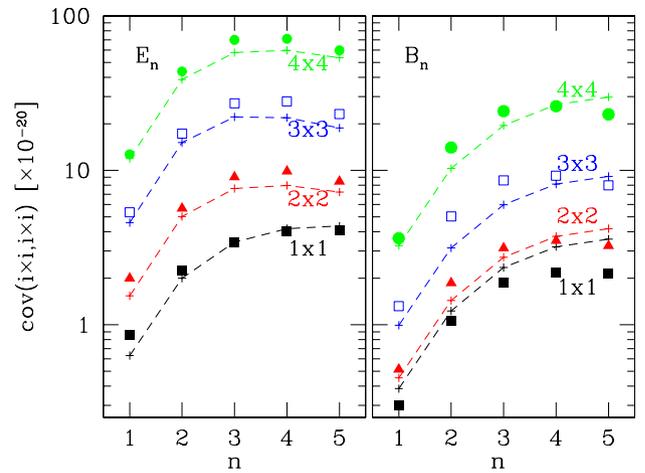}
\end{center}
\caption{Diagonal components of covariance matrices of $E_n$ (left
  panel) and $B_n$ (right panel), to be
  specific $\langle X_n(z_i\times z_j)X_n(z_i\times z_j) \rangle$, 
  are shown for four auto-correlation cases of tomographic redshift bins
  $z_1\times z_1$ (filled squares), 
  $z_2\times z_2$ (filled triangles), $z_3\times z_3$ (open squares),
  and $z_4\times z_4$ (filled circles), from lowest to
  highest , respectively. Theoretical predictions (see text for details)
  are plotted with plus marks connected by dashed lines for visual
  help. \label{fig:cov_en_bn}} 
\end{figure}

Following the methodology of the HSC-Y1 cosmic shear TPCF analysis by
H20, we derive covariance matrices of the COSEBIs measurement using 2268
realizations of mock HSC-Y1 shape catalogs (see
\citet{2019MNRAS.486...52S} for a detailed description of the mock
catalogs).
Since the HSC mock catalogs are constructed based on full-sky lensing
simulation data with galaxy positions, intrinsic shape noise, and
measurement noise taken from the real HSC-Y1 shape catalog, the mock
data naturally have the same survey geometry and the same noise
properties as the real catalog, and include super-survey cosmic shear
signals from these full-sky lensing simulations.
In addition, the effects of nonlinear structure formation
on the lensing shear field are included in the mock data. 
Therefore the covariance matrix computed from the mock catalogs
automatically includes all the contributions, namely, the shape noise,
Gaussian, non-Gaussian, and super-survey covariance with the
survey geometry being naturally taken into account.

We measure the cosmic shear COSEBIs for all 2268 mock catalogs in
exactly the same manner as the real data measurement, and derive
covariances for E- and B-mode COSEBIs.
Results are shown in Figure~\ref{fig:cov2dimg}, in which covariance
matrices normalized by the diagonal components, i.e.,
${\rm Cov}_{E,ij}/\sqrt{{\rm Cov}_{E,ii}{\rm Cov}_{E,jj}}$ are plotted.
We show some diagonal components in Figure~\ref{fig:cov_en_bn} so that the
readers can grasp absolute values of the covariance matrices.
In that figure, for rough comparison, approximate theoretical predictions of
the covariance matrices are shown.
To calculate them, we follow the prescription described in Appendix~A of
\citet{2020A&A...634A.127A}, and we take only major components into
accounts; the Gaussian and shape noise terms for E-mode, and only the
shape noise term for B-mode \citep[see][for contributions from
  non-Gaussian term and super sample
  covariance]{2018JCAP...10..053B,2019MNRAS.486...52S}. 
Overall, reasonably good agreements between 
mock measurements and theoretical
predictions can be seen in Figure~\ref{fig:cov_en_bn}. 
The disagreement between the mock covariance and the theoretical
predictions presumably originates mainly from a simplified assumption to
ignore the effect of survey geometry and masking in estimating the
Gaussian and shape noise terms in the theoretical predictions
\citep[see][]{2020A&A...634A.127A}. We note 
that these effects are properly taken into
account in the mock covariance, which will be used in our cosmological
analysis throughout the paper. 
 
\citet{2019MNRAS.486...52S} studied in detail the accuracy 
on the estimation of the
covariance matrix of cosmic shear TPCFs from the HSC-Y1 mocks, and
investigated the impact of various systematic effects in
the cosmic shear analysis, including photo-$z$ errors and the uncertainty
in the multiplicative bias, on the covariance estimation.
Based on their results, H20 concluded that the
covariance matrix of TPCFs estimated from the mocks is 
calibrated with $<10\%$ accuracy against various systematic effects in
the cosmic shear analysis (see section 4.4 of H20).
Since the COSEBIs are linearly related to TPCFs, a similar level of
accuracy can be expected for the covariance matrix of COSEBIs measurement.
It should be noted that the cosmology dependence of the covariance
cannot be included in our mock based approach, because the HSC-Y1 mock 
catalogs are based on a set of full-sky gravitational lensing ray-tracing
simulations that adopt a specific flat $\Lambda$CDM cosmology
\citep{2017ApJ...850...24T}. 
In the HSC-Y1 cosmic shear PS analysis by H19, the
effect of the cosmology dependence of the covariance on their cosmological
inference was examined by comparing cosmological constraints derived
using the cosmology-dependent
covariance (which is their fiducial model) with those derived 
using a cosmology-independent one (fixed to the best-fit
cosmological model). They found that the best-fit $\Omega_m$ 
and $S_8(=\sigma_8 (\Omega_m/0.3)^\alpha$ with 
$\alpha = 0.45$ or $0.5$) values agree with each other 
within 20\% of the statistical uncertainty.
Based on this result, we assume that the cosmology dependence 
of the covariance matrix does not significantly impact our 
cosmological analysis.

Using the mock $E_n^{ab}$ data vectors and their covariance matrix
obtained above, we examine the Gaussianity of the likelihood function
of cosmic shear COSEBIs as it is a basic assumption in the standard
Bayesian likelihood analysis.
Specifically, we check a {\it necessary condition} of the Gaussianity;
whether a distribution of $\chi^2$ measured from 2268 mock
samples follows the theoretical $\chi^2$ distribution with the same
degrees-of-freedom.
We compute, for each mock data, the standard $\chi^2$,
\begin{equation}
\label{eq:echisq}
\chi_E^2 = \sum_{i,j} (d_{E,i}-\bar{d}_{E,i}) {\rm Cov}_{E,ij}^{-1} (d_{E,j}-\bar{d}_{E,i}),
\end{equation}
where $d_{E,i}$ is the E-mode data vector consisting of 10 tomographic
combinations of $E_n^{ab}$ each with 5 modes and thus having 50
elements,  $\bar{d}_{E,i}$ is its average among 2268 mock samples, and
${\rm Cov}_{E,ij}$ is the E-mode covariance matrix.
We perform a Kolmogorov–Smirnov (KS) test for the distribution of mock
$\chi_E^2$ against the theoretical $\chi^2$ distribution, and obtain the
KS $p$-value of 0.11. Thus we conclude that the mock
$\chi_E^2$ distribution is in a reasonable agreement with the
theoretical one.

\subsection{B-mode null test}
\label{sec:bmode_nulltest}

Here, we quantitatively test the consistency of the B-mode COSEBIs
signals with zero using the standard $\chi^2$ statistics,
\begin{equation}
\label{eq:bchisq}
  \chi_B^2 = \sum_{i,j} d_{B,i} {\rm Cov}_{B,ij}^{-1} d_{B,j},
\end{equation}
where $d_{B,i}$ is the B-mode data vector (presented in
Figure~\ref{fig:cosebis_best}), and ${\rm Cov}_{B,ij}$ is
the B-mode covariance matrix presented in Figure~\ref{fig:cov_en_bn}.
We find $\chi_B^2=66.1$ for 50 degrees-of-freedom, corresponding to the
$p$-value of 0.063. 
Therefore we conclude that no evidence for a significant B-mode signal
is found.

We also compute the B-mode $\chi_B^2$ for each of 2268 mock catalogs.
We compare the distribution of them with the theoretical $\chi^2$
distribution, and find a good agreement between them.
We find that 135 cases out of 2268 mock samples exceed the observed
B-mode $\chi_B^2$ values of 66.1, corresponding to a probability of
0.060 which is in a good agreement with the $p$-value estimated above.

%
%
\section{Cosmological analyses}
\label{sec:analyses}

We employ the standard Bayesian likelihood analysis for the cosmological
inference of measured E-mode COSEBIs. The log-likelihood is given by
\begin{equation}
  \label{eq:log-like}
  -2 \ln {\cal L}(\bm{p}) = \sum_{i,j} \left(d_i-m_i(\bm{p})\right) {\mbox{Cov}}_{ij}^{-1}
  \left(d_j-m_j(\bm{p})\right), 
\end{equation}
where $d_i$ is the E-mode data vector consisting of 10 tomographic
combinations of $E_n^{ab}$ each with 5 modes presented in
Section~\ref{sec:measurements_cosebis},
$m_i(\bm{p})$ is the theoretical model with $\bm{p}$ 
being a set of
parameters detailed in
Section~\ref{sec:model_parameters}, and ${\mbox{Cov}}_{ij}$ is the
covariance matrix presented in Section~\ref{sec:covariance}.
Since our covariance matrix is constructed from 2268 mock 
realizations, its inverse covariance is known to be biased high
\citep[see][and references therein]{Anderson2003, 2007A&A...464..399H}.
When calculating the inverse covariance, we therefore include the so-called
Anderson-Hartlap correction factor
$\alpha = (N_{\mbox{mock}}-N_d-2)/(N_{\mbox{mock}}-1)$, where 
$N_{\mbox{mock}}=2268$ is the number of mock realizations and $N_d=50$ 
is the length of our data vector. 

In order to sample the likelihood efficiently, we employ the multimodal
nested sampling algorithm 
\citep{2008MNRAS.384..449F,2009MNRAS.398.1601F,2019OJAp....2E..10F},
as implemented in the public software {\tt MultiNest}.

\subsection{Model parameters}
\label{sec:model_parameters}

In this subsection, we summarize model parameters and their prior
ranges used in our cosmological analysis.
Prior ranges and choice of parameter set for systematic tests are
summarized in Table~\ref{table:parameters}.

%
%
\begin{table*}
\tbl{Summary of cosmological, astrophysical, and systematics parameters
  used in our cosmological analysis.
  ``flat[$x_1$, $x_2$]'' means a flat prior between $x_1$ and
  $x_2$, whereas ``Gauss($\bar{x}$, $\sigma$)'' means a Gaussian
  prior with the mean $\bar{x}$ and the standard deviation
  $\sigma$. For detail descriptions of parameters, 
  see section \ref{sec:cosmological_parameters} for the cosmological
  parameters, section \ref{sec:astrophysical_parameters} for the
  astrophysical nuisance parameters, and section
  \ref{sec:Systematic_parameters} for the systematics nuisance
  parameters. \label{table:parameters}}
{
\begin{tabular}{lllll}
\hline
  Parameter & \multicolumn{3}{l}{~~~~~~~~~~~~~~~~~~~~~~Prior range} & Section \\
  {}        & Fiducial $\Lambda$CDM & $w$CDM & Systematics tests & {} \\
\hline
  Cosmological    & {} & {} & {} &  \ref{sec:cosmological_parameters} \\
  $\Omega_c$      & flat[0.01, 0.9] & {} & {} & {} \\
  $\sigma_8$ & flat[0.1, 2] & {} & {} & {} \\
  $\Omega_b $   & flat[0.038, 0.053] & {} & {} & {} \\
  $n_s$   & flat[0.87, 1.07] & {} & {} & {} \\
  $h$   & flat[0.64, 0.82] & {} & {} & {} \\
  $\sum m_\nu$ [eV]  & fixed to 0.06  & & {} flat[0, 0.5] for ``$\sum m_\nu$ varied`` & {} \\
  $w$  & fixed to -1  & flat[$-2$, $-1/3$] & {} & {}  \\
  \hline
  Astrophysical    & {} & {} & {} & \ref{sec:astrophysical_parameters} \\
  $A_{\mbox{IA}}$  & flat[-5, 5] & {} & fixed to 0 for ``w/o IA'' & {} \\
  $\eta_{\mbox{IA}}$  & flat[-5, 5] & {} & fixed to 3 for ``IA $\eta_{\mbox{IA}}=3$'' & {} \\
  $A_B$  & fixed to 0 & {} & fixed to 1 for ``$A_B=1$'' or flat[-5, 5] for ``$A_B$ varied'' & {} \\
  \hline
  Systematics    & {} & {} & {} & \ref{sec:Systematic_parameters} \\
  $A_{\mbox{psf}}$  & Gauss(1, 1) & {} & fixed to 0 for ``w/o PSF error'' & Appendix~\ref{sec:cosebis_psf} \\
  $\Delta m$           & Gauss(0, 0.01) & {} & fixed to 0 for ``w/o $\Delta m$'' & {} \\
  $\Delta z_1$           & Gauss(0, 0.0374) & {} & fixed to 0 for ``w/o $p(z)$ error'' & {} \\
  $\Delta z_2$           & Gauss(0, 0.0124) & {} & fixed to 0 for ``w/o $p(z)$ error'' & {} \\
  $\Delta z_3$           & Gauss(0, 0.0326) & {} & fixed to 0 for ``w/o $p(z)$ error'' & {} \\
  $\Delta z_4$           & Gauss(0, 0.0343) & {} & fixed to 0 for ``w/o $p(z)$ error'' & {} \\
  $A_{\langle\gamma\rangle}$  & fixed to 0 & {} & flat[-5, 5] for ``w/ const-$\gamma$'' & Appendix~\ref{sec:cosebis_constant_shear} \\
  \hline
\end{tabular}
}
\begin{tabnote}
{}
\end{tabnote}
\end{table*}

\subsubsection{Cosmological parameters}
\label{sec:cosmological_parameters}

We focus on the flat $\Lambda$CDM cosmological model characterized
by five parameters;
the density parameter of CDM ($\Omega_c$),
the normalization of matter fluctuation ($\sigma_8$),
the density parameter of baryons ($\Omega_b$),
the Hubble parameter ($h$), 
and the scalar spectrum index ($n_s$).
Among those parameters, the cosmic shear COSEBIs are most sensitive to
$\Omega_c$ and $\sigma_8$.
Thus we adopt prior ranges that are sufficiently wide 
for these parameters (see Table~\ref{table:parameters}).
Note that in contrast to H20 in which $A_s$ (the scalar amplitude of the linear matter power
spectrum on $k=0.05$~Mpc$^{-1}$)  was adopted for the
amplitude parameter of the linear CDM power spectrum, we adopt
$\sigma_8$ because for the cosmic shear or more generally low-redshift
probes of large-scale structure, observables are more directly sensitive
to $\sigma_8$. We treat $A_s$ as a derived parameter.
For $\Omega_b$, $n_s$, and $h$, which are only 
weakly constrained with cosmic shear COSEBIs, we set prior 
ranges which largely bracket allowed values from external 
experiments (see Table~\ref{table:parameters}). 
For the sum of neutrino mass, we take $\sum m_\nu =0.06$~eV from the
lower bound indicated by
the neutrino oscillation experiments \citep[e.g.,][for a
  review]{2013neco.book.....L} for our fiducial choice.
As a systematics test, we check the impact of neutrino mass on our
conclusions by varying $\sum m_\nu$.

In addition to the fiducial $\Lambda$CDM model, we consider an extended
model by including the time-independent equation-of-state parameter for
the dark energy ($w$), referred to as the $w$CDM model.
We take a flat prior with $-2<w<-1/3$, which excludes the
non-accelerating expansion of the present day Universe, 
and brackets allowed values from external experiments.

\subsubsection{Astrophysical parameters}
\label{sec:astrophysical_parameters}

In theoretical modeling of the COSEBIs signal, we include two
astrophysical effects; 
one is the effect of baryon physics on the nonlinear matter power
spectrum (see Section~\ref{sec:powerspectra}), and the other is the
contribution of the intrinsic alignment  (IA) of galaxy shapes
\citep[see][for recent reviews]{2015SSRv..193..139K,2015PhR...558....1T}.
Below we summarize our treatment of those effects in our cosmological
analyses.

In modeling baryonic effect, we follow the methodology of
\citet{2017MNRAS.471.4412K}, in which a
modification of the dark matter power spectrum due to the AGN feedback
is modeled by the fitting function derived by
\citet{2015MNRAS.450.1212H}, but an additional parameter ($A_B$) that
controls the strength of the feedback is introduced 
\citep[see Section~5.1.2 of][for the explicit 
expression]{2017MNRAS.471.4412K}. 
We note that H19 and H20 employed the same methodology.
However, since we impose the scale-cut of cosmic shear 
COSEBIs conservatively so that the baryon effects do not have a 
significant impact  on our analysis (see Section~\ref{sec:scalecut}), 
we do not include the baryon effect in our fiducial model, 
but check its impact in our systematics tests; 
one fixing $A_B=1$ that corresponds to 
the original AGN feedback model, and the other in which $A_B$ is a free
parameter (see Table~\ref{table:parameters}).

The IA comes both from the correlation between intrinsic shapes of 
two physically associated galaxies in the same
local field (referred to as the II-term) and from the cross
correlation between lensing shear of background galaxies and the intrinsic
shape of foreground galaxies (referred to as the GI-term).
We employ the standard theoretical framework for these terms, namely, the
nonlinear modification of the tidal alignment model
\citep{2004PhRvD..70f3526H,2007NJPh....9..444B,2011A&A...527A..26J}.
In this model, the E-mode COSEBIs originating from II and GI terms
are given in a similar manner as the cosmic shear COSEBIs, 
equation~(\ref{eq:cosebis-ps}), but with IA power spectra
$P_{\mbox{II}}^{ab}(\ell)$ and $P_{\mbox{GI}}^{ab}(\ell)$ (see
equations~(12) and (13) of H20 for explicit expressions for those
terms).
Following H20, we adopt the standard parametric IA model with two
parameters, the amplitude parameter $A_{\mbox{IA}}$ and the power-law
redshift dependence parameter $\eta_{\mbox{IA}}$ that represents the
effective redshift evolution of the IA amplitude.
Following recent cosmic shear studies
e.g., \citet{2017MNRAS.465.1454H}, \citet{2018PhRvD..98d3528T}, and
H19, we adopt wide prior ranges for these
parameters (see Table~\ref{table:parameters}).

\subsubsection{Systematic parameters}
\label{sec:Systematic_parameters}

Our treatment of systematic effects in the cosmological analysis largely
follows that in H20.
To summarize, in our fiducial model we take account of systematic 
effects from PSF leakage and PSF modeling errors, the uncertainty 
in the shear multiplicative bias correction, and 
uncertainties in the source galaxy redshift distributions.
In addition, in systematics tests we check the impact of the uncertainty of 
the constant shear over fields.
Below we summarize our modeling of those systematic effects, and choices
for prior ranges on nuisance parameters in these models.

Our models for the PSF leakage and PSF modeling errors are described in
Appendix~\ref{sec:cosebis_psf}. We apply the correction for these
systematics by equation~(\ref{eq:En-psf-model}), which has one nuisance
parameter $A_{\rm psf}$.
We adopt a Gaussian prior with $({\rm mean}, \sigma)=(1,1)$ for it and
include them in our fiducial model.

Regarding the uncertainty in the shear multiplicative bias correction,
we follow H20 (see Section~5.2.3 of H20).
In short, we introduce the nuisance parameter $\Delta m$, which
represents the residual multiplicative bias, and modifies the
theoretical prediction for the COSEBIs to
\begin{equation}
  \label{eq:delta_m}
  E_n^{ab} \rightarrow (1+\Delta m)^2 E_n^{ab}.
\end{equation}
A Gaussian prior with $({\rm mean}, \sigma)=(0,0.01)$ is taken for
$\Delta m$  based on the calibration of the HSC-Y1 shear catalog
\citep{2018MNRAS.481.3170M}. 

Regarding uncertainties in the redshift distributions of source
galaxies, we again follow H20.
Introducing a nuisance parameter $\Delta z_a$ for each tomographic
redshift bin, the source redshift distribution, $p^a(z)$, is shifted by 
$p^a(z) \rightarrow  p^a(z+\Delta z_a)$.
Gaussian priors are taken for $\Delta z_a$ with the same mean and
$\sigma$ as those adopted in H20 (see Table~\ref{table:parameters}).

Finally, as described in Appendix~\ref{sec:cosebis_constant_shear}, 
we model a contribution to E-mode COSEBIs arising from a constant shear
by equation~(\ref{eq:En-c-model}), which has one nuisance
parameter $A_{\langle\gamma\rangle}$.
We assume a redshift-independent constant shear for simplicity.
Given that we have not found a strong evidence of the existence of 
the residual constant shear (see Appendix~1 of H20), we do
not include it in our fiducial model, but check its impact as a
systematics test, in which we take a flat prior of
$-5<A_{\langle\gamma\rangle}<5$.

%
%
\section{Results}
\label{sec:results}

We first present cosmological constraints from our
cosmic shear COSEBIs analysis in the fiducial flat $\Lambda$CDM model.
We then discuss the robustness of the results against various
systematics, and present results from internal 
consistency checks among different choices of
tomographic redshift bins and scale-cuts.
We compare our cosmological constraints with other cosmic shear results
and {\it Planck} CMB result.
We also compare our results with those from the HSC-Y1 cosmic shear PS
and TPCF analyses, and examine the consistency among them using mock HSC-Y1
catalogs.

%
%
\begin{figure}
\begin{center}
  \includegraphics[height=82mm,angle=-90]{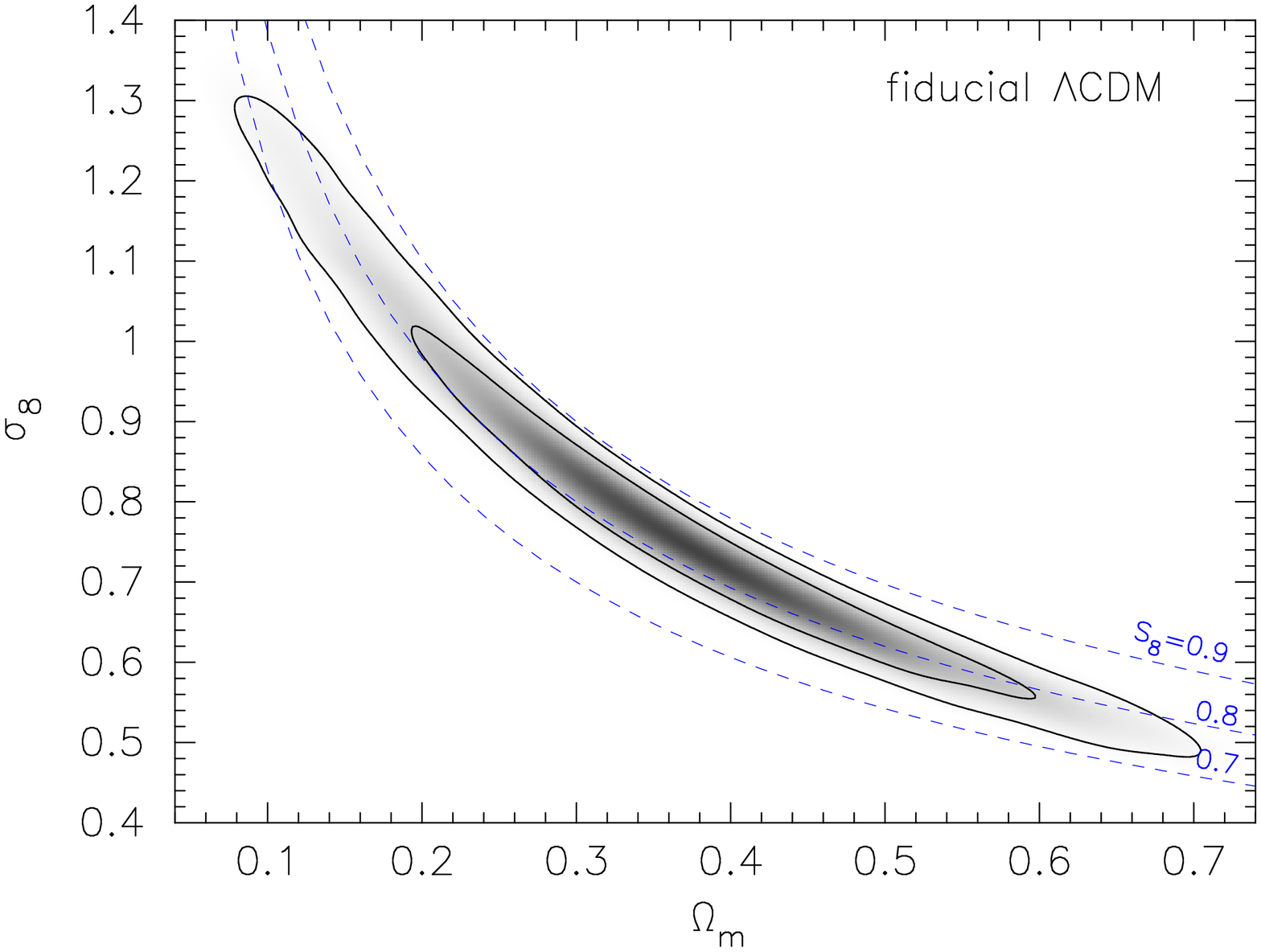}\\
  \includegraphics[height=82mm,angle=-90]{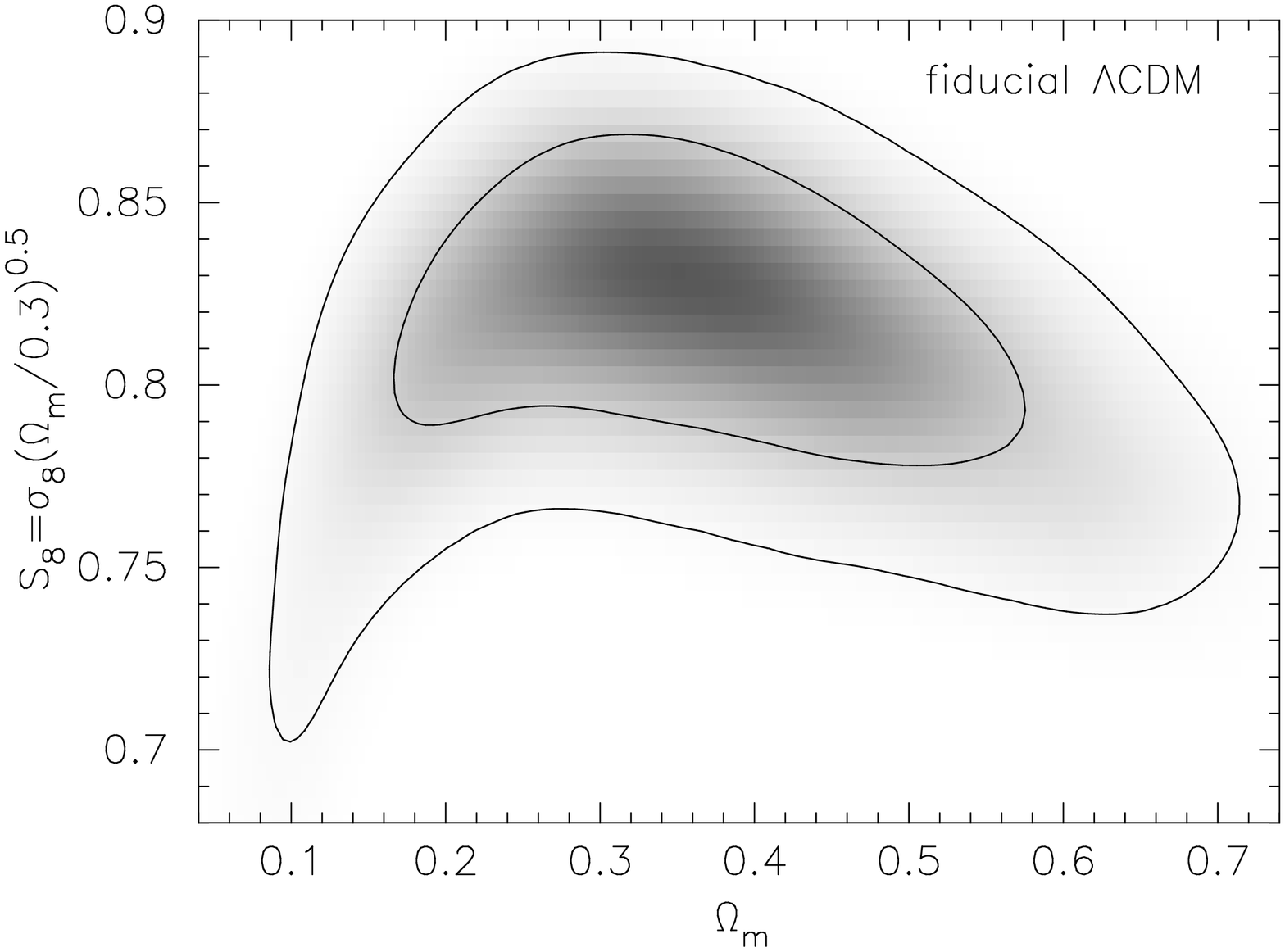}
\end{center}
\caption{Marginalized posterior contours (68\% and 95\%
  confidence levels) with the gray scale for a visual overview of the
  distribution (darker for higher posterior)
  in the $\Omega_m$-$\sigma_8$ plane (top panel) and in the
  $\Omega_m$-$S_8$ plane (bottom panel), where
  $S_8=\sigma_8\sqrt{\Omega_m/0.3}$, in the fiducial flat $\Lambda$CDM
  model.
  In the $\Omega_m$-$\sigma_8$ panel, three constant $S_8$ loci
  ($S_8=0.7$, 0.8 and 0.9 from bottom to top, respectively) are shown by
  dashed blue curves.
  \label{fig:om_sig8_S8}}
\end{figure}

%
%
\begin{figure*}
\begin{center}
  \includegraphics[height=150mm,angle=-90]{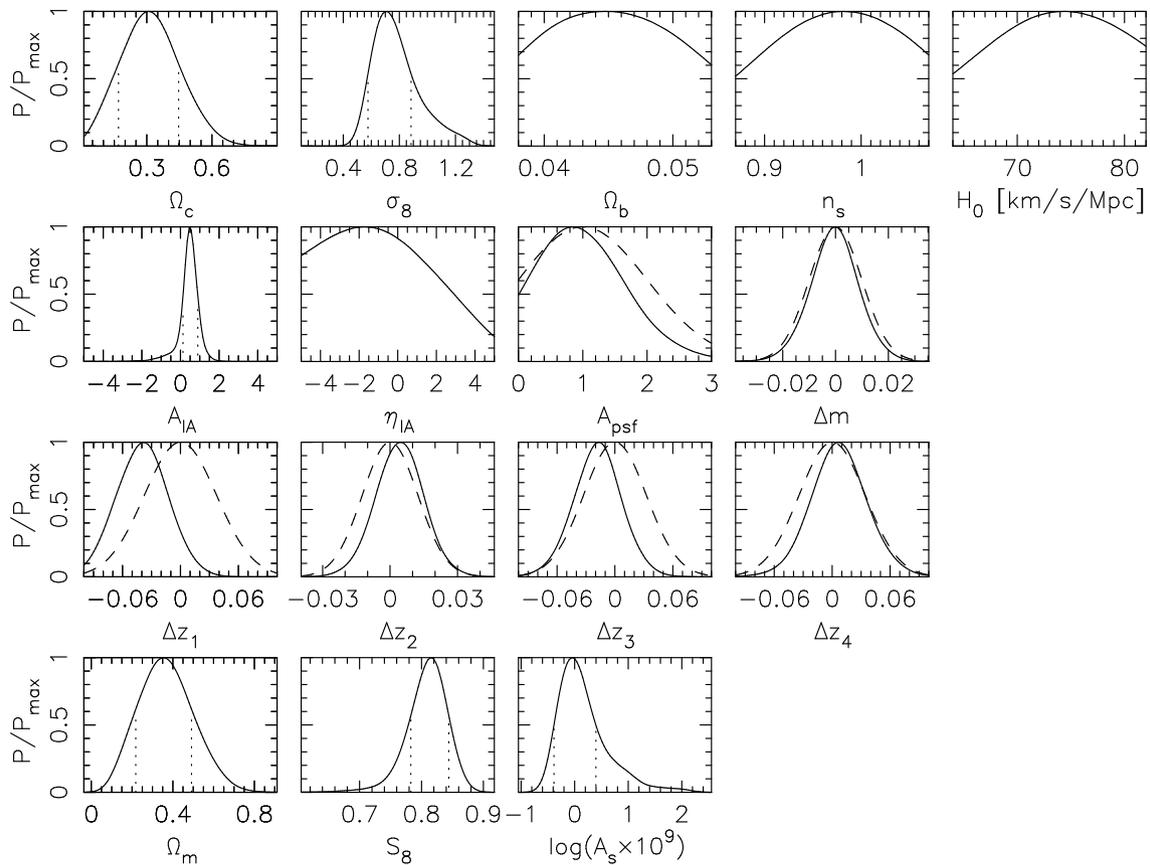}
\end{center}
\caption{The upper three rows show the marginalized one-dimensional
  posterior distributions of model 
  parameters in the fiducial flat $\Lambda$CDM model;
  the cosmological parameters are in the top-row, and the
  astrophysical and systematics parameters are in two middle rows.
  Three panels in the bottom-row are for derived parameters.
  For parameters with flat prior ranges, the plotted range of the
  horizontal-axis indicates its prior range.
  For parameters with Gaussian priors, Gaussian priors are shown 
  by the dashed curves.
  Dotted vertical lines represent the approximate 68\% credible
  intervals, which are not shown 
  for poorly constrained parameters.
  \label{fig:posterior16}}
\end{figure*}

%
%
\subsection{Cosmological constraints in the fiducial flat $\Lambda$CDM model}
\label{sec:fiducial_cosmology}

%
%
\begin{table}
\tbl{Means and 68\% confidence intervals of marginalized posterior
  distributions for well constraint cosmological parameters. \label{table:constraints}}
{
\begin{tabular}{lll}
\hline
Parameter & Mean & 68\% confidence interval \\
\hline
$\Omega_c$ & 0.319 & 0.169 - 0.447\\
$\Omega_m$ & 0.365 & 0.218 - 0.492\\
$A_s\times 10^{9} $ & 1.46 & 0.411 - 2.50\\
$\sigma_8$ & 0.780 & 0.576 - 0.883\\
$S_8$ & 0.809 & 0.783 - 0.844\\
\hline
\end{tabular}
}
\end{table}

First, we present results for our fiducial flat $\Lambda$CDM model.
Marginalized posterior 
contours in the $\Omega_m$-$\sigma_8$ and $\Omega_m$-$S_8$ planes are
shown in Figure~\ref{fig:om_sig8_S8}, and marginalized one-dimensional
posterior distributions of 13 model parameters
and 3 derived parameters 
are shown in Figure~\ref{fig:posterior16}.
We find marginalized 68\% confidence intervals of
$0.218<\Omega_m<0.492$,
$0.576<\sigma_8<0.883$, and
$0.783<S_8<0.844$ 
(see also Table \ref{table:constraints}).

From the posterior distributions shown in
Figure~\ref{fig:posterior16}, it can be seen that the 
HSC-Y1 cosmic shear COSEBIs alone cannot place useful constraints on
3 out of 5 input cosmological parameters ($\Omega_b$, $n_s$ and $H_0$),
but we confirm that the constraint on $S_8$ is not strongly
affected by uncertainties in these parameters as long as they are
restricted within the prior ranges considered in this study.

It is also found from two middle rows of Figure~\ref{fig:posterior16}
that except for $A_{\mbox{IA}}$, one-dimensional posteriors of astrophysical  
and systematics parameters are dominated by priors.
In the following subsections, we discuss effects of these nuisance
parameters on the cosmological inference by changing the parameter setup.

In Figure~\ref{fig:cosebis_best}, we compare the measured cosmic shear
COSEBIs signals with the theoretical model with best-fit parameter values for
the fiducial flat $\Lambda$CDM model.
In these plots, error bars represent the square-root of the diagonal 
elements of the covariance matrix.
We find that our model with the fiducial parameter setup
reproduces the observed signals quite well.
The $\chi^2$ value for the best-fit parameter set is
$\chi^2=40.7$ for the {\it effective}
degree-of-freedom\footnote{Although the total number of model parameters
is 13 for our fiducial case, only three of them ($\Omega_c$, $\sigma_8$,
and $A_{\mbox{IA}}$) are constrained by the data with much
narrower posterior distributions than with priors.
Therefore, the standard definition of degree-of-freedom
($N_d-N_p=50-13$ for our fiducial case) 
would most-likely be underestimated.
A conservative choice of the effective number of free parameters
should account for only these three parameters. 
See \citet{2019PhRvD..99d3506R} and Section~6.1 of \citet{2019PASJ...71...43H} 
for a more mathematically robust way to define the effective number of 
free parameters.} of
$50-3=47$, resulting in a $p$-value of 0.73.

%
%
\subsection{Systematics tests}
\label{sec:systematics_tests}

%
%
\begin{figure}
\begin{center}
  \includegraphics[height=82mm,angle=-90]{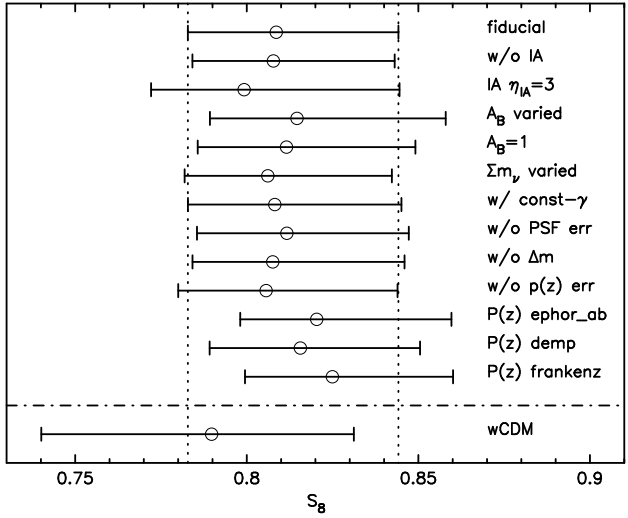}
\end{center}
\caption{Means and 68\% credible intervals of
  marginalized one-dimensional constraints on $S_8=\sigma_8\sqrt{\Omega_m/0.3}$.
  The fiducial case (top) is compared with different setups to check the 
  robustness of our result.
  Vertical dotted lines show the 68\% credible interval of the
  fiducial case.
  See Figure~\ref{fig:om_sig8_sys}, for comparisons of constraints
    in the $\Omega_m$-$\sigma_8$ plane.
  \label{fig:s8ranges}}
\end{figure}

In our cosmological analysis, we have a number of astrophysical and
systematics nuisance parameters that
are marginalized over. 
Also, we have four model parameters that
are fixed to the fiducial values in our fiducial setup
(namely, $\sum m_{\nu}$, $w$, $A_B$, and $A_{\langle \gamma \rangle}$, see Table
\ref{table:parameters}),  but may have, in principle, 
an impact on the cosmological inference.
Below, we discuss effects of these nuisance parameters on the 
cosmological inference by changing the parameter setup
(see Table
  \ref{table:parameters} for parameter setups, and following subsections
  for details).
In addition, we also perform an empirical test on robustness of our
cosmological constraints against possible uncertainties in the source
redshift distributions by replacing the default ones derived from the
COSMOS re-weighted method with ones derived from stacked PDFs with three
photo-$z$ methods, {\tt DEmP}, {\tt Ephor AB}, and {\tt FRANKEN-Z}
(see subsection \ref{sec:impact_dz}, and Section~2.2 of H20 for details).

We focus on $S_8$ constraints to assess
  the impact of the nuisance parameters and source
redshift distributions as it is a
  primary parameter to be constrained by cosmic shear two-point statistics.
The results of these systematics tests are summarized in
Figure~\ref{fig:s8ranges},
in which
credible intervals of marginalized one-dimensional posterior
  distributions of $S_8$ derived from systematics tests are compared with that of
  the fiducial setup.
In this comparison, we use $S_8=\sigma_8 (\Omega_m/0.3)^\alpha$ with
$\alpha=0.5$.
In Appendix~\ref{sec:supplementary_figures}, we also present comparisons
of marginalized posterior contours in the $\Omega_m$-$\sigma_8$ plane
because a single choice of $\alpha$ does not always provide an optimal
description for the $\sigma_8$-$\Omega_m$ degeneracy, especially for cases
with broad confidence contours such as ours. 
Overall, we find that no significant impact on our $S_8$ constraint is found
  from the systematics tests, as described in detail in following subsections.

%
%
\subsubsection{Intrinsic galaxy alignment}
\label{sec:impact_IA}

We find that the marginalized one-dimensional constraint on
$A_{\mbox{IA}}$ obtained from our fiducial cosmological inference is
$A_{\mbox{IA}}=0.44_{-0.31}^{+0.46}$ , which is
consistent with the results from the HSC-Y1 TPCF
($A_{\mbox{IA}}=0.91_{-0.32}^{+0.27}$, H20) and PS analyses
($A_{\mbox{IA}}=0.38\pm 0.70$, H19) for their fiducial setup.

In order to test the robustness of the cosmological constraints against
the uncertainty of the intrinsic galaxy alignment, we perform two
cosmological inferences with different IA modeling. In one case, 
the IA contribution is completely ignored i.e., $A_{\mbox{IA}}$ is fixed
to 0 (``w/o IA'' setup), 
and in the other case $\eta_{\mbox{IA}}$ is fixed to 3 while 
$A_{\mbox{IA}}$ is treated as a free parameter (``IA $\eta_{\rm IA}=3$''
setup, see section 5.4 of H19
for the reasoning of this choice).
The results from these settings are compared
with the fiducial one in Figure~\ref{fig:s8ranges} (see panels (a) and
(b) of Figure~\ref{fig:om_sig8_sys} for constraints in
$\sigma_8$-$\Omega_m$ plane). 
We find that in both of the two cases, the changes in cosmological
constraints are not significant.
On those grounds, we conclude that the effects
of uncertainty in IA modeling on our fiducial cosmological constraints
is insignificant.

%
%
\subsubsection{Baryonic feedback}
\label{sec:impact_baryon}

%
%
\begin{figure}
  \begin{center}
    \includegraphics[height=82mm,angle=-90]{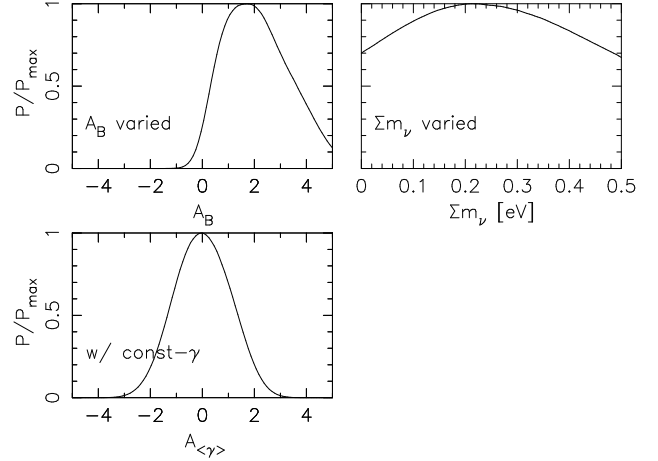}
  \end{center}
  \caption{Marginalized one-dimensional posterior distributions of nuisance
    parameters derived from non-fiducial models. Top-left panel is for the baryon
    feedback model parameter from the ``$A_B$ varied'' setup, top-right
    panel is for the neutrino mass from the ``$\sum m_{\nu}$ varied''
    setup, and the bottom panel is for the
    residual constant shear $A_{\langle\gamma\rangle}$ from the 
    ``w/const-$\gamma$'' setup. 
    \label{fig:posterior_3addedpar}}
\end{figure}

In our fiducial setup, we do not include the effect of the baryonic 
feedback, but instead we impose the conservative scale-cut so that the
baryon effects do not have a significant impact on our analysis (see
Section~\ref{sec:scalecut}).
It is therefore expected that the baryonic feedback effect does not
strongly affect our cosmological constraints.
We check this explicitly by employing an empirical model described in
Section~\ref{sec:astrophysical_parameters} (see Section~4.1.1 of H20 for
details).
Following H20, we consider two cases; the original 
AGN feedback model by \citet{2015MNRAS.450.1212H}, which corresponds 
to fixing the baryon feedback parameter $A_B=1$ (``$A_B=1$'' setup), and 
a more flexible model in which $A_B$ is allowed to vary with a flat prior 
in the range $-5<A_B<5$ (``$A_B$ varied'' setup).
From the results shown in Figure~\ref{fig:s8ranges}
(and see panels (c) and (d) of Figure~\ref{fig:om_sig8_sys} for constraints in
$\sigma_8$-$\Omega_m$ plane), 
we find that in both of the two cases, the changes in cosmological
constraints are not significant.
The marginalized one-dimensional posterior
distribution of $A_B$ obtained from the ``$A_B$ varied'' setup is shown
in Figure~\ref{fig:posterior_3addedpar}, from which it is found that the
constraint on $A_B$ is very weak with $A_B=2.0\pm 1.2$ (mean and $\sigma$).
We also find that the correlation between $A_B$ and $S_8$ is very week.
Based on these results, we conclude that the effect of baryonic feedback on
our fiducial cosmological constraints is insignificant.

%
%
\subsubsection{Neutrino mass}
\label{sec:impact_neutrino}

In our fiducial setup, the neutrino mass is
fixed at $\sum m_{\nu}=0.06$~eV.
Since the non-zero neutrino mass leads to a redshift-dependent
suppression of the matter power spectrum on small scales, it has, in
principle, an impact  on our cosmological inference.
However, the HSC-Y1 cosmic shear analyses are expected to be 
insufficient to place a useful constraint on the neutrino mass due to
the current measurement precision and the scale-cuts on small scales, and
this is indeed the cases for HSC-Y1 PS (H19) and TPCF analyses (H20). 
We find that this is also the case for our COSEBIs analysis;
Figure~\ref{fig:posterior_3addedpar} shows the one-dimensional posterior 
distribution of $\sum m_{\nu}$ obtained from the ``$\sum m_{\nu}$
varied'' setup in which the neutrino mass is allowed
to vary with a flat prior in the range $0<\sum m_{\nu}<0.5$~eV.
The credible interval on $S_8$ is compared with the
fiducial case in Figure~\ref{fig:s8ranges} (see panel (e) of
Figure~\ref{fig:om_sig8_sys} for $\sigma_8$-$\Omega_m$ constraints).
It is found from this comparison result that the non-zero neutrino mass
indeed has little impact on our cosmological constraints. 

%
%
\subsubsection{Residual constant shear}
\label{sec:impact_const_shear}

Here we check the robustness of our fiducial cosmological constraints
against the residual constant shear that is not included in our
fiducial model.
To do so, we test the same setup as 
in the fiducial case
but including a contribution from residual constant shear
by equation~(\ref{eq:En-c-model}) with a
nuisance parameter $A_{\langle \gamma \rangle}$ that controls its
amplitude (``w/ const-$\gamma$'' setup, see
Appendix~\ref{sec:cosebis_constant_shear} for details). 
We adopt a flat prior in the range $-5<A_{\langle \gamma \rangle}<5$.
The derived $S_8$ constraint is compared with the fiducial case in
Figure~\ref{fig:s8ranges} (see also panel (f) of
Figure~\ref{fig:om_sig8_sys}).
We find that the resulting changes in the cosmological constraints are
very small. 
The marginalized one-dimensional posterior distribution of $A_{\langle \gamma \rangle}$ 
is shown in Fig~\ref{fig:posterior_3addedpar}, and its mean and $\sigma$
are found to be $A_{\langle \gamma \rangle}=0.0\pm 1.1$, which is
consistent with the constant shear expected from the cosmic shear that
is coherent over the field (see Appendix~1 of H20).

%
%
\subsubsection{PSF leakage and PSF modeling errors}
\label{sec:impact_psf}

Following the previous HSC-Y1 cosmic shear analyses (H19 and H20), we
employ the conventional model for the PSF leakage and PSF
modeling errors given by equation~(\ref{eq:g_psf}).
We then derive an empirical model for a contribution to the measured
E-mode COSEBIs by equation~(\ref{eq:En-psf-model}) with a nuisance
parameter $A_{\rm psf}$ that controls its amplitude (see
Appendix~\ref{sec:cosebis_psf} for details), for which we adopt a
Gaussian prior of $({\rm mean}, \sigma)=(1,1)$.
Marginalized one-dimensional posterior distributions of these parameters 
from our fiducial analysis are shown in Figure~\ref{fig:posterior16}.
We find that the posterior is largely determined by the prior.
We also find that the marginalized constraint on $A_{\rm psf}$ is not
strongly correlated with either $\Omega_m$, $\sigma_8$, or $S_8$.

In order to check the robustness of our cosmological constraints against
these systematics, we test the same setup as the fiducial case but
ignoring them i.e., setting $A_{\rm psf} =0$ (``w/o PSF error'' setup).
The result is shown in Figure~\ref{fig:s8ranges} (see also panel (g) of
Figure~\ref{fig:om_sig8_sys}).
We find that the changes in the cosmological constraints are very
small as a natural consequence of PSF leakage and PSF modeling
errors 
being smaller than the size of errors on the HSC-Y1 cosmic shear COSEBIs.

%
%
\subsubsection{Shear calibration error}
\label{sec:impact_deltam}

In our fiducial analysis we also take account of the uncertainty 
in the shear multiplicative bias correction using a simple model, 
equation~(\ref{eq:delta_m}), with a Gaussian prior corresponding to 
a 1\% uncertainty (see Section~\ref{sec:Systematic_parameters}).
The marginalized one-dimensional posterior distribution of the model 
parameter $\Delta m$ from our fiducial analysis is shown in 
Figure~\ref{fig:posterior16}, which indicates that the posterior 
is dominated by the prior.

In order to check the effect of this residual calibration bias on our
cosmological constraints, we test the same setup as the fiducial case but
ignoring the nuisance parameter i.e., setting $\Delta m =0$ (``w/o
$\Delta m$'' setup).
The result is shown in Figure~\ref{fig:s8ranges} (see also panel (h) of
Figure~\ref{fig:om_sig8_sys}).
We find that the changes in the cosmological constraints are very small.

%
%
\subsubsection{Source redshift distribution errors}
\label{sec:impact_dz}

In our fiducial analysis, we introduce nuisance parameters $\Delta z_a$,
which represent a shift of each of the source redshift distributions
To be specific, the source redshift distribution, $p^a(z)$, is shifted by 
$p^a(z) \rightarrow  p^a(z+\Delta z_a)$ (see
  Section~\ref{sec:Systematic_parameters}).
Marginalized one-dimensional posterior distributions of 
these parameters from our fiducial analysis are shown in 
Figure~\ref{fig:posterior16}.
Although peak positions of these posteriors show shifts from the
peak the prior distributions, the sizes of the shifts are within
1$\sigma$ of the Gaussian priors, and thus are not statistically
significant. 
We note that a relatively large shift seen for the lowest
redshift bin is also seen in HSC-Y1 TPCF analysis with
a similar size and direction (H20), and thus it may
indicate an unknown bias in estimation of the source redshift
distribution that is not captured in the prior knowledge.

In order to check the robustness of our cosmological constraints against
these uncertainties, we test the same setup as the fiducial analysis but
ignoring these parameters (``w/o $p(z)$ error'' setup).
The result is shown in Figure~\ref{fig:s8ranges} (see also panel (i) of
Figure~\ref{fig:om_sig8_sys}).
We find that the changes in the cosmological constraints are very small.

In addition, following H20, we also perform an empirical test; we replace
the default source redshift distributions derived from the COSMOS
re-weighted method with ones derived from stacked PDFs with three
photo-$z$ methods, {\tt DEmP}, {\tt Ephor AB}, and {\tt FRANKEN-Z}
(see Section~2.2 of H20 for details).
Other settings remain the same as the fiducial setup. 
The results are shown in Figure~\ref{fig:s8ranges} (labeled as
``$P(z)$~{\it method}'', see also panels (j),
(k), and (l) of Figure~\ref{fig:om_sig8_sys}).
Again, we find that the changes in the
cosmological constraints are not significant, and thus we conclude that
no additional systematics are identified from this test.

%
%
\subsection{Internal consistency}
\label{sec:internal_consistency}

%
%
\begin{figure}
\begin{center}
  \includegraphics[height=82mm,angle=-90]{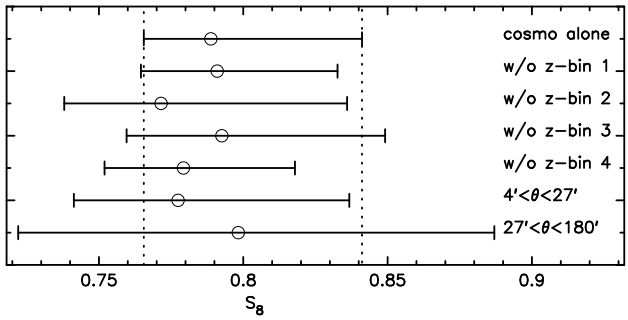}
\end{center}
\caption{Means and 68\% credible intervals of
  marginalized one-dimensional constraints on $S_8$.
  The ``cosmology alone'' case (top) is compared with different setups 
  for internal consistency checks.
  Vertical dotted lines show the 68\% credible interval of
  the cosmology alone case.
  See Figure~\ref{fig:om_sig8_sys3}, for comparisons of constraints
    in the $\Omega_m$-$\sigma_8$ plane. 
  \label{fig:s8ranges_ic}}
\end{figure}

Here we present results of internal consistency checks in which 
we derive cosmological constraints from subsets of the data vector, and
from data vectors generated with smaller-/larger-half scale-cuts 
than the original scale-cut range, and then
we compare derived cosmological constraints 
with 
the ones from a reference setup.
In doing so, following H20, we do not use the fiducial results as the
reference, but instead we adopt the results from the ``cosmology alone''
setup in which we include neither systematics nor astrophysical
parameters but only five cosmological parameters are included as a
baseline for comparison. 
The reason for this choice is to avoid undesirable changes in nuisance
parameters, which may add or cancel out shifts in parameter constraints.
Of course, for this test to be meaningful, the reference setup must
provide cosmological constraints that are consistent with ones from
fiducial case, which we explicitly confirmed.

%
%
\subsubsection{Tomographic redshift bins}
\label{sec:internalcheck_redshiftbin}

First, we exclude one of the four redshift bins, and perform
the cosmological inference with 
the remaining three tomographic bins.
The resulting marginalized constraints on $S_8$ are shown in
Figure~\ref{fig:s8ranges_ic} (see also panels of (a) to (d) of
Figure~\ref{fig:om_sig8_sys3} for 
marginalized constraints in $\Omega_m$-$\sigma_8$ plane). 
We find that constraints on $S_8$ from test setups are consistent within 
1$\sigma$ of the reference result.
Also it is seen in Figure~\ref{fig:om_sig8_sys3} that 68\% credible
contours in the $\Omega_m$-$\sigma_8$ plane in these cases largely
overlap with the reference contour.  
Thus we conclude that no significant internal inconsistency is found
from this test.

%
%
\subsubsection{Different scale-cuts}
\label{sec:internalcheck_scalecuts}

Next, we check the internal consistency among different angular ranges
by splitting the original scale-cut ($4\arcmin<\theta<180\arcmin$) into two
scale-cuts ($4\arcmin<\theta<27\arcmin$ and
$27\arcmin<\theta<180\arcmin$). 
We generate the data vectors and covariance
matrices for these two scale-cuts 
following the same
procedure as for the fiducial scale-cut.
The resulting marginalized constraints on $S_8$ are shown in
Figure~\ref{fig:s8ranges_ic}. 
As expected, the credible intervals are larger than the reference
case. We see that the smaller-half scale-cut has more constraining power
than the larger-half scale-cut.
In addition, some more information on this can be seen in the
comparison plots in the $\Omega_m$-$\sigma_8$ plane (panels of (e) and
(f) of Figure~\ref{fig:om_sig8_sys3}): The smaller-half scale-cut places
constraint contours which are as tight as the reference contours in
the $S_8$ direction (perpendicular to the $\Omega_m$-$\sigma_8$
degeneracy direction), but are very elongated in the degeneracy
direction. Thus from these results we see that a
wider angular range is effective in placing a tighter constraint in
$\Omega_m$-$\sigma_8$ plane.
Overall, the results from the larger/smaller-half scale-cuts are consistent
with the reference results.
Thus we again conclude that no significant internal inconsistency is found
from this test.

%
%
\subsection{$w$CDM model}
\label{sec:wcdm}

%
%
\begin{figure}
\begin{center}
  \includegraphics[height=82mm,angle=-90]{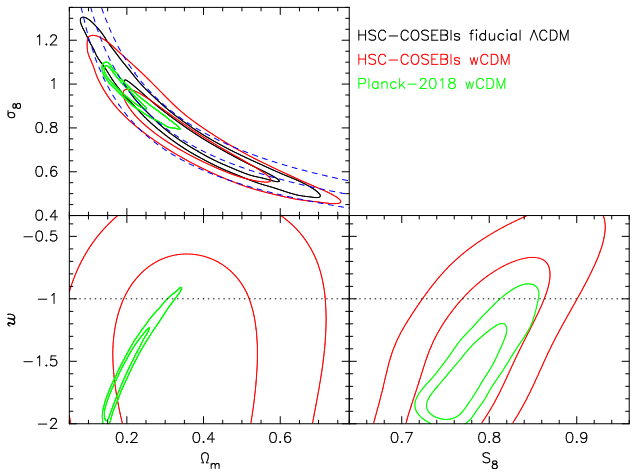}
\end{center}
\caption{Marginalized posterior contours (68\% and 95\%
  confidence levels) in the $\Omega_m$-$\sigma_8$ plane (top), the
  $\Omega_m$-$w$ plane (bottom left) and the $S_8$-$w$ plane (bottom
  right) in the $w$CDM model are shown by red contours. Constraints from
  the fiducial $\Lambda$CDM model are shown by
  black contours, and
  {\it Planck} 2018 results for the $w$CDM model
  \citep[][TT+TE+EE+lowE]{2020A&A...641A...6P} are also shown by green
  contours. 
  In the $\Omega_m$-$\sigma_8$ panel, three constant $S_8$ loci
  ($S_8=0.7$, 0.8 and 0.9 from bottom to top, respectively) are shown by
  dashed blue curves.
  \label{fig:om_sig8_w}}
\end{figure}

In addition to the fiducial $\Lambda$CDM model, we test one 
extended
model by including the time-independent dark energy equation of state
parameter $w$, although it was found in H19 and H20 that HSC-Y1 cosmic
shear two-point statistics alone cannot place a useful constraint on $w$.
We allow $w$ to vary with a flat prior in the range
$-2<w<-1/3$. The setup of the other parameters are 
the same as the fiducial $\Lambda$CDM model.

The marginalized constraints in the $\Omega_m$-$\sigma_8$, $\Omega_m$-$w$, and 
$S_8$-$w$ planes are shown in Figure~\ref{fig:om_sig8_w}, along with
constraints from the fiducial $\Lambda$CDM model and the {\it Planck}
2018 results for the  $w$CDM model
\citep[][TT+TE+EE+lowE]{2020A&A...641A...6P}.
As can be seen from the Figure, our constraints on those parameters
  are consistent with the {\it Planck}'s results, although {\it
    Planck}'s constraints are much tighter than our constraints.
Marginalized one-dimensional constraint range of $S_8$ is shown in
Figure~\ref{fig:s8ranges}.
It is found that the derived two-dimensional marginalized posterior
distributions are similar to those obtained from the HSC-Y1 cosmic shear
TPCF 
(see figure 14 of H20)
and PS (see figure 16 of H19) analyses, as expected. 

%
%
\subsection{Comparison to other constraints from the literature}
\label{sec:comparison}

%
%
\begin{figure}
\begin{center}
  \includegraphics[height=82mm,angle=-90]{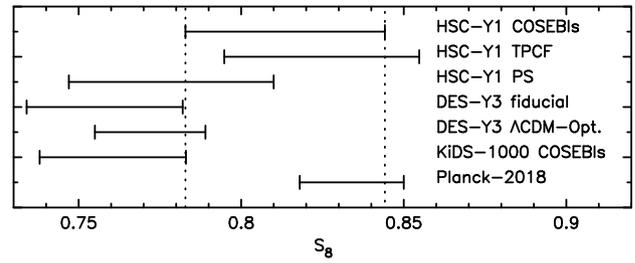}
\end{center}
\caption{68\% credible intervals of marginalized posterior distributions
  of $S_8=\sigma_8\sqrt{\Omega_m/0.3}$ or that of the projected joint
  highest posterior density (PJ-HPD) for KiDS-1000. 
  Our result from the fiducial
  $\Lambda$CDM model is compared with other results in the 
  literature, HSC first year (HSC-Y1)
  cosmic shear TPCF (H20), PS (H19), DES-Y3
  cosmic shear TPCFs \citep[their fiducial and $\Lambda$CDM-Optimized
    models are shown, see][ for details]{2022PhRvD.105b3514A}, KiDS-1000 cosmic
  shear COSEBIs \citep{2021A&A...645A.104A}, and {\it Planck} 2018 CMB
  \citep[][TT+TE+EE+lowE]{2020A&A...641A...6P}.
  Since different studies adopt different definitions of the
  central values (mean, median or peak of the marginalized posterior
  distribution, or the multivariate maximum posterior), central values
  are not shown to avoid possible misunderstanding.
Note that since different studies adopt different priors
  and modeling, part of the differences in the credible intervals may be
  due to the different choices of priors and modeling.
See Figure~\ref{fig:om_sig8_others} for comparisons of constraints from
those studies in the $\Omega_m$-$\sigma_8$ plane.
  \label{fig:s8ranges_others}}
\end{figure}

Next, we compare the cosmological constraints from our fiducial
$\Lambda$CDM model with results from other cosmic shear and CMB 
measurements.
Figure~\ref{fig:s8ranges_others} compares the 68\% credible
intervals of $S_8=\sigma_8\sqrt{\Omega_m/0.3}$, where results of other projects 
are taken from the literature.
See also Figure~\ref{fig:om_sig8_others} for comparison to other results
in $\Omega_m$-$\sigma_8$ plane.

DES-Y3 \citep{2022PhRvD.105b3514A} covers 
a much larger area
(4143~deg$^2$) than the HSC-Y1, yielding tighter constraints
than our fiducial results.
Note that in their cosmological analyses, the neutrino mass density is
allowed to vary.
They provide results from two models; their fiducial model adopts a very
conservative small-scale cut yielding a broader constraint 
than the one from
their $\Lambda$CDM-Optimized model in which smaller scale information are
used safely \citep[see][]{2022PhRvD.105b3514A}.
In both cases, their $S_8$ constraints are lower than ours, and their
$\Lambda$CDM-Optimized model is in about 1.0$\sigma$ difference\footnote{In
estimating a statistical significance of the difference between two measurements of $S_8$, we
adopt a conventional method, $\delta =
(\bar{A}-\bar{B})/\sqrt{\sigma_A^2+\sigma_B^2})$.} from our fiducial result
(see Figure~\ref{fig:s8ranges_others}). 

KiDS-1000 covers an effective area of 777.4~deg$^2$
\citep{2021A&A...645A.104A}.
They employed three two-points statistics; the TPCF, band power spectra, and
COSEBIs, and found cosmological results from these three to be in
excellent agreement.
Their $S_8$ value from COSEBIs analysis is lower than ours and is
in about 
1.4$\sigma$ difference from our fiducial result (see
Figure~\ref{fig:s8ranges_others}). 

The credible interval of $S_8$ from {\it Planck} 2018
\citep[][we take TT+TE+EE+lowE without CMB lensing]{2020A&A...641A...6P}
is consistent with our result.
It is also found from Figure~\ref{fig:om_sig8_others} that the credible
contours in the $\Omega_m$-$\sigma_8$ plane from the 
{\it Planck} 2018 CMB result overlap well with our result.
We therefore conclude that there is no tension between
{\it Planck} 2018 constraints and HSC-Y1 cosmic shear COSEBIs constraints
as far as $S_8$, $\Omega_m$, and $\sigma_8$ are concerned.

%
%
\subsection{Comparison with HSC-Y1
  cosmic shear PS and TPCF results}
\label{sec:comparison_Hikage}

%
%
\begin{figure}
\begin{center}
  \includegraphics[width=82mm]{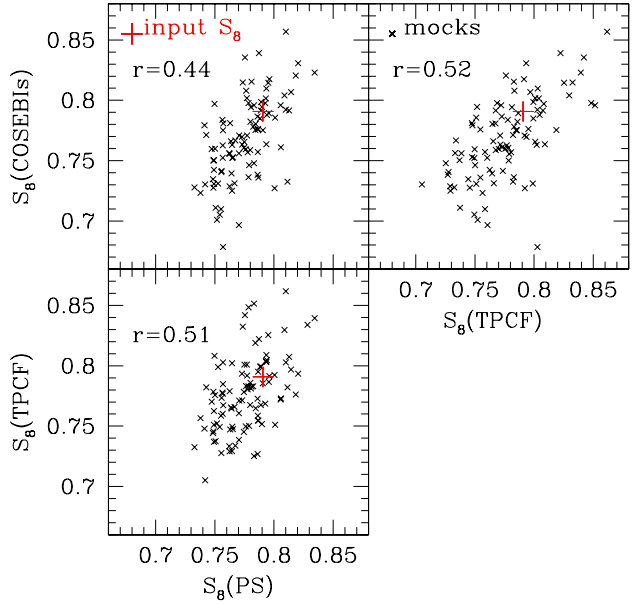}
\end{center}
\caption{Scatter plots showing median values of marginalized
  one-dimensional posterior distributions of $S_8$ derived from
  cosmological analyses on 100 mock catalogs.
  In each of three panels, results from any one of the PS, TPCF, or COSEBIs
  analysis are compared to those from either of the rest. The red
  cross shows the value of $S_8$ adopted in generating the mock
  catalogs. The cross correlation coefficient defined by
  equation~(\ref{eq:covariance_cosmopar}) is
  denoted in each panel. \label{fig:s8median_cl_cosebis_tpcf}}
\end{figure}

%
%
\begin{figure}
\begin{center}
  \includegraphics[width=82mm]{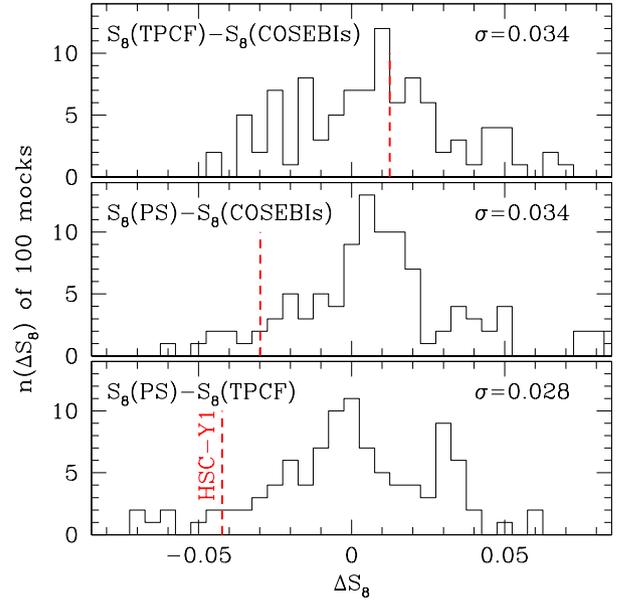}
\end{center}
\caption{Frequency distributions of the differences in median $S_8$ values
  derived from any two of the PS, TPCF and COSEBIs analyses on 100 mock
  catalogs. The root-mean-square value of the distribution ($\sigma$) is
  denoted in each panel. Corresponding results from the real HSC-Y1
  cosmic shear analyses are shown by the vertical red dashed line in
  each panel. 
  \label{fig:diff_s8_median}}
\end{figure}

Finally, we compare the cosmological constraints from our fiducial
model with HSC-Y1 cosmic shear PS (H19) and TPCF (H20) results.
As seen in Figure~\ref{fig:s8ranges_others}, our constraint
on $S_8$ is in good agreement with TPCF
results (H20), and is in mild agreement with PS results (H19).
Similar comparison results are found for two-dimensional constraint
contours in $\Omega_m$-$\sigma_8$ plane presented in panel (a) of
Figure~\ref{fig:om_sig8_others}, although
our contours are much broader than theirs.
Although no strong evidence of inconsistency among these results is seen
from those results, we examine the consistency 
between them,
paying particular attention to the fact that those analyses share the
same HSC-Y1 data set with the same tomographic redshift binning and
adopt a similar analysis setup.

We perform the same cross-correlation analysis as one done in H20 (see
Section~6.7 of H20), in which cosmological inferences on the 100
realistic mock catalogs \citep[see also][]{2018PASJ...70S..26O} are performed with
the same fiducial setups of 
HSC-Y1 PS and TPCF analyses except for ignoring PSF modeling errors because no
PSF modeling error is added in the mock data, and then derived cosmological
constraints are compared.
We apply our COSEBIs analysis on the same mock catalogs but ignoring
PSF modeling errors as well.
We present the scatter plots comparing $S_8$ values (we take the median
of a posterior distribution) derived from these
three cosmological analyses on the same mock catalogs in
Figure~\ref{fig:s8median_cl_cosebis_tpcf}. 
We compute the standard cross-correlation coefficient of those distributions,
\begin{equation}
  \label{eq:covariance_cosmopar}
  r(S_8)={ {\mbox{Cov}(S_8^A,S_8^B)} \over
    { {\mbox{Cov}(S_8^A,S_8^A)^{1/2}}
      {\mbox{Cov}(S_8^B,S_8^B)^{1/2}}}},
\end{equation}
where the superscripts $A$ and $B$ stand for an analysis method (i.e., PS, TPCF or
COSEBIs). We show the results in each panel of
Figure~\ref{fig:s8median_cl_cosebis_tpcf}.
We find $r(S_8) \sim 0.5$ for all the three cases, which indicates that 
the correlations between derived cosmological constraints among the three
analyses are mild.
Although qualitative, this result can be intuitively understood as each
analysis method is sensitive to different multipole ranges, as seen in
Figure~\ref{fig:dcosen_dl}. 
In Figure~\ref{fig:diff_s8_median}, we present frequency distributions
of the differences in the $S_8$ values ($\Delta S_8 = S_8^A-S_8^B$)
derived from any pair of the PS, TPCF and COSEBIs analyses along with corresponding
results from the real HSC-Y1 cosmic shear analyses (shown by the
vertical red dashed line). We also compute the root-mean-square values
of these distributions ($\sigma$) which are denoted in each panel.
We find that derived $\sigma$ values for $\Delta S_8$ are only slightly
smaller than ones for parents $S_8$ (0.022, 0.031, and 0.038 for PS,
TPCF and COSEBIs, respectively), which confirms mild
correlations between derived cosmological constraints among the three
analyses.
We note that similar results but for the COSEBIs, band power spectra, and
TPCF analyses are found in KiDS-1000 cosmic shear study by
\citet{2021A&A...645A.104A}. 
It follows from the above results that 
the $\Delta S_8$ values from three real HSC-Y1 cosmic shear analyses are
explained by statistical fluctuations below 1.5$\sigma$.

%
%
\section{Summary and conclusions}
\label{sec:summary}

We have presented a cosmological analysis of the cosmic shear COSEBIs 
measured from the HSC-Y1 data, covering 136.9~deg$^2$.
Photometric redshifts derived from the HSC five-band photometry are
adopted to select galaxies and divided into four tomographic
redshift bins ranging from $z=0.3$ to $1.5$ with equal widths of 
$\Delta z =0.3$. 
The total number of selected galaxies is 9.6 million.
We compute tomographic E/B-mode COSEBIs from measured cosmic shear TPCFs
on the angular range of $4\arcmin<\theta<180\arcmin$.

In addition to the HSC-Y1 data set, we have used HSC-Y1 mock shape
catalogs constructed from full-sky gravitational lensing simulations
\citep{2017ApJ...850...24T} that fully
take account of the survey geometry and measurement noise
\citep{2019MNRAS.486...52S}.
Using 2268 mock realizations, we have derived the E/B-mode covariance
matrices adopted in our cosmological analysis.
The mock catalogs are also used to assess the statistical significance of
some of our results.

We have quantitatively tested the consistency of the B-mode COSEBIs
signals with zero using the standard $\chi^2$ statistics, and found
$\chi^2=66.1$ for 50 degrees-of-freedom, corresponding to the $p$-value of 0.063.
We thus conclude that no evidence of significant B-mode shear is found.

We have performed a standard Bayesian likelihood analysis for 
the cosmological inference of the measured E-mode COSEBIs.
Our fiducial $\Lambda$CDM model consists of five cosmological
parameters and includes contributions from intrinsic alignment of
galaxies (with 2 parameters for the nonlinear alignment IA model) as
well as six nuisance parameters
(1 for PSF errors, 1 for shear calibration error, and 4 for source redshift
distribution errors).
We have found that our fiducial model fits the measured E-mode COSEBIs
signals very well with a minimum $\chi^2$ of 40.7 for 47 effective
degrees-of-freedom.
Derived marginalized one-dimensional constraint on $S_8$ is
$S_8=0.809_{-0.026}^{+0.036}$ (mean and 68\% credible interval).

We have carefully checked the robustness of our cosmological results
against astrophysical uncertainties in modeling and systematics
uncertainties in measurements.
The former includes the intrinsic alignment of galaxies and the baryonic
feedback effect on the nonlinear matter power spectrum, and the 
latter includes PSF errors, shear calibration error, errors in the 
estimation of source redshift distributions, and a residual constant 
shear over fields. 
We have tested the validity of our treatment of those uncertainties by
changing parameter setups or by adopting empirical or extreme models for
them. 
We have found that none of these uncertainties has a significant impact
on the cosmological constraints.
We have also confirmed the internal consistency of our results among
different choices of tomographic redshift bins and different scale-cuts.

We have examined, using mock HSC-Y1 catalogs, the consistency in 
the $S_8$ constraints
derived from the three HSC-Y1 cosmic shear two-point statistics, the PS
(H19), TPCF (H20), and COSEBIs (this study), which share the same HSC-Y1
data set with the same tomographic redshift binning and adopt a similar
analysis setup. 
We have found that correlations between derived $S_8$ constraints
among the three analyses are mild, most likely because each analysis
method is sensitive to different multipole ranges of cosmic shear power
spectra. 
We have also found that differences in the derived $S_8$ values between
different analyses are explained by statistical fluctuations
below 1.5$\sigma$.

We have compared our $S_8$ constraint with those obtained from
DES-Y3 \citep{2022PhRvD.105b3514A} and KiDS-1000 \citep{2021A&A...645A.104A}.
We have found that our $S_8$ value is higher than theirs such that 68\%
credible intervals only slightly overlap with each other at the
edges. 
Quantitatively, $S_8$ constraint from DES-Y3 (their $\Lambda$CDM-Optimized model)
is in about 1.0$\sigma$ difference, and that of KiDS-1000 (their COSEBIs
result) is in about 1.4$\sigma$ difference from our fiducial result (see
Figure~\ref{fig:s8ranges_others}).
We have also found that the $S_8$ constraint from  {\it Planck}
\citep{2020A&A...641A...6P} is consistent with our result.

%
%
\begin{ack}
We would like to thank the anonymous referee for many constructive comments
on the earlier manuscript which improved the paper.
We would like to thank members of the HSC weak lensing working group for
useful discussions.
We would like to thank R.~Takahashi for
making full-sky gravitational lensing simulation data publicity
available. 
We would like to thank Martin Kilbinger for making the software {\tt
  Athena} publicly available, Antony Lewis and Anthony Challinor for
making the software {\tt CAMB} publicly available, {\tt MultiNest}
developers for {\tt MultiNest} publicly available, HEALPix team for
HEALPix software publicity available, and Nick Kaiser for making the
software {\tt imcat} publicly available.

This work was supported in part by JSPS KAKENHI
Grant Number JP15H05892, JP17K05457, JP20H05856, JP20H00181 and 22K03655.

Data analysis were in part carried out on PC cluster at Center for
Computational Astrophysics, National Astronomical Observatory of
Japan. Numerical computations were in part carried out on Cray XC30 and
XC50 at Center for Computational Astrophysics, National Astronomical
Observatory of Japan, and also on Cray XC40 at YITP in Kyoto
University.

The Hyper Suprime-Cam (HSC) collaboration includes the astronomical
communities of Japan and Taiwan, and Princeton University.  The HSC
instrumentation and software were developed by the National Astronomical
Observatory of Japan (NAOJ), the Kavli Institute for the Physics and
Mathematics of the Universe (Kavli IPMU), the University of Tokyo, the
High Energy Accelerator Research Organization (KEK), the Academia Sinica
Institute for Astronomy and Astrophysics in Taiwan (ASIAA), and
Princeton University.  Funding was contributed by the FIRST program from
the Japanese Cabinet Office, the Ministry of Education, Culture, Sports,
Science and Technology (MEXT), the Japan Society for the Promotion of
Science (JSPS), Japan Science and Technology Agency  (JST), the Toray
Science  Foundation, NAOJ, Kavli IPMU, KEK, ASIAA, and Princeton
University. 

This paper is based on data collected at the Subaru Telescope and
retrieved from the HSC data archive system, which is operated by Subaru
Telescope and Astronomy Data Center (ADC) at NAOJ. Data analysis was in
part carried out with the cooperation of Center for Computational
Astrophysics (CfCA) at NAOJ.  We are honored and grateful for the
opportunity of observing the Universe from Maunakea, which has the
cultural, historical and natural significance in Hawaii. 

This paper makes use of software developed for Vera C. Rubin
Observatory. We thank the Rubin Observatory for making their code
available as free software at \url{http://pipelines.lsst.io/}. 
 
The Pan-STARRS1 Surveys (PS1) and the PS1 public science archive have
been made possible through contributions by the Institute for Astronomy,
the University of Hawaii, the Pan-STARRS Project Office, the Max Planck
Society and its participating institutes, the Max Planck Institute for
Astronomy, Heidelberg, and the Max Planck Institute for Extraterrestrial
Physics, Garching, The Johns Hopkins University, Durham University, the
University of Edinburgh, the Queen’s University Belfast, the
Harvard-Smithsonian Center for Astrophysics, the Las Cumbres Observatory
Global Telescope Network Incorporated, the National Central University
of Taiwan, the Space Telescope Science Institute, the National
Aeronautics and Space Administration under grant No. NNX08AR22G issued
through the Planetary Science Division of the NASA Science Mission
Directorate, the National Science Foundation grant No. AST-1238877, the
University of Maryland, Eotvos Lorand University (ELTE), the Los Alamos
National Laboratory, and the Gordon and Betty Moore Foundation.
\end{ack}

\bibliographystyle{apj}

\begin{thebibliography}{}
\expandafter\ifx\csname natexlab\endcsname\relax\def\natexlab#1{#1}\fi

\bibitem[{{Aihara} {et~al.}(2018{\natexlab{a}}){Aihara}, {Armstrong},
  {Bickerton}, {Bosch}, {Coupon}, {Furusawa}, {Hayashi}, {Ikeda}, {Kamata},
  {Karoji}, {Kawanomoto}, {Koike}, {Komiyama}, {Lang}, {Lupton}, {Mineo},
  {Miyatake}, {Miyazaki}, {Morokuma}, {Obuchi}, {Oishi}, {Okura}, {Price},
  {Takata}, {Tanaka}, {Tanaka}, {Tanaka}, {Uchida}, {Uraguchi}, {Utsumi},
  {Wang}, {Yamada}, {Yamanoi}, {Yasuda}, {Arimoto}, {Chiba}, {Finet},
  {Fujimori}, {Fujimoto}, {Furusawa}, {Goto}, {Goulding}, {Gunn}, {Harikane},
  {Hattori}, {Hayashi}, {He{\l}miniak}, {Higuchi}, {Hikage}, {Ho}, {Hsieh},
  {Huang}, {Huang}, {Imanishi}, {Iwata}, {Jaelani}, {Jian}, {Kashikawa},
  {Katayama}, {Kojima}, {Konno}, {Koshida}, {Kusakabe}, {Leauthaud}, {Lee},
  {Lin}, {Lin}, {Mandelbaum}, {Matsuoka}, {Medezinski}, {Miyama}, {Momose},
  {More}, {More}, {Mukae}, {Murata}, {Murayama}, {Nagao}, {Nakata}, {Niida},
  {Niikura}, {Nishizawa}, {Oguri}, {Okabe}, {Ono}, {Onodera}, {Onoue}, {Ouchi},
  {Pyo}, {Shibuya}, {Shimasaku}, {Simet}, {Speagle}, {Spergel}, {Strauss},
  {Sugahara}, {Sugiyama}, {Suto}, {Suzuki}, {Tait}, {Takada}, {Terai}, {Toba},
  {Turner}, {Uchiyama}, {Umetsu}, {Urata}, {Usuda}, {Yeh}, \&
  {Yuma}}]{2018PASJ...70S...8A}
{Aihara}, H., {Armstrong}, R., {Bickerton}, S., {et~al.} 2018{\natexlab{a}},
  \pasj, 70, S8

\bibitem[{{Aihara} {et~al.}(2018{\natexlab{b}}){Aihara}, {Arimoto},
  {Armstrong}, {Arnouts}, {Bahcall}, {Bickerton}, {Bosch}, {Bundy}, {Capak},
  {Chan}, {Chiba}, {Coupon}, {Egami}, {Enoki}, {Finet}, {Fujimori}, {Fujimoto},
  {Furusawa}, {Furusawa}, {Goto}, {Goulding}, {Greco}, {Greene}, {Gunn},
  {Hamana}, {Harikane}, {Hashimoto}, {Hattori}, {Hayashi}, {Hayashi},
  {He{\l}miniak}, {Higuchi}, {Hikage}, {Ho}, {Hsieh}, {Huang}, {Huang},
  {Ikeda}, {Imanishi}, {Inoue}, {Iwasawa}, {Iwata}, {Jaelani}, {Jian},
  {Kamata}, {Karoji}, {Kashikawa}, {Katayama}, {Kawanomoto}, {Kayo}, {Koda},
  {Koike}, {Kojima}, {Komiyama}, {Konno}, {Koshida}, {Koyama}, {Kusakabe},
  {Leauthaud}, {Lee}, {Lin}, {Lin}, {Lupton}, {Mandelbaum}, {Matsuoka},
  {Medezinski}, {Mineo}, {Miyama}, {Miyatake}, {Miyazaki}, {Momose}, {More},
  {More}, {Moritani}, {Moriya}, {Morokuma}, {Mukae}, {Murata}, {Murayama},
  {Nagao}, {Nakata}, {Niida}, {Niikura}, {Nishizawa}, {Obuchi}, {Oguri},
  {Oishi}, {Okabe}, {Okamoto}, {Okura}, {Ono}, {Onodera}, {Onoue}, {Osato},
  {Ouchi}, {Price}, {Pyo}, {Sako}, {Sawicki}, {Shibuya}, {Shimasaku},
  {Shimono}, {Shirasaki}, {Silverman}, {Simet}, {Speagle}, {Spergel},
  {Strauss}, {Sugahara}, {Sugiyama}, {Suto}, {Suyu}, {Suzuki}, {Tait},
  {Takada}, {Takata}, {Tamura}, {Tanaka}, {Tanaka}, {Tanaka}, {Tanaka},
  {Terai}, {Terashima}, {Toba}, {Tominaga}, {Toshikawa}, {Turner}, {Uchida},
  {Uchiyama}, {Umetsu}, {Uraguchi}, {Urata}, {Usuda}, {Utsumi}, {Wang}, {Wang},
  {Wong}, {Yabe}, {Yamada}, {Yamanoi}, {Yasuda}, {Yeh}, {Yonehara}, \&
  {Yuma}}]{2018PASJ...70S...4A}
{Aihara}, H., {Arimoto}, N., {Armstrong}, R., {et~al.} 2018{\natexlab{b}},
  \pasj, 70, S4

\bibitem[{{Amon} {et~al.}(2022){Amon}, {Gruen}, {Troxel}, {MacCrann},
  {Dodelson}, {Choi}, {Doux}, {Secco}, {Samuroff}, {Krause}, {Cordero},
  {Myles}, {DeRose}, {Wechsler}, {Gatti}, {Navarro-Alsina}, {Bernstein},
  {Jain}, {Blazek}, {Alarcon}, {Fert{\'e}}, {Lemos}, {Raveri}, {Campos},
  {Prat}, {S{\'a}nchez}, {Jarvis}, {Alves}, {Andrade-Oliveira}, {Baxter},
  {Bechtol}, {Becker}, {Bridle}, {Camacho}, {Carnero Rosell}, {Carrasco Kind},
  {Cawthon}, {Chang}, {Chen}, {Chintalapati}, {Crocce}, {Davis}, {Diehl},
  {Drlica-Wagner}, {Eckert}, {Eifler}, {Elvin-Poole}, {Everett}, {Fang},
  {Fosalba}, {Friedrich}, {Gaztanaga}, {Giannini}, {Gruendl}, {Harrison},
  {Hartley}, {Herner}, {Huang}, {Huff}, {Huterer}, {Kuropatkin}, {Leget},
  {Liddle}, {McCullough}, {Muir}, {Pandey}, {Park}, {Porredon}, {Refregier},
  {Rollins}, {Roodman}, {Rosenfeld}, {Ross}, {Rykoff}, {Sanchez},
  {Sevilla-Noarbe}, {Sheldon}, {Shin}, {Troja}, {Tutusaus}, {Tutusaus},
  {Varga}, {Weaverdyck}, {Yanny}, {Yin}, {Zhang}, {Zuntz}, {Aguena}, {Allam},
  {Annis}, {Bacon}, {Bertin}, {Bhargava}, {Brooks}, {Buckley-Geer}, {Burke},
  {Carretero}, {Costanzi}, {da Costa}, {Pereira}, {De Vicente}, {Desai},
  {Dietrich}, {Doel}, {Ferrero}, {Flaugher}, {Frieman}, {Garc{\'\i}a-Bellido},
  {Gaztanaga}, {Gerdes}, {Giannantonio}, {Gschwend}, {Gutierrez}, {Hinton},
  {Hollowood}, {Honscheid}, {Hoyle}, {James}, {Kron}, {Kuehn}, {Lahav}, {Lima},
  {Lin}, {Maia}, {Marshall}, {Martini}, {Melchior}, {Menanteau}, {Miquel},
  {Mohr}, {Morgan}, {Ogando}, {Palmese}, {Paz-Chinch{\'o}n}, {Petravick},
  {Pieres}, {Romer}, {Sanchez}, {Scarpine}, {Schubnell}, {Serrano}, {Smith},
  {Soares-Santos}, {Tarle}, {Thomas}, {To}, {Weller}, \& {DES
  Collaboration}}]{2022PhRvD.105b3514A}
{Amon}, A., {Gruen}, D., {Troxel}, M.~A., {et~al.} 2022, \prd, 105, 023514

\bibitem[{{Anderson}(2003)}]{Anderson2003}
{Anderson}, T.~W. 2003, An introduction to multivariate statistical analysis,
  3rd edn. (Wiley-Interscience)

\bibitem[{{Asgari} {et~al.}(2017){Asgari}, {Heymans}, {Blake},
  {Harnois-Deraps}, {Schneider}, \& {Van Waerbeke}}]{2017MNRAS.464.1676A}
{Asgari}, M., {Heymans}, C., {Blake}, C., {et~al.} 2017, \mnras, 464, 1676

\bibitem[{{Asgari} {et~al.}(2012){Asgari}, {Schneider}, \&
  {Simon}}]{2012A&A...542A.122A}
{Asgari}, M., {Schneider}, P., \& {Simon}, P. 2012, \aap, 542, A122

\bibitem[{{Asgari} {et~al.}(2020){Asgari}, {Tr{\"o}ster}, {Heymans},
  {Hildebrandt}, {van den Busch}, {Wright}, {Choi}, {Erben}, {Joachimi},
  {Joudaki}, {Kannawadi}, {Kuijken}, {Lin}, {Schneider}, \&
  {Zuntz}}]{2020A&A...634A.127A}
{Asgari}, M., {Tr{\"o}ster}, T., {Heymans}, C., {et~al.} 2020, \aap, 634, A127

\bibitem[{{Asgari} {et~al.}(2021){Asgari}, {Lin}, {Joachimi}, {Giblin},
  {Heymans}, {Hildebrandt}, {Kannawadi}, {St{\"o}lzner}, {Tr{\"o}ster}, {van
  den Busch}, {Wright}, {Bilicki}, {Blake}, {de Jong}, {Dvornik}, {Erben},
  {Getman}, {Hoekstra}, {K{\"o}hlinger}, {Kuijken}, {Miller}, {Radovich},
  {Schneider}, {Shan}, \& {Valentijn}}]{2021A&A...645A.104A}
{Asgari}, M., {Lin}, C.-A., {Joachimi}, B., {et~al.} 2021, \aap, 645, A104

\bibitem[{{Barreira} {et~al.}(2018){Barreira}, {Krause}, \&
  {Schmidt}}]{2018JCAP...10..053B}
{Barreira}, A., {Krause}, E., \& {Schmidt}, F. 2018, \jcap, 2018, 053

\bibitem[{{Bird} {et~al.}(2012){Bird}, {Viel}, \&
  {Haehnelt}}]{2012MNRAS.420.2551B}
{Bird}, S., {Viel}, M., \& {Haehnelt}, M.~G. 2012, \mnras, 420, 2551

\bibitem[{{Bridle} \& {King}(2007)}]{2007NJPh....9..444B}
{Bridle}, S., \& {King}, L. 2007, New Journal of Physics, 9, 444

\bibitem[{{Camacho} {et~al.}(2021){Camacho}, {Andrade-Oliveira}, {Troja},
  {Rosenfeld}, {Faga}, {Gomes}, {Doux}, {Fang}, {Lima}, {Miranda}, {Eifler},
  {Friedrich}, {Gatti}, {Bernstein}, {Blazek}, {Bridle}, {Choi}, {Davis},
  {DeRose}, {Gaztanaga}, {Gruen}, {Hartley}, {Hoyle}, {Jarvis}, {MacCrann},
  {Prat}, {Rau}, {Samuroff}, {S{\'a}nchez}, {Sheldon}, {Troxel}, {Vielzeuf},
  {Zuntz}, {Abbott}, {Aguena}, {Allam}, {Annis}, {Bacon}, {Bertin}, {Brooks},
  {Burke}, {Carnero Rosell}, {Carrasco Kind}, {Carretero}, {Castander},
  {Cawthon}, {Costanzi}, {da Costa}, {Pereira}, {De Vicente}, {Desai}, {Diehl},
  {Doel}, {Everett}, {Evrard}, {Ferrero}, {Flaugher}, {Fosalba}, {Friedel},
  {Frieman}, {Garc{\'\i}a-Bellido}, {Gerdes}, {Gruendl}, {Gschwend},
  {Gutierrez}, {Hinton}, {Hollowood}, {Honscheid}, {Huterer}, {James}, {Kuehn},
  {Kuropatkin}, {Lahav}, {Maia}, {Marshall}, {Melchior}, {Menanteau}, {Miquel},
  {Morgan}, {Paz-Chinch{\'o}n}, {Petravick}, {Pieres}, {Plazas Malag{\'o}n},
  {Reil}, {Rodriguez-Monroy}, {Sanchez}, {Scarpine}, {Schubnell}, {Serrano},
  {Sevilla-Noarbe}, {Smith}, {Soares-Santos}, {Suchyta}, {Tarle}, {Thomas},
  {To}, {Varga}, {Weller}, \& {Wilkinson}}]{2021arXiv211107203C}
{Camacho}, H., {Andrade-Oliveira}, F., {Troja}, A., {et~al.} 2021, arXiv
  e-prints, arXiv:2111.07203

\bibitem[{{Challinor} \& {Lewis}(2011)}]{2011PhRvD..84d3516C}
{Challinor}, A., \& {Lewis}, A. 2011, \prd, 84, 043516

\bibitem[{{Chisari} {et~al.}(2018){Chisari}, {Richardson}, {Devriendt},
  {Dubois}, {Schneider}, {Le Brun}, {Beckmann}, {Peirani}, {Slyz}, \&
  {Pichon}}]{2018MNRAS.480.3962C}
{Chisari}, N.~E., {Richardson}, M.~L.~A., {Devriendt}, J., {et~al.} 2018,
  \mnras, 480, 3962

\bibitem[{{Dark Energy Survey Collaboration} {et~al.}(2016){Dark Energy Survey
  Collaboration}, {Abbott}, {Abdalla}, {Aleksi{\'c}}, {Allam}, {Amara},
  {Bacon}, {Balbinot}, {Banerji}, {Bechtol}, {Benoit-L{\'e}vy}, {Bernstein},
  {Bertin}, {Blazek}, {Bonnett}, {Bridle}, {Brooks}, {Brunner}, {Buckley-Geer},
  {Burke}, {Caminha}, {Capozzi}, {Carlsen}, {Carnero-Rosell}, {Carollo},
  {Carrasco-Kind}, {Carretero}, {Castander}, {Clerkin}, {Collett}, {Conselice},
  {Crocce}, {Cunha}, {D'Andrea}, {da Costa}, {Davis}, {Desai}, {Diehl},
  {Dietrich}, {Dodelson}, {Doel}, {Drlica-Wagner}, {Estrada}, {Etherington},
  {Evrard}, {Fabbri}, {Finley}, {Flaugher}, {Foley}, {Fosalba}, {Frieman},
  {Garc{\'{\i}}a-Bellido}, {Gaztanaga}, {Gerdes}, {Giannantonio}, {Goldstein},
  {Gruen}, {Gruendl}, {Guarnieri}, {Gutierrez}, {Hartley}, {Honscheid}, {Jain},
  {James}, {Jeltema}, {Jouvel}, {Kessler}, {King}, {Kirk}, {Kron}, {Kuehn},
  {Kuropatkin}, {Lahav}, {Li}, {Lima}, {Lin}, {Maia}, {Makler}, {Manera},
  {Maraston}, {Marshall}, {Martini}, {McMahon}, {Melchior}, {Merson}, {Miller},
  {Miquel}, {Mohr}, {Morice-Atkinson}, {Naidoo}, {Neilsen}, {Nichol}, {Nord},
  {Ogando}, {Ostrovski}, {Palmese}, {Papadopoulos}, {Peiris}, {Peoples},
  {Percival}, {Plazas}, {Reed}, {Refregier}, {Romer}, {Roodman}, {Ross},
  {Rozo}, {Rykoff}, {Sadeh}, {Sako}, {S{\'a}nchez}, {Sanchez}, {Santiago},
  {Scarpine}, {Schubnell}, {Sevilla-Noarbe}, {Sheldon}, {Smith}, {Smith},
  {Soares-Santos}, {Sobreira}, {Soumagnac}, {Suchyta}, {Sullivan}, {Swanson},
  {Tarle}, {Thaler}, {Thomas}, {Thomas}, {Tucker}, {Vieira}, {Vikram},
  {Walker}, {Wechsler}, {Weller}, {Wester}, {Whiteway}, {Wilcox}, {Yanny},
  {Zhang}, \& {Zuntz}}]{2016MNRAS.460.1270D}
{Dark Energy Survey Collaboration}, {Abbott}, T., {Abdalla}, F.~B., {et~al.}
  2016, \mnras, 460, 1270

\bibitem[{{de Jong} {et~al.}(2013){de Jong}, {Verdoes Kleijn}, {Kuijken}, \&
  {Valentijn}}]{2013ExA....35...25D}
{de Jong}, J.~T.~A., {Verdoes Kleijn}, G.~A., {Kuijken}, K.~H., \& {Valentijn},
  E.~A. 2013, Experimental Astronomy, 35, 25

\bibitem[{{Feroz} \& {Hobson}(2008)}]{2008MNRAS.384..449F}
{Feroz}, F., \& {Hobson}, M.~P. 2008, \mnras, 384, 449

\bibitem[{{Feroz} {et~al.}(2009){Feroz}, {Hobson}, \&
  {Bridges}}]{2009MNRAS.398.1601F}
{Feroz}, F., {Hobson}, M.~P., \& {Bridges}, M. 2009, \mnras, 398, 1601

\bibitem[{{Feroz} {et~al.}(2019){Feroz}, {Hobson}, {Cameron}, \&
  {Pettitt}}]{2019OJAp....2E..10F}
{Feroz}, F., {Hobson}, M.~P., {Cameron}, E., \& {Pettitt}, A.~N. 2019, The Open
  Journal of Astrophysics, 2, 10

\bibitem[{{Hamana} {et~al.}(2020){Hamana}, {Shirasaki}, {Miyazaki}, {Hikage},
  {Oguri}, {More}, {Armstrong}, {Leauthaud}, {Mandelbaum}, {Miyatake},
  {Nishizawa}, {Simet}, {Takada}, {Aihara}, {Bosch}, {Komiyama}, {Lupton},
  {Murayama}, {Strauss}, \& {Tanaka}}]{2020PASJ...72...16H}
{Hamana}, T., {Shirasaki}, M., {Miyazaki}, S., {et~al.} 2020, \pasj, 72, 16

\bibitem[{{Harnois-D{\'e}raps} {et~al.}(2015){Harnois-D{\'e}raps}, {van
  Waerbeke}, {Viola}, \& {Heymans}}]{2015MNRAS.450.1212H}
{Harnois-D{\'e}raps}, J., {van Waerbeke}, L., {Viola}, M., \& {Heymans}, C.
  2015, \mnras, 450, 1212

\bibitem[{{Hartlap} {et~al.}(2007){Hartlap}, {Simon}, \&
  {Schneider}}]{2007A&A...464..399H}
{Hartlap}, J., {Simon}, P., \& {Schneider}, P. 2007, \aap, 464, 399

\bibitem[{{Hellwing} {et~al.}(2016){Hellwing}, {Schaller}, {Frenk}, {Theuns},
  {Schaye}, {Bower}, \& {Crain}}]{2016MNRAS.461L..11H}
{Hellwing}, W.~A., {Schaller}, M., {Frenk}, C.~S., {et~al.} 2016, \mnras, 461,
  L11

\bibitem[{{Hikage} {et~al.}(2019){Hikage}, {Oguri}, {Hamana}, {More},
  {Mandelbaum}, {Takada}, {K{\"o}hlinger}, {Miyatake}, {Nishizawa}, {Aihara},
  {Armstrong}, {Bosch}, {Coupon}, {Ducout}, {Ho}, {Hsieh}, {Komiyama},
  {Lanusse}, {Leauthaud}, {Lupton}, {Medezinski}, {Mineo}, {Miyama},
  {Miyazaki}, {Murata}, {Murayama}, {Shirasaki}, {Sif{\'o}n}, {Simet},
  {Speagle}, {Spergel}, {Strauss}, {Sugiyama}, {Tanaka}, {Utsumi}, {Wang}, \&
  {Yamada}}]{2019PASJ...71...43H}
{Hikage}, C., {Oguri}, M., {Hamana}, T., {et~al.} 2019, \pasj, 71, 43

\bibitem[{{Hildebrandt} {et~al.}(2017){Hildebrandt}, {Viola}, {Heymans},
  {Joudaki}, {Kuijken}, {Blake}, {Erben}, {Joachimi}, {Klaes}, {Miller},
  {Morrison}, {Nakajima}, {Verdoes Kleijn}, {Amon}, {Choi}, {Covone}, {de
  Jong}, {Dvornik}, {Fenech Conti}, {Grado}, {Harnois-D{\'e}raps}, {Herbonnet},
  {Hoekstra}, {K{\"o}hlinger}, {McFarland}, {Mead}, {Merten}, {Napolitano},
  {Peacock}, {Radovich}, {Schneider}, {Simon}, {Valentijn}, {van den Busch},
  {van Uitert}, \& {Van Waerbeke}}]{2017MNRAS.465.1454H}
{Hildebrandt}, H., {Viola}, M., {Heymans}, C., {et~al.} 2017, \mnras, 465, 1454

\bibitem[{{Hildebrandt} {et~al.}(2020){Hildebrandt}, {K{\"o}hlinger}, {van den
  Busch}, {Joachimi}, {Heymans}, {Kannawadi}, {Wright}, {Asgari}, {Blake},
  {Hoekstra}, {Joudaki}, {Kuijken}, {Miller}, {Morrison}, {Tr{\"o}ster},
  {Amon}, {Archidiacono}, {Brieden}, {Choi}, {de Jong}, {Erben}, {Giblin},
  {Mead}, {Peacock}, {Radovich}, {Schneider}, {Sif{\'o}n}, \&
  {Tewes}}]{2020A&A...633A..69H}
{Hildebrandt}, H., {K{\"o}hlinger}, F., {van den Busch}, J.~L., {et~al.} 2020,
  \aap, 633, A69

\bibitem[{{Hinshaw} {et~al.}(2013){Hinshaw}, {Larson}, {Komatsu}, {Spergel},
  {Bennett}, {Dunkley}, {Nolta}, {Halpern}, {Hill}, {Odegard}, {Page}, {Smith},
  {Weiland}, {Gold}, {Jarosik}, {Kogut}, {Limon}, {Meyer}, {Tucker}, {Wollack},
  \& {Wright}}]{2013ApJS..208...19H}
{Hinshaw}, G., {Larson}, D., {Komatsu}, E., {et~al.} 2013, \apjs, 208, 19

\bibitem[{{Hirata} \& {Seljak}(2003)}]{2003MNRAS.343..459H}
{Hirata}, C., \& {Seljak}, U. 2003, \mnras, 343, 459

\bibitem[{{Hirata} \& {Seljak}(2004)}]{2004PhRvD..70f3526H}
{Hirata}, C.~M., \& {Seljak}, U. 2004, \prd, 70, 063526

\bibitem[{{Ilbert} {et~al.}(2009){Ilbert}, {Capak}, {Salvato}, {Aussel},
  {McCracken}, {Sanders}, {Scoville}, {Kartaltepe}, {Arnouts}, {Le Floc'h},
  {Mobasher}, {Taniguchi}, {Lamareille}, {Leauthaud}, {Sasaki}, {Thompson},
  {Zamojski}, {Zamorani}, {Bardelli}, {Bolzonella}, {Bongiorno}, {Brusa},
  {Caputi}, {Carollo}, {Contini}, {Cook}, {Coppa}, {Cucciati}, {de la Torre},
  {de Ravel}, {Franzetti}, {Garilli}, {Hasinger}, {Iovino}, {Kampczyk},
  {Kneib}, {Knobel}, {Kovac}, {Le Borgne}, {Le Brun}, {F{\`e}vre}, {Lilly},
  {Looper}, {Maier}, {Mainieri}, {Mellier}, {Mignoli}, {Murayama}, {Pell{\`o}},
  {Peng}, {P{\'e}rez-Montero}, {Renzini}, {Ricciardelli}, {Schiminovich},
  {Scodeggio}, {Shioya}, {Silverman}, {Surace}, {Tanaka}, {Tasca}, {Tresse},
  {Vergani}, \& {Zucca}}]{2009ApJ...690.1236I}
{Ilbert}, O., {Capak}, P., {Salvato}, M., {et~al.} 2009, \apj, 690, 1236

\bibitem[{{Joachimi} {et~al.}(2011){Joachimi}, {Mandelbaum}, {Abdalla}, \&
  {Bridle}}]{2011A&A...527A..26J}
{Joachimi}, B., {Mandelbaum}, R., {Abdalla}, F.~B., \& {Bridle}, S.~L. 2011,
  \aap, 527, A26

\bibitem[{{Kilbinger}(2015)}]{2015RPPh...78h6901K}
{Kilbinger}, M. 2015, Reports on Progress in Physics, 78, 086901

\bibitem[{{Kirk} {et~al.}(2015){Kirk}, {Brown}, {Hoekstra}, {Joachimi},
  {Kitching}, {Mandelbaum}, {Sif{\'o}n}, {Cacciato}, {Choi}, {Kiessling},
  {Leonard}, {Rassat}, \& {Sch{\"a}fer}}]{2015SSRv..193..139K}
{Kirk}, D., {Brown}, M.~L., {Hoekstra}, H., {et~al.} 2015, \ssr, 193, 139

\bibitem[{{K{\"o}hlinger} {et~al.}(2017){K{\"o}hlinger}, {Viola}, {Joachimi},
  {Hoekstra}, {van Uitert}, {Hildebrandt}, {Choi}, {Erben}, {Heymans},
  {Joudaki}, {Klaes}, {Kuijken}, {Merten}, {Miller}, {Schneider}, \&
  {Valentijn}}]{2017MNRAS.471.4412K}
{K{\"o}hlinger}, F., {Viola}, M., {Joachimi}, B., {et~al.} 2017, \mnras, 471,
  4412

\bibitem[{{Laigle} {et~al.}(2016){Laigle}, {McCracken}, {Ilbert}, {Hsieh},
  {Davidzon}, {Capak}, {Hasinger}, {Silverman}, {Pichon}, {Coupon}, {Aussel},
  {Le Borgne}, {Caputi}, {Cassata}, {Chang}, {Civano}, {Dunlop}, {Fynbo},
  {Kartaltepe}, {Koekemoer}, {Le F{\`e}vre}, {Le Floc'h}, {Leauthaud}, {Lilly},
  {Lin}, {Marchesi}, {Milvang-Jensen}, {Salvato}, {Sanders}, {Scoville},
  {Smolcic}, {Stockmann}, {Taniguchi}, {Tasca}, {Toft}, {Vaccari}, \&
  {Zabl}}]{2016ApJS..224...24L}
{Laigle}, C., {McCracken}, H.~J., {Ilbert}, O., {et~al.} 2016, \apjs, 224, 24

\bibitem[{{Lesgourgues} {et~al.}(2013){Lesgourgues}, {Mangano}, {Miele}, \&
  {Pastor}}]{2013neco.book.....L}
{Lesgourgues}, J., {Mangano}, G., {Miele}, G., \& {Pastor}, S. 2013, {Neutrino
  Cosmology}

\bibitem[{{Mandelbaum}(2018)}]{2018ARA&A..56..393M}
{Mandelbaum}, R. 2018, \araa, 56, 393

\bibitem[{{Mandelbaum} {et~al.}(2018{\natexlab{a}}){Mandelbaum}, {Miyatake},
  {Hamana}, {Oguri}, {Simet}, {Armstrong}, {Bosch}, {Murata}, {Lanusse},
  {Leauthaud}, {Coupon}, {More}, {Takada}, {Miyazaki}, {Speagle}, {Shirasaki},
  {Sif{\'o}n}, {Huang}, {Nishizawa}, {Medezinski}, {Okura}, {Okabe}, {Czakon},
  {Takahashi}, {Coulton}, {Hikage}, {Komiyama}, {Lupton}, {Strauss}, {Tanaka},
  \& {Utsumi}}]{2018PASJ...70S..25M}
{Mandelbaum}, R., {Miyatake}, H., {Hamana}, T., {et~al.} 2018{\natexlab{a}},
  \pasj, 70, S25

\bibitem[{{Mandelbaum} {et~al.}(2018{\natexlab{b}}){Mandelbaum}, {Lanusse},
  {Leauthaud}, {Armstrong}, {Simet}, {Miyatake}, {Meyers}, {Bosch}, {Murata},
  {Miyazaki}, \& {Tanaka}}]{2018MNRAS.481.3170M}
{Mandelbaum}, R., {Lanusse}, F., {Leauthaud}, A., {et~al.} 2018{\natexlab{b}},
  \mnras, 481, 3170

\bibitem[{{McCarthy} {et~al.}(2017){McCarthy}, {Schaye}, {Bird}, \& {Le
  Brun}}]{2017MNRAS.465.2936M}
{McCarthy}, I.~G., {Schaye}, J., {Bird}, S., \& {Le Brun}, A. M.~C. 2017,
  \mnras, 465, 2936

\bibitem[{{Mead} {et~al.}(2015){Mead}, {Peacock}, {Heymans}, {Joudaki}, \&
  {Heavens}}]{2015MNRAS.454.1958M}
{Mead}, A.~J., {Peacock}, J.~A., {Heymans}, C., {Joudaki}, S., \& {Heavens},
  A.~F. 2015, \mnras, 454, 1958

\bibitem[{{Oguri} {et~al.}(2018){Oguri}, {Miyazaki}, {Hikage}, {Mandelbaum},
  {Utsumi}, {Miyatake}, {Takada}, {Armstrong}, {Bosch}, {Komiyama},
  {Leauthaud}, {More}, {Nishizawa}, {Okabe}, \& {Tanaka}}]{2018PASJ...70S..26O}
{Oguri}, M., {Miyazaki}, S., {Hikage}, C., {et~al.} 2018, \pasj, 70, S26

\bibitem[{{Planck Collaboration} {et~al.}(2020{\natexlab{a}}){Planck
  Collaboration}, {Aghanim}, {Akrami}, {Ashdown}, {Aumont}, {Baccigalupi},
  {Ballardini}, {Banday}, {Barreiro}, {Bartolo}, {Basak}, {Battye}, {Benabed},
  {Bernard}, {Bersanelli}, {Bielewicz}, {Bock}, {Bond}, {Borrill}, {Bouchet},
  {Boulanger}, {Bucher}, {Burigana}, {Butler}, {Calabrese}, {Cardoso},
  {Carron}, {Challinor}, {Chiang}, {Chluba}, {Colombo}, {Combet}, {Contreras},
  {Crill}, {Cuttaia}, {de Bernardis}, {de Zotti}, {Delabrouille}, {Delouis},
  {Di Valentino}, {Diego}, {Dor{\'e}}, {Douspis}, {Ducout}, {Dupac}, {Dusini},
  {Efstathiou}, {Elsner}, {En{\ss}lin}, {Eriksen}, {Fantaye}, {Farhang},
  {Fergusson}, {Fernandez-Cobos}, {Finelli}, {Forastieri}, {Frailis},
  {Fraisse}, {Franceschi}, {Frolov}, {Galeotta}, {Galli}, {Ganga},
  {G{\'e}nova-Santos}, {Gerbino}, {Ghosh}, {Gonz{\'a}lez-Nuevo}, {G{\'o}rski},
  {Gratton}, {Gruppuso}, {Gudmundsson}, {Hamann}, {Handley}, {Hansen},
  {Herranz}, {Hildebrandt}, {Hivon}, {Huang}, {Jaffe}, {Jones}, {Karakci},
  {Keih{\"a}nen}, {Keskitalo}, {Kiiveri}, {Kim}, {Kisner}, {Knox},
  {Krachmalnicoff}, {Kunz}, {Kurki-Suonio}, {Lagache}, {Lamarre}, {Lasenby},
  {Lattanzi}, {Lawrence}, {Le Jeune}, {Lemos}, {Lesgourgues}, {Levrier},
  {Lewis}, {Liguori}, {Lilje}, {Lilley}, {Lindholm}, {L{\'o}pez-Caniego},
  {Lubin}, {Ma}, {Mac{\'\i}as-P{\'e}rez}, {Maggio}, {Maino}, {Mandolesi},
  {Mangilli}, {Marcos-Caballero}, {Maris}, {Martin}, {Martinelli},
  {Mart{\'\i}nez-Gonz{\'a}lez}, {Matarrese}, {Mauri}, {McEwen}, {Meinhold},
  {Melchiorri}, {Mennella}, {Migliaccio}, {Millea}, {Mitra},
  {Miville-Desch{\^e}nes}, {Molinari}, {Montier}, {Morgante}, {Moss}, {Natoli},
  {N{\o}rgaard-Nielsen}, {Pagano}, {Paoletti}, {Partridge}, {Patanchon},
  {Peiris}, {Perrotta}, {Pettorino}, {Piacentini}, {Polastri}, {Polenta},
  {Puget}, {Rachen}, {Reinecke}, {Remazeilles}, {Renzi}, {Rocha}, {Rosset},
  {Roudier}, {Rubi{\~n}o-Mart{\'\i}n}, {Ruiz-Granados}, {Salvati}, {Sandri},
  {Savelainen}, {Scott}, {Shellard}, {Sirignano}, {Sirri}, {Spencer},
  {Sunyaev}, {Suur-Uski}, {Tauber}, {Tavagnacco}, {Tenti}, {Toffolatti},
  {Tomasi}, {Trombetti}, {Valenziano}, {Valiviita}, {Van Tent}, {Vibert},
  {Vielva}, {Villa}, {Vittorio}, {Wandelt}, {Wehus}, {White}, {White},
  {Zacchei}, \& {Zonca}}]{2020A&A...641A...6P}
{Planck Collaboration}, {Aghanim}, N., {Akrami}, Y., {et~al.}
  2020{\natexlab{a}}, \aap, 641, A6

\bibitem[{{Planck Collaboration} {et~al.}(2020{\natexlab{b}}){Planck
  Collaboration}, {Aghanim}, {Akrami}, {Ashdown}, {Aumont}, {Baccigalupi},
  {Ballardini}, {Banday}, {Barreiro}, {Bartolo}, {Basak}, {Benabed}, {Bernard},
  {Bersanelli}, {Bielewicz}, {Bock}, {Bond}, {Borrill}, {Bouchet}, {Boulanger},
  {Bucher}, {Burigana}, {Calabrese}, {Cardoso}, {Carron}, {Challinor},
  {Chiang}, {Colombo}, {Combet}, {Crill}, {Cuttaia}, {de Bernardis}, {de
  Zotti}, {Delabrouille}, {Di Valentino}, {Diego}, {Dor{\'e}}, {Douspis},
  {Ducout}, {Dupac}, {Efstathiou}, {Elsner}, {En{\ss}lin}, {Eriksen},
  {Fantaye}, {Fernandez-Cobos}, {Finelli}, {Forastieri}, {Frailis}, {Fraisse},
  {Franceschi}, {Frolov}, {Galeotta}, {Galli}, {Ganga}, {G{\'e}nova-Santos},
  {Gerbino}, {Ghosh}, {Gonz{\'a}lez-Nuevo}, {G{\'o}rski}, {Gratton},
  {Gruppuso}, {Gudmundsson}, {Hamann}, {Handley}, {Hansen}, {Herranz}, {Hivon},
  {Huang}, {Jaffe}, {Jones}, {Karakci}, {Keih{\"a}nen}, {Keskitalo}, {Kiiveri},
  {Kim}, {Knox}, {Krachmalnicoff}, {Kunz}, {Kurki-Suonio}, {Lagache},
  {Lamarre}, {Lasenby}, {Lattanzi}, {Lawrence}, {Le Jeune}, {Levrier}, {Lewis},
  {Liguori}, {Lilje}, {Lindholm}, {L{\'o}pez-Caniego}, {Lubin}, {Ma},
  {Mac{\'\i}as-P{\'e}rez}, {Maggio}, {Maino}, {Mandolesi}, {Mangilli},
  {Marcos-Caballero}, {Maris}, {Martin}, {Mart{\'\i}nez-Gonz{\'a}lez},
  {Matarrese}, {Mauri}, {McEwen}, {Melchiorri}, {Mennella}, {Migliaccio},
  {Miville-Desch{\^e}nes}, {Molinari}, {Moneti}, {Montier}, {Morgante}, {Moss},
  {Natoli}, {Pagano}, {Paoletti}, {Partridge}, {Patanchon}, {Perrotta},
  {Pettorino}, {Piacentini}, {Polastri}, {Polenta}, {Puget}, {Rachen},
  {Reinecke}, {Remazeilles}, {Renzi}, {Rocha}, {Rosset}, {Roudier},
  {Rubi{\~n}o-Mart{\'\i}n}, {Ruiz-Granados}, {Salvati}, {Sandri}, {Savelainen},
  {Scott}, {Sirignano}, {Sunyaev}, {Suur-Uski}, {Tauber}, {Tavagnacco},
  {Tenti}, {Toffolatti}, {Tomasi}, {Trombetti}, {Valiviita}, {Van Tent},
  {Vielva}, {Villa}, {Vittorio}, {Wandelt}, {Wehus}, {White}, {White},
  {Zacchei}, \& {Zonca}}]{2020A&A...641A...8P}
---. 2020{\natexlab{b}}, \aap, 641, A8

\bibitem[{{Raveri} \& {Hu}(2019)}]{2019PhRvD..99d3506R}
{Raveri}, M., \& {Hu}, W. 2019, \prd, 99, 043506

\bibitem[{{Schaye} {et~al.}(2010){Schaye}, {Dalla Vecchia}, {Booth}, {Wiersma},
  {Theuns}, {Haas}, {Bertone}, {Duffy}, {McCarthy}, \& {van de
  Voort}}]{2010MNRAS.402.1536S}
{Schaye}, J., {Dalla Vecchia}, C., {Booth}, C.~M., {et~al.} 2010, \mnras, 402,
  1536

\bibitem[{{Schneider} {et~al.}(2010){Schneider}, {Eifler}, \&
  {Krause}}]{2010A&A...520A.116S}
{Schneider}, P., {Eifler}, T., \& {Krause}, E. 2010, \aap, 520, A116

\bibitem[{{Schneider} {et~al.}(2002){Schneider}, {van Waerbeke}, {Kilbinger},
  \& {Mellier}}]{2002A&A...396....1S}
{Schneider}, P., {van Waerbeke}, L., {Kilbinger}, M., \& {Mellier}, Y. 2002,
  \aap, 396, 1

\bibitem[{{Shirasaki} {et~al.}(2019){Shirasaki}, {Hamana}, {Takada},
  {Takahashi}, \& {Miyatake}}]{2019MNRAS.486...52S}
{Shirasaki}, M., {Hamana}, T., {Takada}, M., {Takahashi}, R., \& {Miyatake}, H.
  2019, \mnras, 486, 52

\bibitem[{{Smith} {et~al.}(2003){Smith}, {Peacock}, {Jenkins}, {White},
  {Frenk}, {Pearce}, {Thomas}, {Efstathiou}, \&
  {Couchman}}]{2003MNRAS.341.1311S}
{Smith}, R.~E., {Peacock}, J.~A., {Jenkins}, A., {et~al.} 2003, \mnras, 341,
  1311

\bibitem[{{Springel} {et~al.}(2018){Springel}, {Pakmor}, {Pillepich},
  {Weinberger}, {Nelson}, {Hernquist}, {Vogelsberger}, {Genel}, {Torrey},
  {Marinacci}, \& {Naiman}}]{2018MNRAS.475..676S}
{Springel}, V., {Pakmor}, R., {Pillepich}, A., {et~al.} 2018, \mnras, 475, 676

\bibitem[{{Takahashi} {et~al.}(2017){Takahashi}, {Hamana}, {Shirasaki},
  {Namikawa}, {Nishimichi}, {Osato}, \& {Shiroyama}}]{2017ApJ...850...24T}
{Takahashi}, R., {Hamana}, T., {Shirasaki}, M., {et~al.} 2017, \apj, 850, 24

\bibitem[{{Takahashi} {et~al.}(2012){Takahashi}, {Sato}, {Nishimichi},
  {Taruya}, \& {Oguri}}]{2012ApJ...761..152T}
{Takahashi}, R., {Sato}, M., {Nishimichi}, T., {Taruya}, A., \& {Oguri}, M.
  2012, \apj, 761, 152

\bibitem[{{Tanaka} {et~al.}(2018){Tanaka}, {Coupon}, {Hsieh}, {Mineo},
  {Nishizawa}, {Speagle}, {Furusawa}, {Miyazaki}, \&
  {Murayama}}]{2018PASJ...70S...9T}
{Tanaka}, M., {Coupon}, J., {Hsieh}, B.-C., {et~al.} 2018, \pasj, 70, S9

\bibitem[{{Troxel} \& {Ishak}(2015)}]{2015PhR...558....1T}
{Troxel}, M.~A., \& {Ishak}, M. 2015, \physrep, 558, 1

\bibitem[{{Troxel} {et~al.}(2018){Troxel}, {MacCrann}, {Zuntz}, {Eifler},
  {Krause}, {Dodelson}, {Gruen}, {Blazek}, {Friedrich}, {Samuroff}, {Prat},
  {Secco}, {Davis}, {Fert{\'e}}, {DeRose}, {Alarcon}, {Amara}, {Baxter},
  {Becker}, {Bernstein}, {Bridle}, {Cawthon}, {Chang}, {Choi}, {De Vicente},
  {Drlica-Wagner}, {Elvin-Poole}, {Frieman}, {Gatti}, {Hartley}, {Honscheid},
  {Hoyle}, {Huff}, {Huterer}, {Jain}, {Jarvis}, {Kacprzak}, {Kirk}, {Kokron},
  {Krawiec}, {Lahav}, {Liddle}, {Peacock}, {Rau}, {Refregier}, {Rollins},
  {Rozo}, {Rykoff}, {S{\'a}nchez}, {Sevilla-Noarbe}, {Sheldon}, {Stebbins},
  {Varga}, {Vielzeuf}, {Wang}, {Wechsler}, {Yanny}, {Abbott}, {Abdalla},
  {Allam}, {Annis}, {Bechtol}, {Benoit-L{\'e}vy}, {Bertin}, {Brooks},
  {Buckley-Geer}, {Burke}, {Carnero Rosell}, {Carrasco Kind}, {Carretero},
  {Castander}, {Crocce}, {Cunha}, {D'Andrea}, {da Costa}, {DePoy}, {Desai},
  {Diehl}, {Dietrich}, {Doel}, {Fernandez}, {Flaugher}, {Fosalba},
  {Garc{\'{\i}}a-Bellido}, {Gaztanaga}, {Gerdes}, {Giannantonio}, {Goldstein},
  {Gruendl}, {Gschwend}, {Gutierrez}, {James}, {Jeltema}, {Johnson}, {Johnson},
  {Kent}, {Kuehn}, {Kuhlmann}, {Kuropatkin}, {Li}, {Lima}, {Lin}, {Maia},
  {March}, {Marshall}, {Martini}, {Melchior}, {Menanteau}, {Miquel}, {Mohr},
  {Neilsen}, {Nichol}, {Nord}, {Petravick}, {Plazas}, {Romer}, {Roodman},
  {Sako}, {Sanchez}, {Scarpine}, {Schindler}, {Schubnell}, {Smith}, {Smith},
  {Soares-Santos}, {Sobreira}, {Suchyta}, {Swanson}, {Tarle}, {Thomas},
  {Tucker}, {Vikram}, {Walker}, {Weller}, {Zhang}, \& {DES
  Collaboration}}]{2018PhRvD..98d3528T}
{Troxel}, M.~A., {MacCrann}, N., {Zuntz}, J., {et~al.} 2018, \prd, 98, 043528

\bibitem[{{van Daalen} {et~al.}(2011){van Daalen}, {Schaye}, {Booth}, \& {Dalla
  Vecchia}}]{2011MNRAS.415.3649V}
{van Daalen}, M.~P., {Schaye}, J., {Booth}, C.~M., \& {Dalla Vecchia}, C. 2011,
  \mnras, 415, 3649

\end{thebibliography}

\appendix

%
%
\section{Test of the measurements method of COSEBIs using mock catalogs}
\label{sec:test_measurement}
%

%
%
\begin{figure}
  \begin{center}
  \includegraphics[width=82mm]{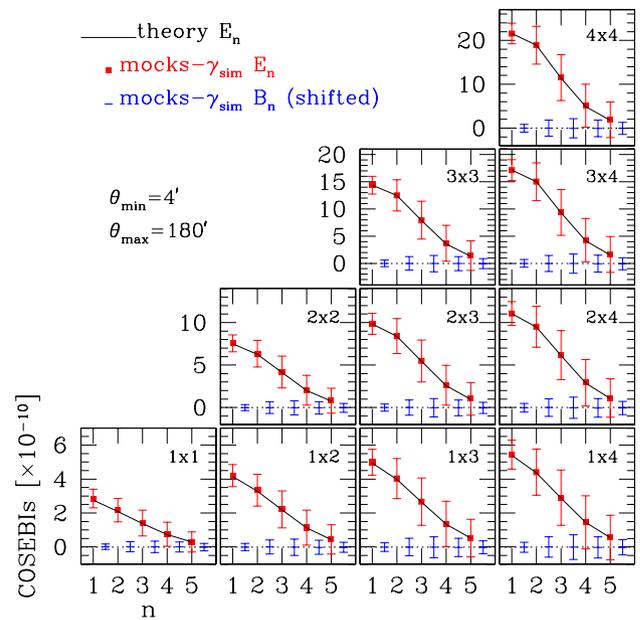}
\end{center}
\caption{E/B-mode COSEBIs of the same redshift bin combinations as
  the HSC-Y1 measurements (shown in Figure~\ref{fig:cosebis_best})
  measured from HSC-Y1 mock catalogs are compared with the theoretical
  predictions.
  In these mock measurements, the cosmic shear signals without
  shape-noise components are used.
  Means and root-mean-squares (RMSs) of COSEBIs signals computed from
  2268 HSC-Y1 mock catalogs are shown by red filled squares and blue bars
  for E- and B-mode, respectively.
  B-mode signals are shifted to right by 0.5 for clarify.
  The black lines show the theoretical predictions of E-mode COSEBIs
  based on WMAP9 cosmological model \citep{2013ApJS..208...19H} which
  is adopted in the ray-tracing simulation \citep{2017ApJ...850...24T}.
  \label{fig:mock_en_bn_gsim}}
\end{figure}

Here we present results of an accuracy test of our method of
  measuring COSEBIs signals, which is described in
section~\ref{sec:measurements_cosebis}.
To do this, we use 2268 HSC-Y1 mock catalogs but
ignoring shape-noise \citep{2019MNRAS.486...52S}.
We measure E/B-mode COSEBIs from each of 2268 mock catalogs using the
same method as the real HSC-Y1 COSEBIs measurements (see
section~\ref{sec:measurements_cosebis}), and compute means and root-mean-squares
(RMSs) of them.
Since the shape-noise is not included in this mock analysis, the RMSs
represent an expected sample variance for the HSC-Y1 field.
The results are shown in Figure~\ref{fig:mock_en_bn_gsim} along with the
theoretical predictions of E-mode COSEBIs based on WMAP9 cosmological
model \citep{2013ApJS..208...19H}, which is adopted in the ray-tracing
simulation \citep{2017ApJ...850...24T}. It is seen from the figure that
the averaged $E_n$ signals agree with the theoretical predictions well
within the RMSs, and $B_n$ signals are consistent with zero.
Therefore we conclude that the measurement method is accurate enough for
this study.

%
%
\section{Estimations of COSEBIs signals from possible systematics}
\label{sec:cosebis_systematics}
%
 
\subsection{COSEBIs from shapes of PSF and residuals}
\label{sec:cosebis_psf}

%
%
\begin{figure}
  \begin{center}
  \includegraphics[width=82mm]{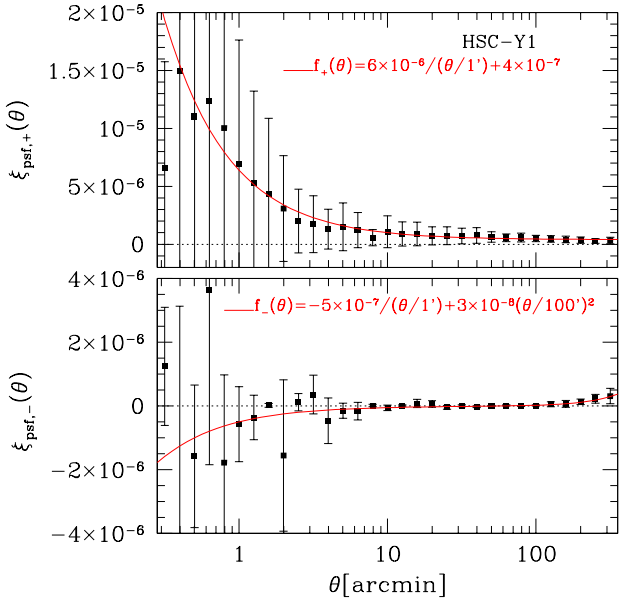}
\end{center}
\caption{Symbols with error-bars show $\xi_{\mbox{psf},\pm}$ defined in
  equation~(\ref{eq:xi_psf}) with measured  $\xi_\pm^{pp}$,
  $\xi_\pm^{pq}$ and $\xi_\pm^{qq}$ from HSC-Y1 data set (see Appendix 2
  of H20). 
    Here $\alpha_{\mbox{psf}}=0.029\pm 0.010$ and
    $\beta_{\mbox{psf}}=-1.42\pm 1.11$ are adopted. Errors shown are
    computed from those of $\alpha_{\mbox{psf}}$ and $\beta_{\mbox{psf}}$.
    The solid curve show the fitting functions (see text for details).
  \label{fig:xi_psf}}
\end{figure}

Here we describe our scheme to estimate COSEBIs signals arising from
residuals in the correction for the point spread function (PSF)
anisotropy in galaxy shapes \citep[see][for a
  review]{2018ARA&A..56..393M}, and present the result.
In fact, systematic tests of the HSC-Y1 shape catalog showed small
residual correlations between galaxy 
shears and PSF shapes \citep{2018PASJ...70S..25M,2018PASJ...70S..26O}, 
which may bias the cosmic shear COSEBIs and our cosmological analysis.

We first estimate TPCFs arising from PSF residuals following the scheme
adopted in H20, which is based on the simple model used by H19 \citep[see
  also][]{2018PhRvD..98d3528T}.
The model assumes that PSF residuals are added to the shear linearly
\begin{equation}
  \label{eq:g_psf}
  \gamma^{\mbox{sys}} =\alpha_{\mbox{psf}} \gamma^p + \beta_{\mbox{psf}} \gamma^q,
\end{equation}
where $\gamma^P$ is the shear\footnote{``Shears'' of stars and
PSFs are converted from the measured distortion using the relation
between them for intrinsically round objects ($\gamma = e/2$).
See \citet{2018PASJ...70S..25M} for the definition of distortion of
star images.} 
of the shape of the model PSF, and $\gamma^q$
is the difference in shears between the PSF model and the true PSF, 
as estimated from the shapes of individual stars, $\gamma^\ast$, i.e.,
$\gamma^q = \gamma^p-\gamma^\ast$.
The first and second terms of the right hand side of 
equation~(\ref{eq:g_psf}) represent the residual
PSF effects from the deconvolution error and the imperfect PSF model,
respectively.
With these terms added to the measured shear $\hat{\gamma}$, the
contributions from these terms to observed TPCFs are written as 
\begin{equation}
  \label{eq:xi_psf}
  \xi_{\mbox{psf},\pm}(\theta) = \alpha_{\mbox{psf}}^2 \xi_\pm^{pp}(\theta)
  +2 \alpha_{\mbox{psf}}\beta_{\mbox{psf}} \xi_\pm^{pq}(\theta)
  +\beta_{\mbox{psf}}^2 \xi_\pm^{qq}(\theta),
\end{equation}
where $\xi_\pm^{pp}$ and $\xi_\pm^{qq}$ represent the
auto-TPCFs of $\gamma^p$ and $\gamma^q$, respectively, and
$\xi_\pm^{pq}$ are the cross-TPCFs of $\gamma^p$ and $\gamma^q$.
Those TPCFs measured from HSC-Y1 data are presented in Figures~20 and 23
of H20, for $\xi_+$ and $\xi_-$, respectively.
The model parameters, $\alpha_{\mbox{psf}}$ and
$\beta_{\mbox{psf}}$, can be estimated by the cross correlation
functions between $\gamma^{p,q}$ and galaxy shears, $\xi^{gp,gq}=\langle
\hat{\gamma} \gamma^{p,q} \rangle$, which are related to $\xi_\pm^{pp,pq,qq}$ as
\begin{eqnarray}
\label{eq:xi_gp_gq}
\xi_{\pm}^{gp}&=&\alpha_{\mbox{psf}}\xi_\pm^{pp}+\beta_{\mbox{psf}}\xi_\pm^{pq},\\
\label{eq:xi_gp_gq2}
\xi_{\pm}^{gq}&=&\alpha_{\mbox{psf}}\xi_\pm^{pq}+\beta_{\mbox{psf}}\xi_\pm^{qq}.
\end{eqnarray}
In measuring these quantities, we use the combined galaxy catalog of the
four tomographic redshift bins, because the measurement of
$\xi_-^{gp,gq}$ is very noisy as shown in Figure~19 of H20. 
As a consequence, we do not take into account possible redshift
dependence of $\alpha_{\mbox{psf}}$ and $\beta_{\mbox{psf}}$.
For $\xi_+$ component, the derived $\alpha_{\mbox{psf}}$ and
$\beta_{\mbox{psf}}$ with equations~(\ref{eq:xi_gp_gq}) and
(\ref{eq:xi_gp_gq2}) are shown in Figure~21 of H20, and their means and
standard deviations were found to be 
$\alpha_{\mbox{psf}}=0.029\pm 0.010$ and
$\beta_{\mbox{psf}}=-1.42\pm 1.11$ (H20).
For $\xi_-$ component, due to very poor signal-to-noise ratios of the 
TPCFs (see Figures~19 and 23 of H20), useful estimated values of
$\alpha_{\mbox{psf}}$ and $\beta_{\mbox{psf}}$ could not be derived (see
Appendix 2 of H20), and thus in the following analysis we adopt the values
obtained from $\xi_+$ component for $\xi_-$ component.
In Figure~\ref{fig:xi_psf}, $\xi_{\mbox{psf},\pm}$ defined in equation~(\ref{eq:xi_psf})
with $\alpha_{\mbox{psf}}=0.029\pm 0.010$ and
$\beta_{\mbox{psf}}=-1.42\pm 1.11$ are shown, where 
error bars are computed from those of
$\alpha_{\mbox{psf}}$ and $\beta_{\mbox{psf}}$.
We derived the fitting functions of those estimates over the
$\theta$-range of our interest as (plotted as solid curves in
Figure~\ref{fig:xi_psf})
\begin{eqnarray}
\label{eq:xip_psf_fit}
\xi_{\mbox{psf},+}^{\rm fit}(\theta) &=& 6\times 10^{-6}
   \left({\theta\over{1'}}\right)^{-1} + 4\times 10^{-7}, \\
\label{eq:xim_psf_fit}
\xi_{\mbox{psf},-}^{\rm fit}(\theta) &=&  -5\times 10^{-7}
   \left({\theta\over{1'}}\right)^{-1} + 3\times 10^{-8} \left({\theta\over{100'}}\right)^2.
\end{eqnarray}
%

%
%
\begin{figure}
\begin{center}
  \includegraphics[width=82mm]{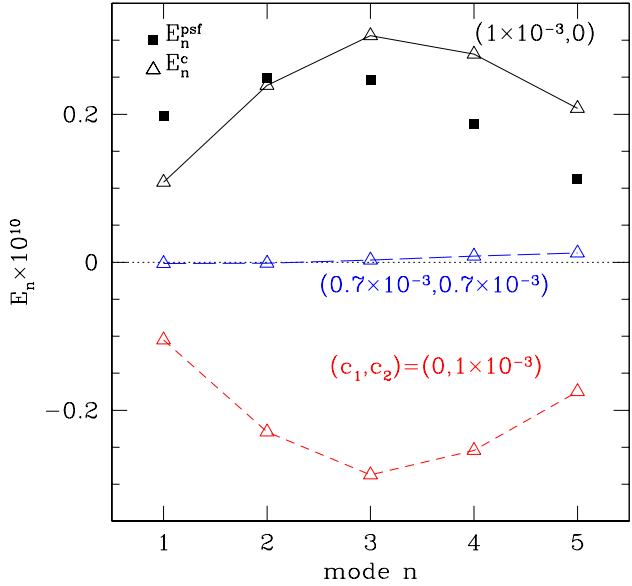}
\end{center}
\caption{Filled squares show E-mode COSEBIs signals expected from PSF
  residuals in HSC-Y1 shape catalog (see Appendix
  \ref{sec:cosebis_psf} for details). Crosses show E-mode COSEBIs
  signals arising from a constant shear of $|c|=10^{-3}$ on positions of
  galaxies used in this study (see Appendix
  \ref{sec:cosebis_constant_shear} for details). Results for three
  cases are plotted; 
  $(c_1,c_2) = (10^{-3},0)$, $(0, 10^{-3})$,
  and  $(1/\sqrt{2}\times10^{-3},1/\sqrt{2}\times10^{-3})$ for the
  solid, dashed, and long-dashed line, respectively. Here results of 
  tomographic redshift bins of $z_3\times z_3$ are shown, but results
  are insensitive to a choice of bins. 
  \label{fig:en_psf_constgam_4t180_s16a}}
\end{figure}

Using the above fitting functions,
we derived E-mode COSEBIs expected 
from PSF residuals using equation~(\ref{eq:cosebis-e}) with the same
scale-cut ($4\arcmin<\theta<180\arcmin$) as one adopted in our cosmological analysis.
The result is shown in Figure~\ref{fig:en_psf_constgam_4t180_s16a} as
filled squares.
Although the derived E-mode COSEBIs should be considered as a rough 
estimate because it is based on the simple linear model,
equations~(\ref{eq:g_psf}) and (\ref{eq:xi_psf}) with
$\alpha_{\mbox{psf}}$ and $\beta_{\mbox{psf}}$ being not well
determined, we may expect that it captures its characteristic shape. 
Therefore, in our cosmological analysis, we take this result as a
reference model, $E_n^{\rm psf-ref}$, and introduce an amplitude
parameter as
\begin{equation}
\label{eq:En-psf-model}
E_n^{\rm psf} = A_{\rm psf} E_n^{\rm psf-ref}.
\end{equation}
We add this term to the theoretical prediction for the E-mode COSEBIs.
We treat $A_{\rm psf}$ as a nuisance parameter (see
Section~\ref{sec:Systematic_parameters}).

\subsection{COSEBIs from the constant shear over a field}
\label{sec:cosebis_constant_shear}

The value of the shear averaged over a field is not expected to be zero
due to the presence of the cosmic shear signal on scales larger than a field. 
However, it could also be non-zero due to residual systematics in the shear 
estimation and/or data reduction process.
The latter, if present, may bias the cosmological inference.
\citet{2018PASJ...70S..25M} examined a mean shear of the HSC-Y1 shape
catalog, and found no evidence of a mean shear above 
that expected from large-scale cosmic shear.
H20 reexamined it for each tomographic galaxy sample and for
each field, and found that measured mean shears are consistent with 
that expected from large-scale cosmic shear (see Appendix 1 of H20). 
Although, as just mentioned above, no clear evidence of an excess mean
shear was found, we check the impact of a constant shear over a field on
our cosmological analysis by modeling it as a redshift-independent
constant shear for simplicity, which we describe below.

COSEBIs signals arising from a constant shear were studies in detail in 
Appendix D of \citet{2021A&A...645A.104A} that we follow here.
We denote two components of a constant shear in the Cartesian coordinates
as $(c_1, c_2)$.
A constant shear generates a separation-independent constant $\xi_+$
term (to be specific, $\xi_+^c=c_1^2+c_2^2$), and a separation-dependent
$\xi_-^c$ term, whose shape
is determined by a field geometry and masks (see equation~(D.3) of
\citet{2021A&A...645A.104A}).
In the transformation from TPCFs to COSEBIs,
equation~(\ref{eq:cosebis-e}), the constant $\xi_+^c$ term is filtered out,
only $\xi_-^c$ term remains, and E-mode COSEBIs is given by equation~(D.6) of
\citet{2021A&A...645A.104A}, 
\begin{eqnarray}
\label{eq:cosebis-e-c}
E_n^c &=& {1\over 2}(c_1^2 - c_2^2) \int_{\theta_{\rm min}}^{\theta_{\rm max}}
d\theta~\theta T_{-n} \langle \cos(4\phi)\rangle(\theta)\nonumber \\
{}&& + c_1 c_2 \int_{\theta_{\rm min}}^{\theta_{\rm max}}
d\theta~\theta T_{-n} \langle \sin(4\phi)\rangle(\theta),
\end{eqnarray}
where $\phi$ is the polar angle of the vector connecting two galaxies in
the Cartesian coordinates, and $\langle ... \rangle(\theta)$ denotes the 
average taken over all pairs of galaxies with a separation $\theta$.

We estimate actual $E_n^c$ signals for HSC-Y1 shear catalog with $|c|=10^{-3}$
(which is a typical value of HSC-Y1 shear catalog, see Figure~18 of H20)
by artificially assigning a constant shear of $(c_1, c_2)$ (but in the sky coordinates, see a
discussion below on this point) to all the galaxies, and doing the same
COSEBIs measurement procedure as done for the real data.
The results are shown in Figure~\ref{fig:en_psf_constgam_4t180_s16a}
for three cases; $(c_1,c_2) = (10^{-3},0)$, $(0, 10^{-3})$,
and  $(1/\sqrt{2}\times10^{-3},1/\sqrt{2}\times10^{-3})$ for the
solid, dashed, and long-dashed line, respectively.
In the plot, results of tomographic redshift bins of $z_3\times z_3$ are shown,
but are insensitive to a choice of bins. 
It is found from these results that the first term of
equation (\ref{eq:cosebis-e-c}) is much larger than the second
term except for cases of $|c_1| \simeq |c_2|$ though $E_n^c$ itself is very
small in such cases. 
Note that from equation~(\ref{eq:cosebis-e-c}), one may expect that
$|E_n^c|$ for cases of $(c_1,c_2) = (10^{-3},0)$ and $(0, 10^{-3})$
should be equal, but not exactly in those actual measurements as can be seen in
Figure~\ref{fig:en_psf_constgam_4t180_s16a}.
The reason for this is our use of the sky coordinates, instead of  the
Cartesian coordinates, in assigning a constant shear; we
simply replace shears of HSC shape catalog which are defined in the sky
coordinates with a constant shear $(c_1, c_2)$.
In this case, equation~(\ref{eq:cosebis-e-c}) does not hold, but is
still useful to understand a basic behavior of $E_n^c$ as each of 6
HSC-Y1 fields is not very wide, and 5 out of 6 fields have a strip-shaped
geometry on the equator (see Figure~1 of \citet{2018PASJ...70S..25M}).

Our modeling of E-mode COSEBIs arising from the constant shear, which we
use to check the impact of a constant shear on
our cosmological analysis, is as follows:
We consider a redshift-independent constant shear for simplicity (see
Appendix~1 of H20 for a discussion on this point).
From the above analysis, we find that the first term of
equation~(\ref{eq:cosebis-e-c}) dominates $E_n^c$ signals, and thus it may
be reasonable to adopt the measured $E_n^c$ for the case of 
$(c_1,c_2) = (10^{-3},0)$ as a reference for modeling.
We denote it as $E_n^{c-\rm ref}$, and introduce an amplitude
parameter as
\begin{equation}
\label{eq:En-c-model}
E_n^{\langle\gamma\rangle} = A_{\langle\gamma\rangle} E_n^{c-\rm ref}.
\end{equation}
We add this term to the theoretical prediction for the E-mode COSEBIs.
We treat $A_{\langle\gamma\rangle}$ as a nuisance parameter (see
Section~\ref{sec:Systematic_parameters}).

%
%
\section{Supplementary figures}
\label{sec:supplementary_figures}

%
%
\begin{figure*}
  \begin{center}
    \includegraphics[height=82mm,angle=-90]{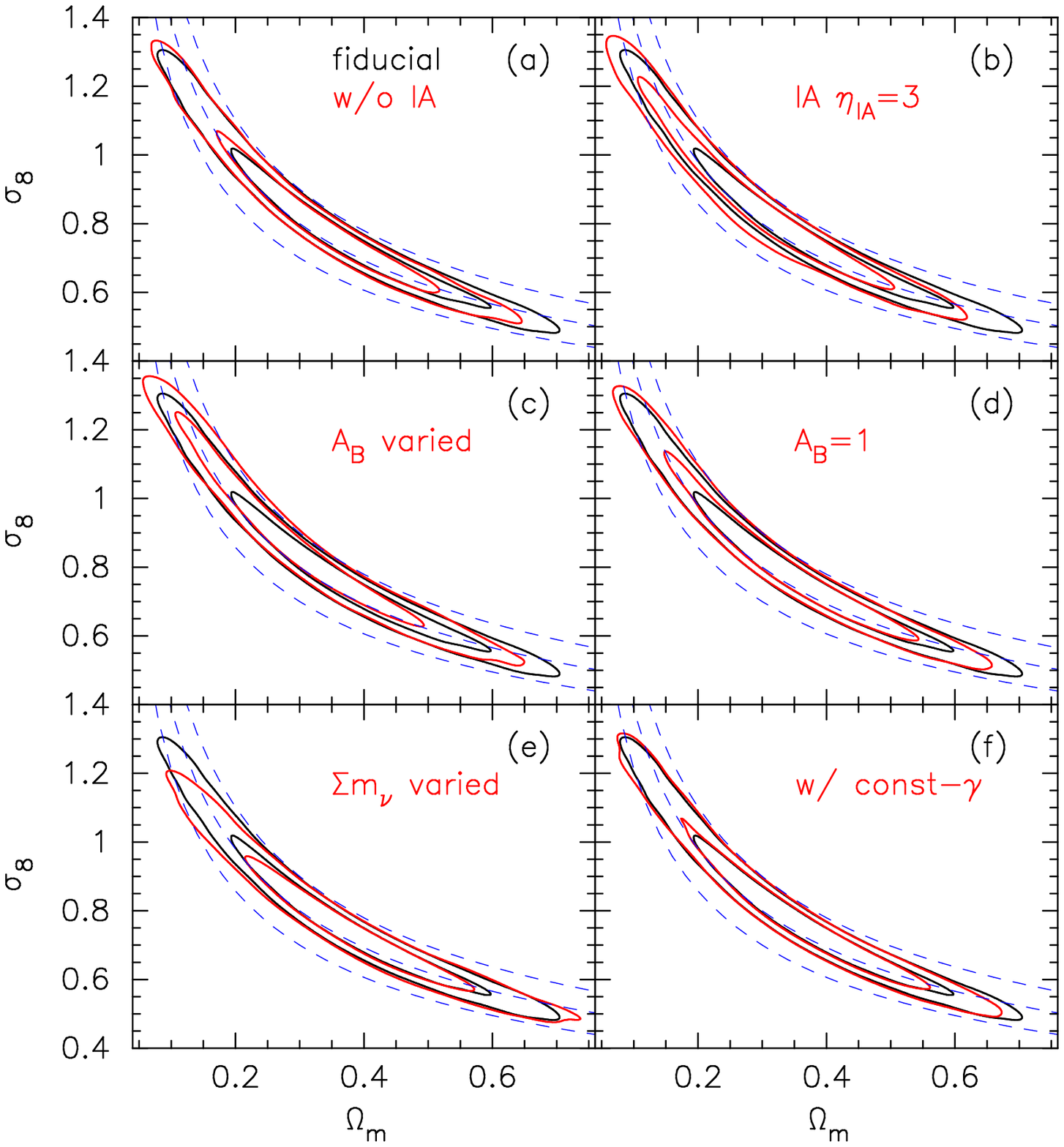}
    \includegraphics[height=82mm,angle=-90]{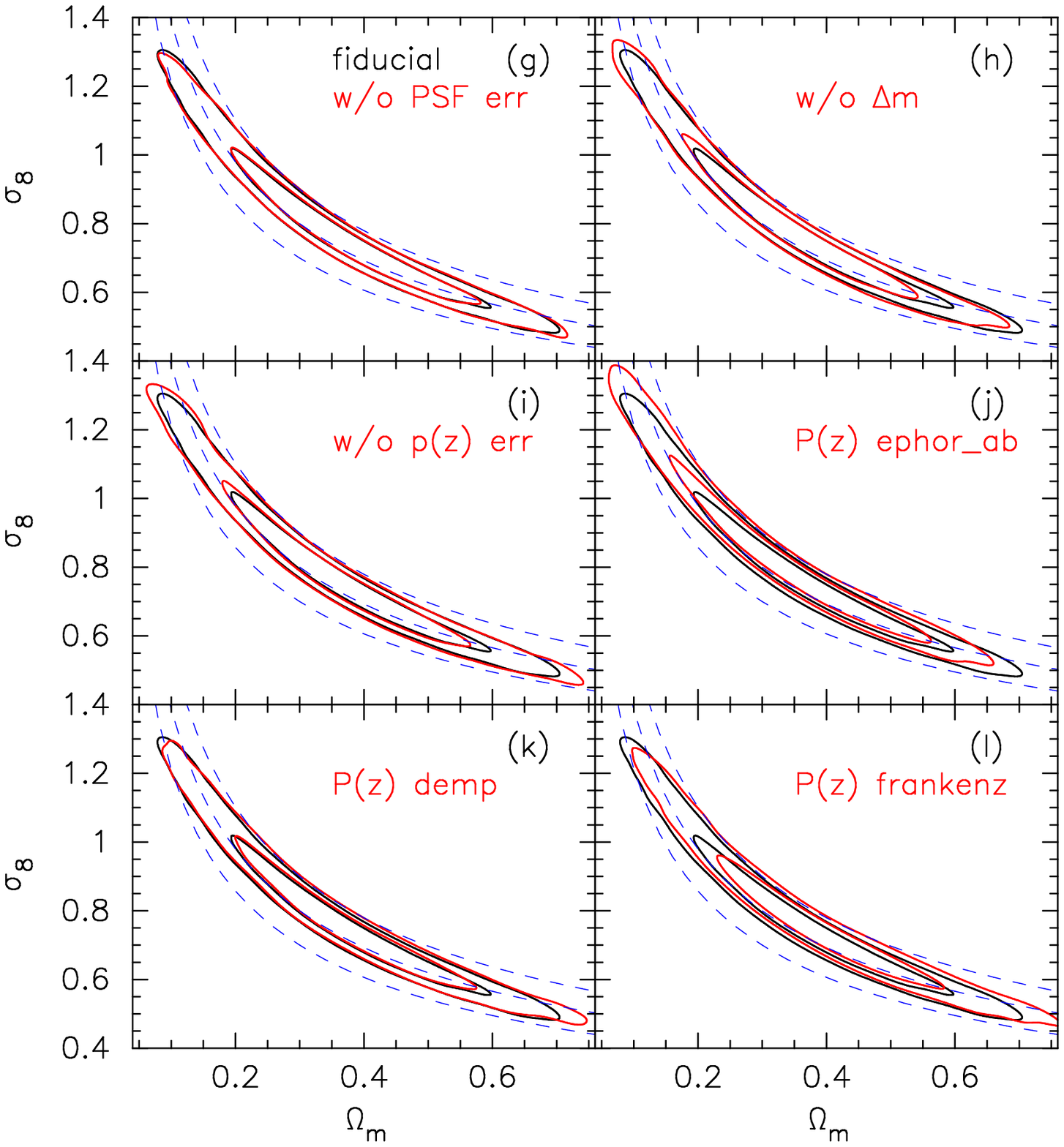}
  \end{center}
  \caption{Comparison of constraints in the
    $\Omega_m$-$\sigma_8$ plane between the fiducial setup (black
    contours) and different assumptions, as described in the text 
    (red contours showing 68\% and 95\% credible levels).
    Dashed blue curves show constant $S_8$ loci 
    ($S_8=0.7$, 0.8 and 0.9 from bottom to top, respectively).
    \label{fig:om_sig8_sys}}
\end{figure*}

%
%
\begin{figure}
\begin{center}
  \includegraphics[height=82mm,angle=-90]{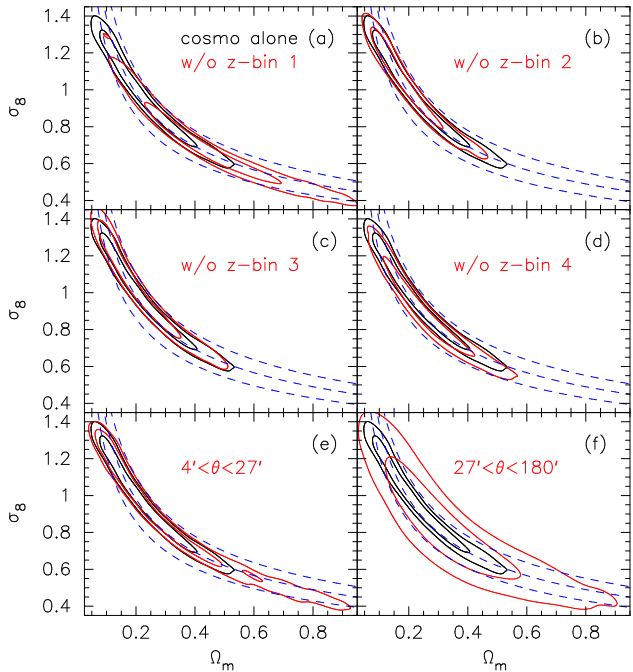}
\end{center}
\caption{Comparison of constraints in the
  $\Omega_m$-$\sigma_8$ plane from the cosmology alone setup (black
  contours) with different setups for internal consistency checks 
  (red contours showing 68\% and 95\% credible levels).
  Dashed blue curves show constant $S_8$ loci 
    ($S_8=0.7$, 0.8 and 0.9 from bottom to top, respectively).
  \label{fig:om_sig8_sys3}}
\end{figure}

%
%
\begin{figure}
\begin{center}
  \includegraphics[height=82mm,angle=-90]{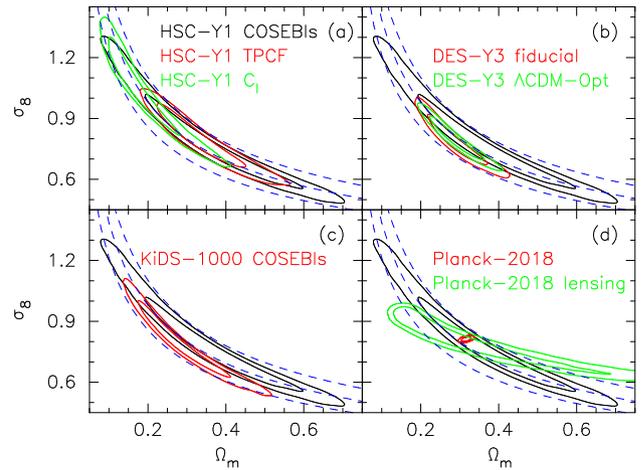}
\end{center}
\caption{Marginalized posterior contours (68\% and 95\%
  credible levels) in the $\Omega_m$-$\sigma_8$ plane.
  Our result from the fiducial $\Lambda$CDM model (black contours) is
  compared with results in the literature (red- or green-line contours):
  (A) HSC-Y1 cosmic shear TPCF (red lines, H20) and PS results
  (green lines, H19).
  (b) Dark Energy Survey Year 1 (DES-Y1) cosmic shear TPCF result
  \citep{2018PhRvD..98d3528T}. Note that in
  Section~\ref{sec:comparison}, we take $S_8$ constraint from DES-Y3
  result \citep{2022PhRvD.105b3514A}, but its chain is not yet available.
  (c) KiDS-1000 cosmic shear COSEBIs result \citep{2021A&A...645A.104A}.
  (d) {\it Planck} 2018 CMB result without
  CMB lensing \citep[][TT+TE+EE+lowE, red lines]{2020A&A...641A...6P},
  and {\it Planck} 2018 lensing result \citep[][green
    lines]{2020A&A...641A...8P}.
  Dashed blue curves show constant $S_8$ loci 
    ($S_8=0.7$, 0.8 and 0.9 from bottom to top, respectively).
  \label{fig:om_sig8_others}}
\end{figure}

In this Appendix, we present supplementary figures for 
systematics tests (Section~\ref{sec:systematics_tests}), internal
consistency checks (Section~\ref{sec:internal_consistency}), and
comparison of our cosmological constraints with ones from other projects
(Section~\ref{sec:comparison}).
Although in those sections we use $S_8=\sigma_8 (\Omega_m/0.3)^\alpha$
with $\alpha=0.5$ for a primary parameter to be constrained by cosmic
shear two-point statistics, 
a single
choice of $\alpha$ does not always provide an optimal description for the
$\sigma_8$-$\Omega_m$ degeneracy, especially for cases with broad
confidence contours such as ours. 
Therefore, in this Appendix, we present marginalized posterior contours 
in the $\Omega_m$-$\sigma_8$ plane.

In Figure~\ref{fig:om_sig8_sys}, we compare constraints from our
fiducial setup with different setups for systematics tests (see
Section~\ref{sec:systematics_tests} for details).
Figure~\ref{fig:om_sig8_sys3} shows the same comparison but between the
cosmology alone setup and different setups for internal consistency
checks (see Section~\ref{sec:internal_consistency}). 

In Figure~\ref{fig:om_sig8_others}, we compare constraints from our
fiducial $\Lambda$CDM model with other results in the literature (see
Section~\ref{sec:comparison}), where
constraints from other studies are derived from publicly available chains.
Note that although different studies adopt different priors, we
do not adjust them to our fiducial setup, but rather use their original
priors. Therefore, part of the difference in the posteriors
may be due to the different choices of priors and modeling.

There is one thing to note about panel (d) of
Figure~\ref{fig:om_sig8_others};
we compare our result  with {\it Planck} 2018 lensing result
\citep{2020A&A...641A...8P}, which is not discussed in the main text
 because it does not provide a useful
constraint on $S_8=\sigma_8(\Omega_m/0.3)^{0.5}$ but provides a tight
constraint on $\sigma_8\Omega_m^{0.25}$ due to a different
degeneracy direction in $\Omega_m$-$\sigma_8$ plane as can be seen from the
panel (d). 
The cosmic microwave background (CMB) lensing is a unique probe
of the distribution of mass in the Universe at intermediate redshifts
(typically $z\sim 2$) and is thus complementary to cosmic shear
statistics as is demonstrated in \citet{2020A&A...641A...8P}.
Such complementarity can also be seen in the panel (d), although a
combined cosmological inference with {\it Planck} results is beyond the
scope of this study. 

\end{document}